\documentclass[amsmath,amssymb,amsfonts,superscriptaddress,nofootinbib,twocolumn,prd]{revtex4-2}

\usepackage{graphicx}
\usepackage{hyperref}
\usepackage{bm}
\usepackage{xcolor}
\usepackage{mathtools}
\usepackage{placeins}
\usepackage{aas_macros}
\usepackage{orcidlink}

\newcommand{\vk}{\mathbf{k}}
\newcommand{\vq}{\mathbf{q}}
\newcommand{\vx}{\mathbf{x}}
\newcommand{\vs}{\mathbf{s}}
\newcommand{\vpsi}{\boldsymbol{\Psi}}
\newcommand{\Nk}{N_{\!k}}

\newcommand{\Nki}{N_{\!k,i}}
\newcommand{\Nkj}{N_{\!k,j}}
\newcommand{\Nsim}{N_{\rm sim}}
\newcommand{\Paa}{P^{aa}}
\newcommand{\Pbb}{P^{bb}}
\newcommand{\Pab}{P^{ab}}
\newcommand{\Cov}{\mathrm{Cov}}
\newcommand{\Var}{\mathrm{Var}}
\newcommand{\beq}{\begin{equation}}
\newcommand{\eeq}{\end{equation}}
\newcommand{\bL}{b_{\!L}}
\newcommand{\AbacusSummit}{\textsc{AbacusSummit}~}

\newcommand{\referee}[1]{\textcolor{black}{#1}}

\begin{document}

\title{Fewer simulations, sharper covariances: Reducing mock covariance noise\\ with Zeldovich approximation control variates}

\author{Boryana Hadzhiyska\orcidlink{0000-0002-2312-3121}}
\affiliation{Institute of Astronomy, Madingley Road, Cambridge, CB3 0HA, UK}
\affiliation{Kavli Institute for Cosmology Cambridge, Madingley Road, Cambridge, CB3 0HA, UK}
\author{Martin White\orcidlink{0000-0001-9912-5070}}
\affiliation{Physics Division, Lawrence Berkeley National Laboratory, Berkeley, CA 94720, USA}
\affiliation{Berkeley Center for Cosmological Physics, Department of Physics,
University of California, Berkeley, CA 94720, USA}

\date{\today}

\begin{abstract}
We present a control-variate method for reducing the variance of power spectrum covariance matrix estimates from simulations of large-scale structure. The key idea is to pair each mock simulation with a cheap Zeldovich-approximation realization sharing the same initial conditions, and to use the known statistical properties of the Zeldovich field to remove correlated sample variance from the covariance estimator. Under a Gaussian disconnected approximation, we derive fully analytic expressions for both the optimal control-variate coefficient, $\beta(k,\ell;k',\ell')$, and the corresponding correlation, $\rho(k,\ell;k',\ell')$, in terms of the auto- and cross-power spectra of the target and control fields. In the monopole case, the correlation takes the particularly simple form $\rho(k,k') = r^2(k),r^2(k')$, where $r(k)$ is the standard cross-correlation coefficient between the target and Zeldovich fields, implying that covariance estimation remains highly efficient whenever the two fields are strongly correlated. 
For masked redshift-space lognormal mocks, resembling Luminous Red Galaxies from the Dark Energy Spectroscopic Instrument (DESI), we find that the control-variate estimator reduces the variance of the covariance matrix by approximately an order of magnitude on large scales, $k \lesssim 0.05\,h\,{\rm Mpc}^{-1}$, precisely where accurate covariance estimation is most challenging.  The gains are smaller for higher $k$ but typically accelerate convergence by a  factor of 2-3, substantially lowering the computational cost of covariance estimation for current and upcoming large-scale structure surveys.
We demonstrate that after a computationally cheap parametric-bootstrap calibration of the precision-matrix bias, the control-variate covariance yields a $\chi^2$ distribution consistent with the expected number of degrees of freedom.
Due to its simplicity, this method is readily implementable in current imaging and spectroscopic surveys (e.g., DESI, \textit{Euclid}, LSST, PFS, \textit{SPHEREx}).
\end{abstract}

\maketitle

\section{Introduction}
\label{sec:intro}

Accurate covariance matrices of summary statistics such as the galaxy power spectrum are essential ingredients in the cosmological inference pipeline.
The covariance matrix enters the likelihood function and directly affects the precision and accuracy of derived cosmological parameters \citep{2013PhRvD..88f3537D,2007A&A...464..399H,2013MNRAS.432.1928T,2014MNRAS.439.2531P,2016MNRAS.456L.132S}.
For current Stage-IV surveys such as DESI \citep{2016arXiv161100036D}, Euclid \citep{2025A&A...697A...1E}, \textit{SPHEREx}\citep{2014arXiv1412.4872D}, Subaru PFS \citep{2014PASJ...66R...1T}, and the Rubin Observatory LSST \citep{2009arXiv0912.0201L}, the statistical power of the data demands covariance estimates with percent-level precision, which in turn requires thousands or even tens of thousands of mock realizations if estimated from simulations.

There are broadly three approaches to covariance estimation.
{Analytic methods} compute the Gaussian and, in some cases, non-Gaussian contributions to the covariance from perturbation theory \citep{1999ApJ...527....1S,1999MNRAS.308.1179M,2017JCAP...11..051B,2020PhRvD.102l3517W,2021PhRvD.103b3501T} or halo model calculations \citep{2002PhR...372....1C,2018A&A...615A...1L}.
These are fast but rely on approximations whose accuracy must be validated, and they struggle to capture the effects of realistic survey geometry or complex procedures such as fiber assignment to targets.
{Sample covariance from simulations} is the most straightforward approach: one runs an ensemble of $\Nsim$ mock catalogs with independent initial conditions, measures the statistic of interest in each, and computes the sample covariance \citep{2009ApJ...700..479T,2019MNRAS.485.2806B,2019MNRAS.482.1786L}.
Provided that the model of the mock catalogs has been tuned properly, this method naturally includes all physical effects (nonlinearity, bias, redshift-space distortions, survey geometry) but converges slowly: the fractional error on each element of the covariance scales as $\sim \sqrt{2/\Nsim}$, and achieving few-percent precision on the full matrix requires $\Nsim \gtrsim \mathcal{O}(10^3\text{--}10^4)$.
{Internal estimators} such as jackknife or bootstrap methods \citep{2009MNRAS.396...19N,2022MNRAS.514.1289M} avoid the need for external mocks but can be biased and are limited by the survey volume.

A number of techniques for reducing the noise of an estimated covariance matrix have been proposed in the literature \cite{2008MNRAS.389..766P,2015MNRAS.454.4326P,2016MNRAS.460.1567P,2017MNRAS.466L..83J,2018MNRAS.473.4150F,2019MNRAS.483..189H,2020PhRvD.102l3521W,2026arXiv260413851F,2026arXiv260618446R}. The {control variate} technique, borrowed from Monte Carlo methods in statistics \citep{10.5555/515699}, offers a way to accelerate the convergence of the sample-covariance approach by exploiting correlations between the target quantity and a related quantity whose expectation value is known. It has potential to yield large improvements at little cost, with full theoretical control.
The basic idea is simple: if $X$ is the estimator of interest and $C$ is a correlated ``control'' estimator with known mean $\mu_C = \langle C \rangle$, then the modified estimator
\beq
Y = X - \beta\,(C - \mu_C)
\eeq
has the same expectation value as $X$ but reduced variance, provided $\beta$ is chosen appropriately and $X$ and $C$ are sufficiently correlated. Intuitively, if we know that $C$ fluctuated above its mean, $\mu_C$, then it is likely that $X$ fluctuated `high' as well, and so it should be corrected down.

This approach was introduced in the cosmological context by Chartier et al. \citep{2021MNRAS.503.1897C,2022MNRAS.509.2220C,2022MNRAS.515.1296C}, who used correlated lower-cost realizations sharing the same initial conditions as control variates for $N$-body simulations.
The analytical version of this method has since been applied to various statistics including the power spectrum multipoles \citep{2022JCAP...09..059K,2023JCAP...02..008D}, bispectrum \citep{2025arXiv251007375K}, post-reconstruction power spectrum \citep{2023OJAp....6E..38H}, and in the context of hybrid theory emulation \citep{2026JCAP...03..078B}.
In all cases, the central challenge is finding a tractable control variate that is highly correlated with the quantity of interest. Determining the control-variate coefficient $\beta$ constitutes an additional challenge. Finally, estimating the mean of the control variate, $\mu_C$, can itself be computationally expensive. Analytical control variate methods can therefore be particularly valuable, since, when available, they provide an effectively noise-free estimate of the expectation value of $C$.

In this work, we develop and test a control-variate framework for reducing the computational cost of estimating large-scale structure covariance matrices from simulations. Our approach uses Zeldovich realizations sharing the same initial conditions as the target mock catalogs in order to remove correlated sample variance from the covariance estimator. We derive analytic expressions for the optimal control-variate coefficient, $\beta$, and the resulting variance reduction factor under a Gaussian disconnected approximation, showing that the performance of the method is governed primarily by the cross-correlation coefficient between the target and control fields (as expected). We validate these predictions against brute-force estimates from large ensembles of simulations and apply the method to a realistic configuration consisting of redshift-space LRG-like lognormal catalogs with a DESI-like survey mask. The resulting target--control cross-correlation closely resembles that measured in realistic DESI LRG mocks, leading to substantial variance reduction on large scales. In particular, we find that the control-variate method reduces the variance of the covariance matrix by approximately an order of magnitude, implying that $\sim 10^3$ paired simulations achieve the statistical precision of $\sim 10^4$ standard realizations.

The structure of this paper is as follows.
Section~\ref{sec:methods} describes our simulation methodology, the control-variate formalism, and the derivation of the analytic $\beta$. In Section~\ref{sec:validation}, we numerically validate our analytical approximations for the CV method. 
Section~\ref{sec:results_rsd_mask} presents our main results in the realistic configuration (redshift space with mask). In Section~\ref{sec:other_metrics}, we study in detail the performance of our CV-reduced covariance matrix and compare it with the standard, non-CV-reduced version.
\referee{Section~\ref{sec:practice} shows, step by step, what an application of the method to an $N$-body mock suite involves.}
We conclude in Section~\ref{sec:conclusions}.

\section{Methods}
\label{sec:methods}

A central goal of this paper is to validate a control-variate (CV) approach for reducing the finite-sample noise of covariance matrices in realistic redshift-space analyses. This validation requires running very large ensembles of paired simulations, since the statistical uncertainty in covariance estimates decreases only slowly with the number of realizations. We therefore use lognormal mocks as a computationally efficient way to generate the high-volume suites needed to test the methodology in a controlled setting.

Although lognormal mocks are an idealization of late-time structure formation, they reproduce the key ingredient for CV performance: the (de)correlation between the control field and the target field. In other words, the level and scale dependence of the decorrelation observed in the lognormal simulations turn out to be quite adequate for our purposes. In fact, in terms of the correlation coefficient $r(k)$, the lognormal mocks have a very similar behavior to what is found for realistic DESI LRG galaxies \citep{2023OJAp....6E..38H}, supporting the use of lognormal realizations as a faithful testbed for covariance-noise reduction.

We emphasize that the lognormal mocks are used here purely as a validation testbed and are not themselves the intended application of the method: their computational cheapness is precisely what allows us to generate the very large ($10^4$) reference ensembles needed to characterize the noise properties of covariance estimators, an exercise that would be prohibitively expensive with $N$-body mocks. In an actual application, the target mocks would be the (expensive) simulations used by a survey, and, as we derive below, the expected performance is governed entirely by the measured target--control cross-correlation coefficient $r(k)$: any target mock suite with a similar $r(k)$ will exhibit similar gains. To further test the sensitivity of the method to the idealizations of the fiducial setup, in Sec.~\ref{sec:2lpt_mocks} we introduce a variant of the target mocks with 2LPT displacements and velocities, breaking the exact correspondence between the coherent large-scale velocity of the mocks and that of the Zeldovich control field.

\subsection{Simulations}
\label{sec:sims}

\subsubsection{Initial conditions}
\label{sec:ics}

In this paper, we generate cheap simulations to test our control variates method on lognormal mocks. Our simulations, both the target lognormal mocks and the control Zeldovich realizations, share common Gaussian initial conditions.
For each realization $s = 1, \ldots, \Nsim$, we generate a Gaussian random field $\delta_L^{(s)}(\vk)$ on a cubic grid of side length $L_{\rm box} = 2000 \ {\rm Mpc}/h$ with $N_{\rm mesh}^3$ grid points, where $N_{\rm mesh}=512$ (corresponding to a Nyquist frequency of $k_{\rm Ny}=\pi N_{\rm mesh}/L_{\rm box}\approx 0.8\,h\,{\rm Mpc}^{-1}$), drawn from the linear matter power spectrum $P_L(k)$ at a specified redshift $z$.
The phases and amplitudes of each Fourier mode are drawn independently as
\beq
\delta_L(\vk) = \sqrt{P_L(k)/V}\;\eta(\vk)\,,
\eeq
where $\eta(\vk)$ is a complex Gaussian random variable with $\langle|\eta|^2\rangle = 1$ and $V = L_{\rm box}^3$ is the box volume. 
The linear power spectrum $P_L(k)$ is computed using the Boltzmann solver \texttt{CLASS} \citep{2011JCAP...07..034B}, with cosmological parameters matching the \AbacusSummit simulations.

We generate a Gaussian random field $\delta(\vx)$ in Fourier space by drawing complex amplitudes with power
spectrum $P_L(k)$, enforcing Hermitian symmetry so that the real-space field
is real, and transforming back via an inverse FFT. From the same Fourier-space
realisation we simultaneously compute the three components of the Zeldovich
displacement field (see below) so that the lognormal density field and the Zeldovich field share identical initial conditions. The Zeldovich displacement field is used when creating both the lognormal mocks, which act as our tracer, as well as the Zeldovich approximation predictions for the covariance of the power spectrum of our tracer.

\subsubsection{Zeldovich approximation}
\label{sec:zeldovich}

The control field is computed from the Zeldovich approximation \citep{1970A&A.....5...84Z}, which provides the leading-order Lagrangian perturbation theory prediction for the evolved density field.
In the Zeldovich approximation, particles initially at Lagrangian position $\vq$ are displaced to Eulerian position
\beq\label{eq:zel}
\vx(\vq) = \vq + \vpsi(\vq)\,,
\eeq
where the displacement field $\vpsi(\vq)$ is given by
\beq\label{eq:psi}
\vpsi(\vk) = \frac{i\vk}{k^2}\,\delta_L(\vk)\,.
\eeq
The Zeldovich density field is obtained by distributing a uniform grid of particles according to Eq.~\eqref{eq:zel} and painting them to a mesh using a mass-assignment scheme (we use triangular-shape cloud, TSC).

For the biased Zeldovich field, we follow the standard Lagrangian advection prescription.
We begin from two Lagrangian source fields defined on the initial (Lagrangian) coordinates $\mathbf{q}$,
\begin{itemize}
    \item the $1\rm cb$ (cdm+baryons) Lagrangian source corresponding to a unit-weight field consisting of ones everywhere, and
    \item the total-matter initial overdensity field, $\delta_{\rm IC}(\mathbf{q})$,
\end{itemize}
and we advect them to Eulerian positions via the Zeldovich displacement at some final redshift, $z$. Note that the advected $1\rm cb$ field corresponds to the Zeldovich predicted cdm+baryons field at redshift $z$.

The key properties of the Zeldovich field are that it correctly describes the large-scale advection that dominates displacements in CDM-like models and, because Eq.~\eqref{eq:psi} is a deterministic linear functional of the initial conditions, the Zeldovich realization is perfectly determined by the initial Gaussian modes --- the same modes that seed the lognormal mock.
This ensures strong correlation between the target and control fields, which is the foundation of the control-variate method.

When working in redshift space, we apply the plane-parallel approximation with the line of sight along the $z$-axis.
For the Zeldovich field, this is straightforward: the redshift-space position is
\beq
\vs(\vq) = \vq + \vpsi(\vq) + \frac{f}{aH}\,\Psi_z(\vq)\,\hat{z}\,,
\eeq
where $f$ is the logarithmic growth rate, $a$ is the scale factor, $H$ is the Hubble parameter, and $\Psi_z$ is the $z$-component of the displacement.
In practice, since we work at the field level, this amounts to an additional displacement along $\hat{z}$ proportional to the velocity field.

Power spectrum multipoles are estimated using the Yamamoto estimator \citep{2003ApJ...595..577Y}, which we implement using the power spectrum code of 
\texttt{abacusutils}\footnote{\url{https://github.com/abacusorg/abacusutils}}.
We measure the monopole ($\ell=0$) and quadrupole ($\ell=2$), which are concatenated into a single data vector of length $2\times N_k$, where $N_k$ is the number of $k$-bins.

\subsubsection{Lognormal mocks}
\label{sec:lognormal}

The target simulations are lognormal mock catalogs \citep{1991MNRAS.248....1C},
which provide a fast way to generate non-Gaussian density fields with
approximately correct one-point statistics and two-point clustering. Our
procedure follows the implementation of \texttt{nbodykit}\footnote{\url{https://nbodykit.readthedocs.io/en/latest/}} and proceeds as
follows. Our chosen redshift, $z = 0.5$, and bias, $b = 2.2$ for the lognormal resembles
the population of Luminous Red Galaxies (LRGs) in DESI \citep{2023AJ....165...58Z,2023JCAP...11..097Z}.

\begin{itemize}
\item {Lognormal density field.} The lognormal overdensity field is
obtained from the Gaussian field $\delta(\vx)$ via
\beq\label{eq:lognormal}
1 + \delta_{\rm LN}(\vx)
= \frac{\exp\!\bigl[b_L\,\delta(\vx)\bigr]}
       {\langle\exp\!\bigl[b_L\,\delta(\vx)\bigr]\rangle}\,,
\eeq
where $b_L = b - 1$ is the Lagrangian bias parameter and $b$ is the Eulerian
linear bias. The normalisation by the spatial mean ensures
$\langle\delta_{\rm LN}\rangle = 0$ exactly. This form of the lognormal
transformation, with the Lagrangian bias applied to the exponent before
normalisation, follows the convention of \texttt{nbodykit}'s
\texttt{LogNormalCatalog}. We use $b = 2.2$ throughout.

\item {Galaxy positions.} Discrete galaxy positions are obtained by
Poisson-sampling the lognormal density field with mean number density
$\bar{n} = 10^{-3}\,(h^{-1}{\rm Mpc})^{-3}$, and then displacing each
galaxy from its grid position by the Zeldovich displacement field. 

\item To incorporate redshift-space distortions, we further displace each galaxy along the line-of-sight (\(z\)) direction by the line-of-sight component of the Zeldovich displacement field,
\[
s_z = x_z + f \Psi_z \, ,
\]
where \(f\) is the linear growth rate and \(\Psi_z\) is the \(z\)-component of the displacement field defined in Eq.~\ref{eq:psi}.

\item To simulate the effect of non-linear structure evolution and Finger-of-God effects, we add a random displacement along the line-of-sight to each galaxy, drawn from a Gaussian distribution centered at $300 \ {\rm  km\, s^{-1}} / (a H)$.

\item The displaced positions are assigned back to a mesh using
triangular-shaped cloud (TSC) mass assignment with window compensation to
correct for the smoothing effect of the assignment scheme \citep{1981csup.book.....H},
\begin{equation}
    W_{\rm TSC}(\mathbf{k})
=\prod_{i=1}^3 \mathrm{sinc}^3\!\left(\frac{k_i\,\Delta x}{2}\right),
\end{equation}
where $\mathrm{sinc}(x)=\sin(x)/x$ and $\Delta x$ is the mesh spacing,
yielding the final lognormal overdensity field $\delta_{\rm LN}(\vx)$.

\item {Power spectra.} From each realisation we measure three power spectra in 255 bins spaced linearly in $k$ from 0 to the Nyquist frequency, $k \approx 0.8\,h\,{\rm Mpc}^{-1}$. The lognormal auto-power spectrum $P^{aa}(k) \equiv P_{\rm LN}(k)$ is measured from $\delta_{\rm LN}$, and plays the role of $P^{aa}$  below.
The (linearly biased) Zeldovich control field is defined explicitly as
\beq\label{eq:za_bias}
\delta_{\rm ZA}(\vx) \equiv \delta_{1\rm cb}(\vx) + \bL\,\delta(\vx)\,,
\eeq
where $\delta_{1\rm cb}$ is the Zeldovich-advected unit-weight (``$1\rm cb$'') field and $\delta$ is the Zeldovich-advected initial overdensity field of Sec.~\ref{sec:zeldovich}; this is the standard first-order Lagrangian bias prescription. Cross- and auto-correlating Eq.~\eqref{eq:za_bias} with the lognormal field then yields Eqs.~\eqref{eq:pab_def} and~\eqref{eq:pbb} below.
The cross-power spectrum between the lognormal field and the (linearly biased) Zeldovich field, 
\beq\label{eq:pab_def}
P^{ab}(k) = P_{\rm LN \times 1cb}(k) + \bL\, P_{\rm LN \times \delta}(k),
\eeq
is measured from the Zeldovich approximation predicted field and $\delta_{\rm LN}(\vx)$, and serves as
$P^{ab}$. The biased Zeldovich auto-power spectrum is then
\begin{align}
\label{eq:pbb}
    \Pbb(k) &= P_{1\rm cb,1cb}(k) + 2\bL\,P_{1\rm cb,\delta}(k)
    \nonumber \\
    &+ \bL^2\,P_{\delta,\delta}(k)\,,
\end{align}
with the Zeldovich operators described in Sec.~\ref{sec:zeldovich}.

\end{itemize}

\subsubsection{Survey mask}
\label{sec:mask}

\begin{figure}
\centering
\includegraphics[width=0.48\textwidth]{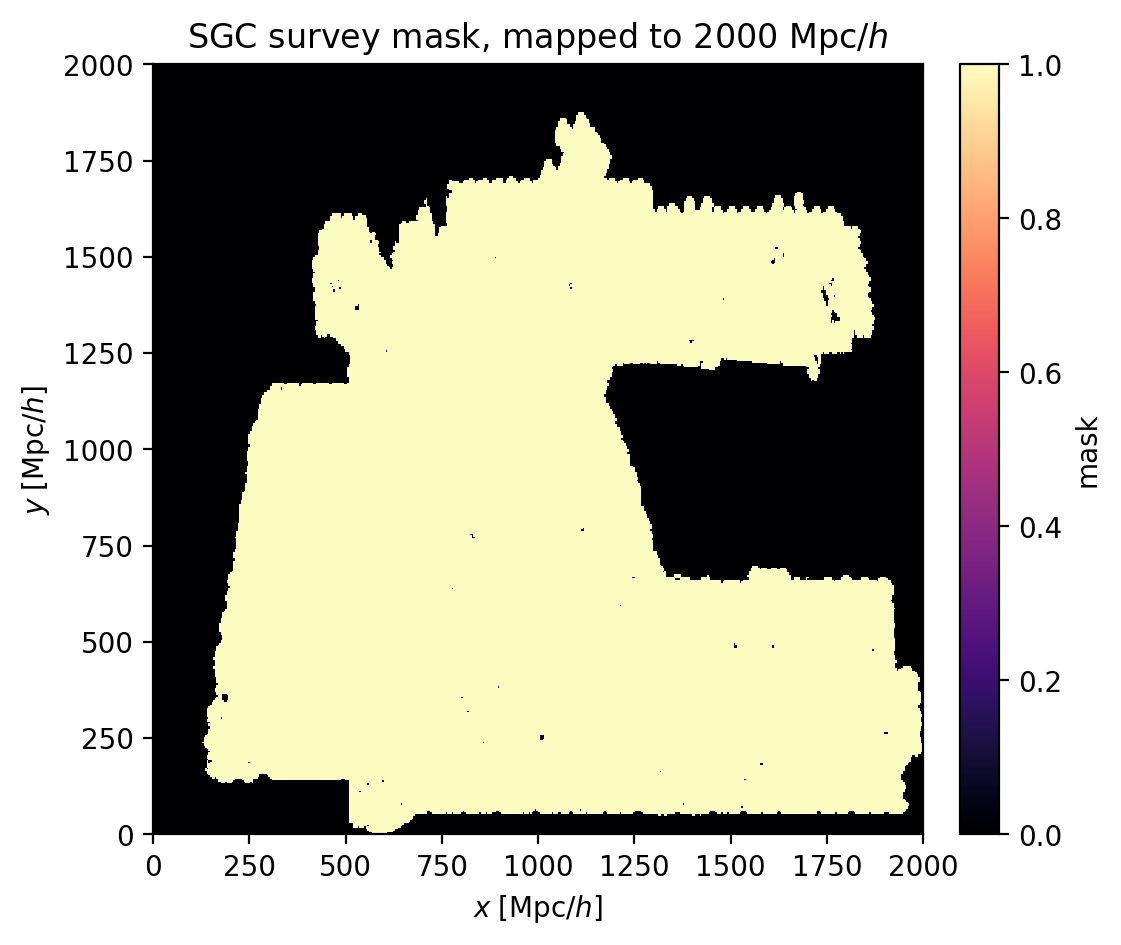}
\caption{%
DESI South Galactic Cap (SGC) survey mask applied to the LRG-like lognormal mocks of this study.  In addition to the irregular boundary there are numerous `holes' in the survey, the largest of which can be seen as black dots within the footprint. }
\label{fig:mask}
\end{figure}

To test the method in a realistic setting, we apply a survey window function modeled on the DESI South Galactic Cap (SGC) geometry at redshift $z = 0.5$.
The mask is a binary angular selection function applied to the simulation box; regions outside the observed region in the $xy$ cross-section are zeroed, and we additionally cut off $200\,h^{-1}$Mpc from each each side of the box in the $z$ direction.
This breaks the periodicity of the box and introduces off-diagonal elements in the covariance matrix (mode coupling), which is one of the key complications the control-variate method must handle.

When the mask is applied, the power spectrum estimator uses the FKP weighting scheme \citep{1994ApJ...426...23F}, and the normalization accounts for the effective survey volume.
We denote masked measurements with the subscript ``mask'' where disambiguation is needed. We show the mask in Fig.~\ref{fig:mask}.

\subsubsection{2LPT target mocks}
\label{sec:2lpt_mocks}

In the fiducial configuration described above, the coherent (Kaiser) part of the large-scale velocity field of the target mocks is, by construction, the Zeldovich velocity itself (the stochastic Finger-of-God dispersion and the lognormal transformation plus Poisson sampling are the only sources of target--control decorrelation). In realistic $N$-body-based mocks, the large-scale velocity field is not perfectly reproduced by the Zeldovich approximation. To assess the robustness of the method to such a velocity-field mismatch, we generate a variant of the target mocks in which both the galaxy displacements and the redshift-space velocities are computed at second order in Lagrangian perturbation theory (2LPT) \citep{1995A&A...296..575B,1998MNRAS.299.1097S,2006MNRAS.373..369C}, while the control field remains pure Zeldovich (1LPT).

Concretely, with all fields normalized at the target redshift ($D_1=1$), the displacement applied to the mock galaxies is $\vpsi = \vpsi^{(1)} + \vpsi^{(2)}$, where $\vpsi^{(1)}$ is the Zeldovich displacement of Eq.~\eqref{eq:psi} and
\beq\label{eq:psi2}
\vpsi^{(2)}(\vk) = \frac{3}{7}\,\Omega_m^{-1/143}(z)\,\frac{i\vk}{k^2}\,S(\vk)\,,
\eeq
with the second-order source $S(\vx) = \sum_{i<j}\left[\phi_{,ii}\phi_{,jj} - \phi_{,ij}^2\right]$ and $\nabla^2\phi = \delta_L$.
The redshift-space displacement along the line of sight is correspondingly $\Delta s_z = f_1\,\Psi^{(1)}_z + f_2\,\Psi^{(2)}_z$, with $f_2 \simeq 2\,\Omega_m^{6/11}(z)$ \citep{1995A&A...296..575B}; the stochastic velocity dispersion is unchanged. This introduces a controlled, physically motivated mismatch between the target and control fields in both real space and, in particular, in the coherent velocity field entering the redshift-space distortions. We present the corresponding results in Sec.~\ref{sec:2lpt_validation}.

\subsection{Control variate formalism}
\label{sec:formalism}

\subsubsection{Setup and notation}
\label{sec:notation}

Let $\hat{P}^{(s)}_\ell(k)$ denote the power spectrum multipole measured from the $s$-th lognormal realization, and let $\hat{Q}^{(s)}_\ell(k)$ denote the corresponding measurement from the paired Zeldovich realization (same initial conditions).
We organize these into data vectors $\mathbf{d}^{(s)}$ and $\mathbf{c}^{(s)}$ by concatenating all $(k, \ell)$ bins:
\beq
d^{(s)}_i = \hat{P}^{(s)}_{\ell_i}(k_i)\,, \qquad c^{(s)}_i = \hat{Q}^{(s)}_{\ell_i}(k_i)\,,
\eeq
where the index $i$ runs over all $k$-bins and multipoles.

The sample covariance matrices of the target and control fields are
\beq
X_{ij} = \frac{1}{\Nsim - 1}\sum_{s=1}^{\Nsim}\left(d^{(s)}_i - \bar{d}_i\right)\left(d^{(s)}_j - \bar{d}_j\right)\,,
\eeq
and similarly $C_{ij}$ for the control field.
The true covariance of the control field,
\beq
\mu_{ij} \equiv \Cov\!\left[\hat{Q}_{\ell_i}(k_i),\,\hat{Q}_{\ell_j}(k_j)\right]\,,
\eeq
can be estimated to high precision from a large ensemble of Zeldovich-only simulations, since these are computationally inexpensive. In our fiducial analysis we estimate $\mu_C$ from $N_{\rm theo}=10,000$ Zeldovich realizations that are passed through the same survey mask as the data, because the mask breaks translational invariance and induces non-trivial mode coupling in $\mu_C$ that is generally cumbersome to evaluate analytically. 

Because $\mu_{ij}$ is estimated from a finite number, $N_{\rm th}$,
of Zeldovich realizations, there is an additional contribution to the variance of
$C_{ij}-\hat\mu_{ij}$. For control simulations
statistically independent of the paired realizations used to compute $X_{ij}$,
\begin{equation}
\Var(C_{ij}-\hat\mu_{ij})=\Var(C_{ij})+\Var(\hat\mu_{ij}),
\end{equation}
with $\Var(\hat\mu_{ij})$ scaling as $\mathcal{O}\!\left((N_{\rm th}-1)^{-1}\right)$.
Therefore, for $N_{\rm th}\gg N_{\rm sim}$ these corrections are
subdominant, and the expressions for $\beta_{ij}$ and $\rho_{ij}$
remain good smooth approximations.

\subsubsection{Control variate estimator}
\label{sec:cv_estimator}

In our notation, $\beta_{ij}$ and $\rho_{ij}$ are defined element-by-element in the $(k,\ell)$ space of the covariance matrix. They are {not} matrices acting on the data vector, but rather scalar coefficients for each element of the covariance matrix. We could thus denote the elements by a single index, e.g., $\alpha \equiv ij$, but we opt to use the standard double-index notation for covariance matrices. In principle, one could also consider a matrix-valued control variate acting on the data vector level, but the element-wise approach is simpler and, as we show, already highly effective.

The control-variate-corrected covariance estimator is
\beq\label{eq:Y}
Y_{ij} = X_{ij} - \beta_{ij}\left(C_{ij} - \mu_{ij}\right)\,,
\eeq
where $\beta_{ij}$ is a coefficient matrix.
Since $\langle C_{ij} \rangle = \mu_{ij}$, we have $\langle Y_{ij} \rangle = \langle X_{ij} \rangle$, so the estimator is unbiased regardless of $\beta$, assuming our estimate of $\mu_C$ is accurate and low-noise.

The variance of $Y_{ij}$ is
\beq\label{eq:varY}
\Var[Y_{ij}] = \Var[X_{ij}] - 2\beta_{ij}\Cov[X_{ij}, C_{ij}] + \beta_{ij}^2\Var[C_{ij}]\,.
\eeq
Minimizing with respect to $\beta_{ij}$ gives the optimal coefficient
\beq\label{eq:beta_opt}
\beta_{ij}^{\rm opt} = \frac{\Cov[X_{ij}, C_{ij}]}{\Var[C_{ij}]}\,,
\eeq
and the resulting minimum variance is
\beq\label{eq:varY_min}
\Var[Y_{ij}]_{\rm min} = \Var[X_{ij}]\left(1 - \rho_{ij}^2\right)\,,
\eeq
where
\beq\label{eq:rho}
\rho_{ij} = \frac{\Cov[X_{ij}, C_{ij}]}{\sqrt{\Var[X_{ij}]\,\Var[C_{ij}]}}
\eeq
is the correlation coefficient between the target and control covariance estimators.
The effective number of simulations saved is $1/(1 - \rho^2)$: a correlation of $\rho = 0.95$ corresponds to a factor of $\sim 10$ reduction, meaning 1{,}000 paired simulations are equivalent to $\sim$10{,}000 unpaired ones.

\subsection{Analytic approximations}
\label{sec:analytic_beta}

The brute-force estimation of $\beta_{ij}$ from simulations requires estimating
$\Cov[X_{ij}, C_{ij}]$ and $\Var[C_{ij}]$ from a large ensemble of realizations,
which is computationally expensive and itself subject to sampling noise.
Here we instead derive closed-form expressions for $\beta$ and $\rho$ under the Gaussian
(disconnected) approximation, assuming the Fourier-space density field
$\delta(\mathbf{k})$ is Gaussian distributed. In the limit of a large number
of modes per bin the power spectrum estimates are also approximately
Gaussian distributed (by the central limit theorem). Below we begin by considering the covariance of the monopole of the power spectrum.  For ease of notation, we denote the monopole of the power spectrum by $P(k) \equiv P_{\ell=0}(k)$, and we refer to the auto-power spectrum as $P^{a}(k) \equiv P^{aa}(k)$. 

In the subsections below, we introduce a sequence of simplifying approximations to make the control-variate construction analytically and computationally tractable.  
Below we also validate these approximations directly against brute-force simulations, demonstrating that they do not introduce an appreciable bias in the resulting covariance (or precision) estimates.  
Accordingly, the role of these approximations is mainly to enable a practical implementation; any residual impact is therefore expected to be limited to a (possibly modest) reduction in the CV optimality rather than a qualitative change in the inferred statistics.

\subsubsection{Setup}
\label{sec:setup}

The target and control quantities entering the control-variate estimator are
elements of sample covariance matrices estimated from $\Nsim$ paired
simulations. 
\beq
X_{ij} = \hat{\Sigma}^{aa}_{ij}\,, \qquad C_{ij} = \hat{\Sigma}^{bb}_{ij}\,,
\eeq
where notation $a$ is used to refer to the lognormal mocks, $b$ to the Zeldovich approximation mocks. The covariance is estimated via
\beq
\hat{\Sigma}^{ab}_{ij} =
\frac{1}{\Nsim-1}\sum_{s=1}^{\Nsim}
\Delta \hat{P}^{a,(s)}_i \,
\Delta \hat{P}^{b,(s)}_j \,,
\eeq
where 
\beq
\Delta \hat{P}^{a,(s)}_i \equiv
\hat{P}^{a,(s)}(k_i) - \bar{P}^a(k_i)\,.
\eeq
and $\bar{P}^a(k_i) = \Nsim^{-1}\sum_s \hat{P}^{a,(s)}(k_i)$ is the
simulation mean. The quantities $X_{ij}$ and $C_{ij}$ are therefore sample
covariance matrix elements, which are quadratic in the fields, and their
covariance over many realizations is a fourth-order quantity in the fields.

\subsubsection{Covariance of the sample covariance matrix}
\label{sec:cov_cov}

Under the Gaussian approximation, the fields are fully characterised by their
two-point functions, and the covariance of two-point estimators follows from
the Isserlis--Wick theorem. For Gaussian fields, the binned power spectrum
estimator satisfies
\beq\label{eq:cov_Pk}
\Cov\bigl[\hat{P}^a(k_i),\, \hat{P}^b(k_j)\bigr]
\simeq \frac{2\,[\Pab(k_i)]^2}{\Nki}\,\delta_{ij}\,,
\eeq
where $\Nki$ is the number of independent Fourier modes in bin $i$ and the Kronecker delta reflects the fact that different $k$-shells are uncorrelated for Gaussian fields. In practice, this diagonal form is only approximate, since survey masks and window functions couple different Fourier modes and induce off-diagonal covariance between $k$-bins. Throughout this work, however, we adopt the diagonal Gaussian approximation above, and later demonstrate that it provides
an adequate description for the purposes of our control variate calculations.

The sample covariance estimator $\hat{\Sigma}^{ab}_{ij}$ is an unbiased
estimator of $\Sigma^{ab}_{ij} \equiv \Cov[\hat{P}^a(k_i), \hat{P}^b(k_j)]$.
For Gaussian fields, $\hat P$,
the distribution of $\hat{\Sigma}$ follows a Wishart
distribution, and the covariance of two sample covariance elements is (assuming $\Sigma$ is dominated by its diagonal components)
\beq\label{eq:wishart}
\Cov\bigl[\hat{\Sigma}^{aa}_{ij},\, \hat{\Sigma}^{bb}_{ij}\bigr]
= \frac{1}{\Nsim - 1}
\left[
\Sigma^{ab}_{ii}\,\Sigma^{ab}_{jj} + (\Sigma^{ab}_{ij})^2
\right]\,,
\eeq
where the multiplication is executed element-wise. 
Since $\Sigma^{ab}_{ij} = \Cov[\hat{P}^a_i, \hat{P}^b_j]
= 2[\Pab(k_i)]^2 \delta_{ij} / \Nki$ is diagonal in $k$, the second term
vanishes for $i \neq j$ and both terms contribute equally for $i = j$.

For Gaussian fields, the Wishart covariance implies an extra factor
$(1+\delta_{ij})$ for the diagonal sample-covariance elements. Under
the assumption that $\Sigma^{ab}_{ij}$ is diagonal in $k$ (and hence
$\Sigma^{ab}_{ij}\propto \delta_{ij}$), one may write
\begin{align}
\Cov\!\left[\hat\Sigma^{aa}_{ij},\,\hat\Sigma^{bb}_{ij}\right]
&=\frac{1+\delta_{ij}}{\Nsim-1}\,\Sigma^{ab}_{ii}\Sigma^{ab}_{jj},\\
\Var\!\left[\hat\Sigma^{bb}_{ij}\right]
&=\frac{1+\delta_{ij}}{\Nsim-1}\,\Sigma^{bb}_{ii}\Sigma^{bb}_{jj},\\
\Var\!\left[\hat\Sigma^{aa}_{ij}\right]
&=\frac{1+\delta_{ij}}{\Nsim-1}\,\Sigma^{aa}_{ii}\Sigma^{aa}_{jj}.
\end{align}
These $(1+\delta_{ij})$ factors cancel in the ratios defining $\beta_{ij}$
and $\rho_{ij}$.

In both cases, substituting Eq.~\eqref{eq:cov_Pk} gives
\beq\label{eq:covXC}
\Cov\bigl[X_{ij},\, C_{ij}\bigr]
= (1+\delta_{ij}) \frac{4\,[\Pab(k_i)]^2\,[\Pab(k_j)]^2}{(\Nsim-1)\,\Nki\,\Nkj}\,,
\eeq
and similarly
\beq\label{eq:varC}
\Var\bigl[C_{ij}\bigr]
= (1+\delta_{ij}) \frac{4\,[\Pbb(k_i)]^2\,[\Pbb(k_j)]^2}{(\Nsim-1)\,\Nki\,\Nkj}\,,
\eeq
\beq\label{eq:varX}
\Var\bigl[X_{ij}\bigr]
= (1+\delta_{ij}) \frac{4\,[\Paa(k_i)]^2\,[\Paa(k_j)]^2}{(\Nsim-1)\,\Nki\,\Nkj}\,.
\eeq

\subsubsection{Analytic expressions for $\beta$ and $\rho$}
\label{sec:beta_rho_analytic}

Taking the ratio of Eqs.~\eqref{eq:covXC} and~\eqref{eq:varC}, the
$(\Nsim-1)$ and $\Nk$ factors cancel exactly, giving the main result:
\begin{eqnarray}
\label{eq:beta_analytic}
\beta_{ij}
&=& \frac{\Cov[X_{ij},\,C_{ij}]}{\Var[C_{ij}]}
= \frac{[\Pab(k_i)]^2\,[\Pab(k_j)]^2}{[\Pbb(k_i)]^2\,[\Pbb(k_j)]^2} .
\end{eqnarray}
Defining
\beq\label{eq:beta1_def}
\alpha(k) \equiv \frac{[\Pab(k)]^2}{[\Pbb(k)]^2}\,,
\eeq
we write this as
\beq
\boxed{\beta_{ij} \equiv \alpha(k_i)\,\alpha(k_j)}
\eeq
The correlation coefficient between $X_{ij}$ and $C_{ij}$ follows
from Eqs.~\eqref{eq:covXC}--\eqref{eq:varX}:
\begin{eqnarray}
\label{eq:rho_analytic}
\rho_{ij}
&=& \frac{\Cov[X_{ij},\,C_{ij}]}{\sqrt{\Var[X_{ij}]\,\Var[C_{ij}]}} \nonumber \\
&=& \frac{[\Pab(k_i)]^2\,[\Pab(k_j)]^2}
       {\Paa(k_i)\,\Pbb(k_i)\,\Paa(k_j)\,\Pbb(k_j)} .
\end{eqnarray}
Defining 
\beq\label{eq:r_def}
r(k) \equiv \frac{\Pab(k)}{\sqrt{\Paa(k)\,\Pbb(k)}}
\eeq
as the standard field-level cross-correlation coefficient, we write this as
\beq
\boxed{\rho_{ij} = r^2(k_i)\,r^2(k_j)}.
\eeq

In the above expressions for $\beta_{ij}=\alpha_i\alpha_j$ and
$\rho_{ij}=r_i^2 r_j^2$, we assume that the
disconnected covariance matrix of the bandpower
estimators is diagonal in the bin index,
$\Sigma^{ab}_{ij}\propto \delta_{ij}$. Any mode coupling induced by the
survey window, bin overlap, or non-Gaussian connected terms would
introduce off-diagonal contributions and in general spoil this
factorization. Ignoring the coupling, both $\beta_{ij}$
and $\rho_{ij}$ factorise into a product of per-bin quantities, a direct
consequence of the diagonal structure of $\Sigma^{ab}_{ij}$ in $k$-space
under the Gaussian approximation. Importantly, both are independent of the
number of simulations $\Nsim$ and the number of modes $\Nki$: these cancel
in the ratio and $\beta$ and $\rho$ depend only on ratios of mean power
spectra.

\begin{figure*}[!t]
\centering
\includegraphics[width=0.48\textwidth]{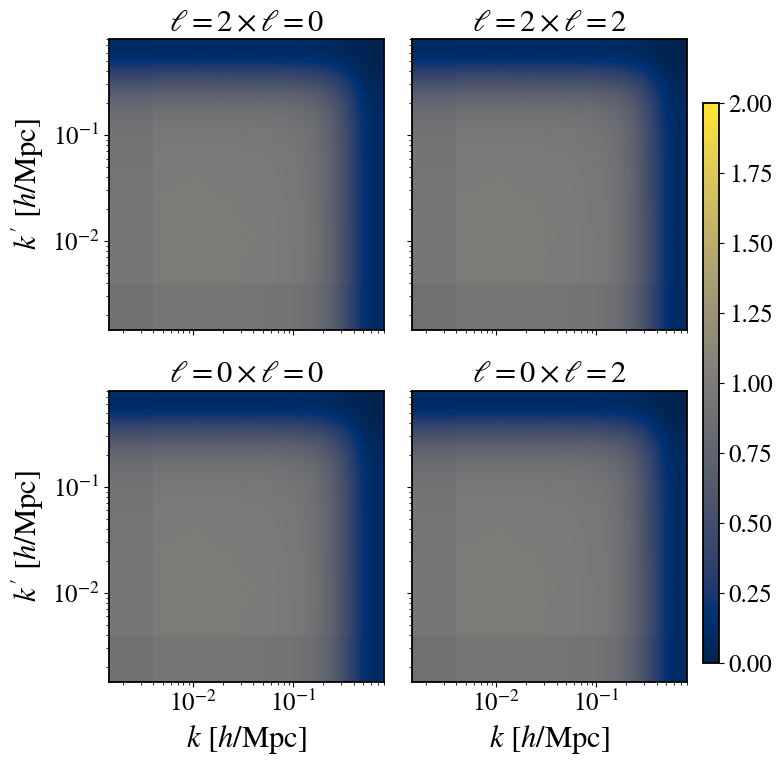}
\hfill
\includegraphics[width=0.48\textwidth]{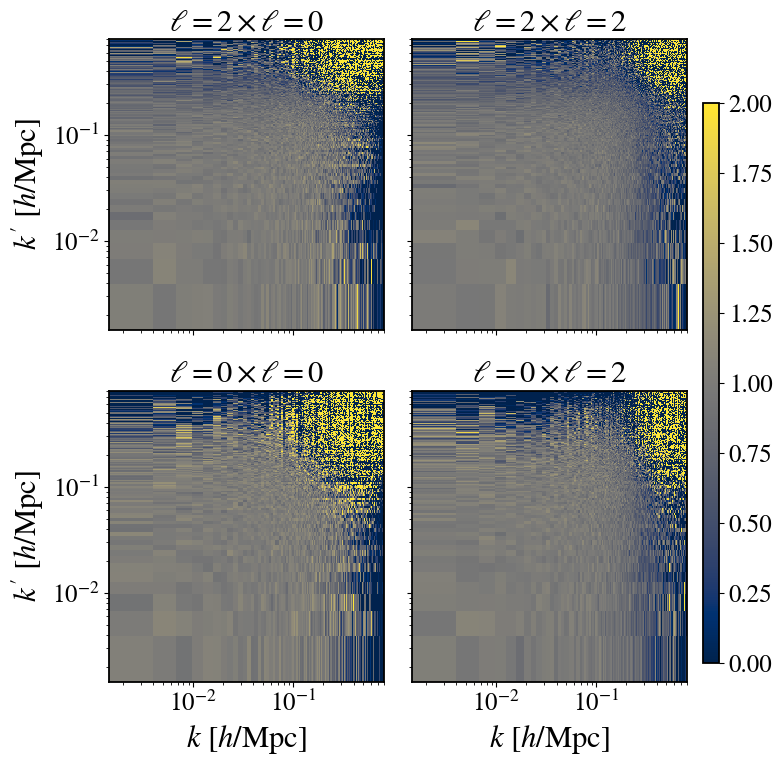}
\caption{%
Comparison of the CV weight matrix $\beta(k,\ell;k',\ell')$ in redshift space with a DESI-like survey mask.
{Left:} analytic $\beta$.
{Right:} brute-force $\beta$ estimated from $10{,}000$ simulations split into $100$ batches of $100$.
Each map is visualized in the $2\times 2$ block form corresponding to multipole pairs $(\ell,\ell')=(0,0),(0,2),(2,0),(2,2)$.
The dominant large-scale structure agrees well; the brute-force realization shows additional small-scale noise, most visibly in the upper-right corner of the right panel.  }
\label{fig:beta_analytic_vs_brute}
\end{figure*}

\subsubsection{Extension to multipoles}
\label{sec:multipoles}

The disconnected derivation above for the monopole relies only on the Gaussian covariance of the underlying bandpower estimators and on the fact that, in the $\mu$-binning used there, different $\mu$-bins can be treated as approximately independent. 
For redshift-space multipoles one can proceed analogously, but the crucial difference is that while the $\mu$-bins remain (approximately) independent at the level of the anisotropic power $P(k,\mu)$, the multipole projection mixes $\mu$ and induces strong correlations between different multipoles.
As a result, the corresponding Gaussian covariance in the $(k,\ell)$ basis is generally not diagonal in $\ell$, and the factorized expressions obtained in the monopole case do not directly carry over.

Concretely, writing the Legendre multipole estimator as
\begin{equation}
\hat P_\ell^a(k)=\frac{2\ell+1}{2}\int_{-1}^{1} d\mu\;
\mathcal{L}_\ell(\mu)\,\hat P^a(k,\mu)\,,
\end{equation}
the Gaussian covariance between multipoles at fixed $k$ has the schematic form
\begin{equation}
\Cov\!\big[\hat P_\ell^a(k),\,\hat P_{\ell'}^b(k')\big]
\simeq
\delta_{kk'}\,\frac{(2\ell+1)(2\ell'+1)}{4\,N_k}\;
\mathcal{K}^{ab}_{\ell\ell'}(k)\,,
\end{equation}
with the kernel
\begin{equation}
\mathcal{K}^{ab}_{\ell\ell'}(k)\equiv
\int_{-1}^{1} d\mu\;\mathcal{L}_\ell(\mu)\,\mathcal{L}_{\ell'}(\mu)\,
\bigl[P_{ab}(k,\mu)\bigr]^2 .
\end{equation}
If one {assumed} $\mathcal{K}^{ab}_{\ell\ell'}(k)\approx 0$ for $\ell\neq \ell'$, then the CV-weight construction would again factorize in a manner analogous to the monopole.
However, in practice multipole blocks are highly correlated, so a fully non-factorized evaluation of $\beta_{(k,\ell),(k',\ell')}$ (and the corresponding $\rho$) requires the off-diagonal kernels $\mathcal{K}_{\ell\ell'}$.
Carrying out this semi-analytic program is therefore considerably more cumbersome than in the monopole case.

Fortunately, guided by numerical experiments, we find that the smooth analytic CV prediction obtained from the monopole block provides an excellent proxy for the mixed and higher multipole blocks relevant for our realistic analysis.
Specifically, the monopole-derived analytic $\beta$ and $\rho$ track the $\ell,\ell'=0,2$ and $2,2$ blocks closely enough that using the monopole-based approximation everywhere yields a stable and effectively unbiased CV-corrected covariance.
A plausible reason is that the covariance in the Gaussian projection is dominated by the largest underlying spectrum (typically the monopole, since it carries most of the signal-to-noise), so the $\ell$--$\ell'$ covariance structure is largely inherited from the monopole even when exact diagonal-in-$\ell$ factorization fails. More generally, the accuracy of the monopole-based approximation should depend on how the effective multipole weights scale with the tracer bias; for sufficiently low-biased tracers, the quadrupole and higher-$\ell$ spectra can become more prominent, which may slightly alter the relative importance of the neglected blocks.

\begin{figure*}
\centering
\includegraphics[width=0.95\textwidth]{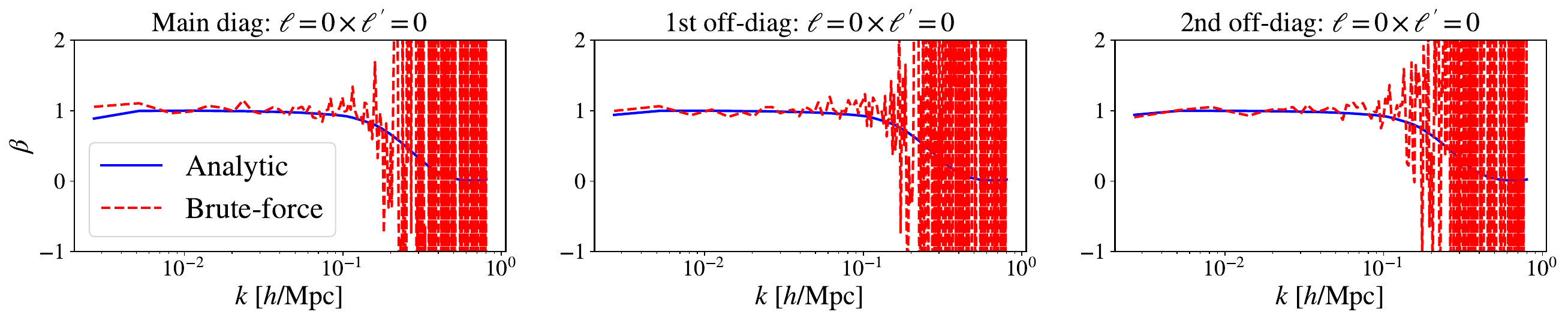}\\
\includegraphics[width=0.95\textwidth]{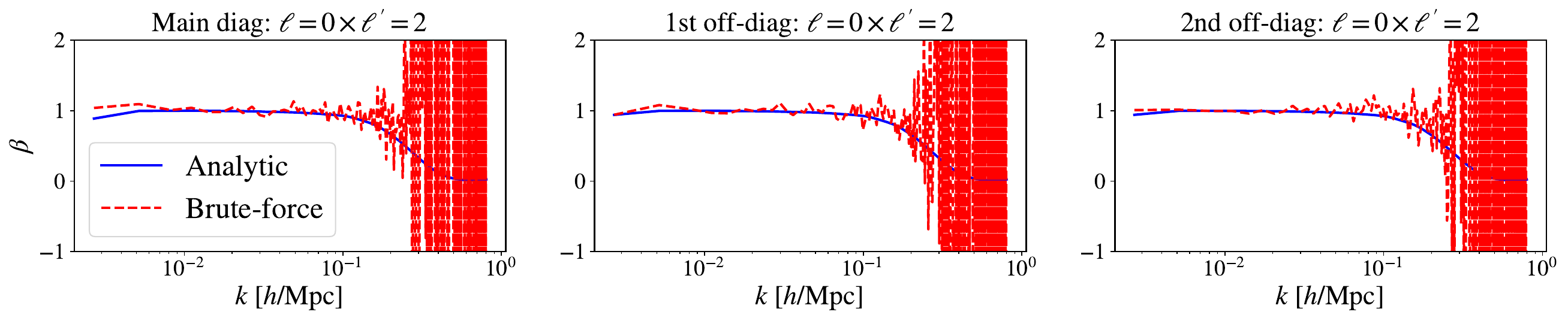}\\
\includegraphics[width=0.95\textwidth]{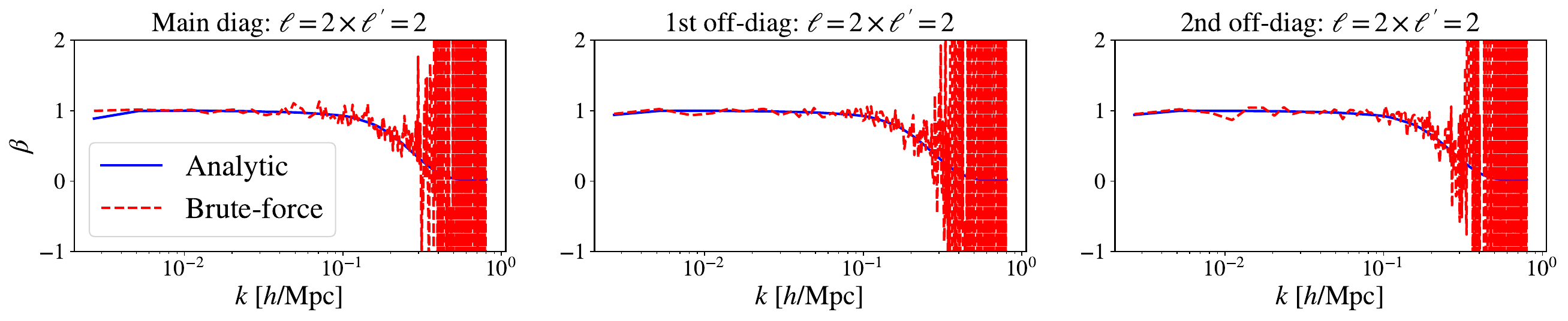}\\
\caption{%
Diagonal comparison of the CV weight $\beta$ between analytic and brute-force constructions in redshift space with mask.
The figure shows three block pairs: $(\ell,\ell')=(0,0)$ (left), $(0,2)$ (middle), and $(2,2)$ (right).
Within each block pair, the top/center/bottom subpanels correspond to the main diagonal, the first off-diagonal, and the second off-diagonal (in $(k_i,k_j)$).
Analytic $\beta$ (smooth) and brute-force $\beta$ (noisier at high $k$) agree well on large scales, motivating the use of the analytic $\beta$ throughout.  }
\label{fig:beta_diagonals}
\end{figure*}

\section{Validation}
\label{sec:validation}

In this Section, we study how well our analytic prediction for the CV weight and the corresponding correlation coefficient agrees with brute-force estimates using numerical simulations.

\subsection{Setup}

In this paper, we consider the following realistic configuration: power spectrum multipoles ($\ell = 0, 2$) measured in redshift space through a DESI SGC-like survey mask.  We have chosen an SGC-like mask because it is quite complex, arising from combining of several observational probes.
This case is of practical relevance to current and future spectroscopic surveys and a valuable test for the control-variate (CV) method, as the mask induces substantial mode coupling and generates a non-diagonal covariance in $(k,\ell)$ space.

With multipoles $\ell\in\{0,2\}$, the covariance forms a $2\times 2$ block matrix in $(k,\ell)$ space. Because the covariance is symmetric, $C_{\ell\ell'}(k_i,k_j)=C_{\ell'\ell}(k_j,k_i)$, and the same symmetry holds for the CV-related matrices, so the resulting CV-corrected covariance $Y$ is symmetric as well. In the figures we display a $2\times2$ subset of block pairs, ordered as $(0,0),(0,2),(2,0),(2,2)$.

Throughout this section, we denote by $X$ the (sample) covariance matrix estimated from $N_{\rm sim}=1000$ paired simulations, and by $Y$ the corresponding CV-corrected covariance,
\begin{equation}
Y \;=\; X \;-\; \beta\,\bigl(C-\mu_C\bigr)\,,
\end{equation}
where $C$ is the covariance of the control field (Zeldovich approximation) evaluated on the same survey geometry, $\mu_C$ is its mean estimated from a large ensemble of Zeldovich realizations (10,000), and $\beta$ is the CV weight matrix.
By construction, if $\mu_C=\mathbb{E}[C]$ is correct, then $Y$ remains unbiased even when $\beta$ is only approximately optimal (the CV then becomes suboptimal but should not introduce a bias in $\mathbb{E}[Y]$).

\begin{figure*}
\centering
\includegraphics[width=0.48\textwidth]{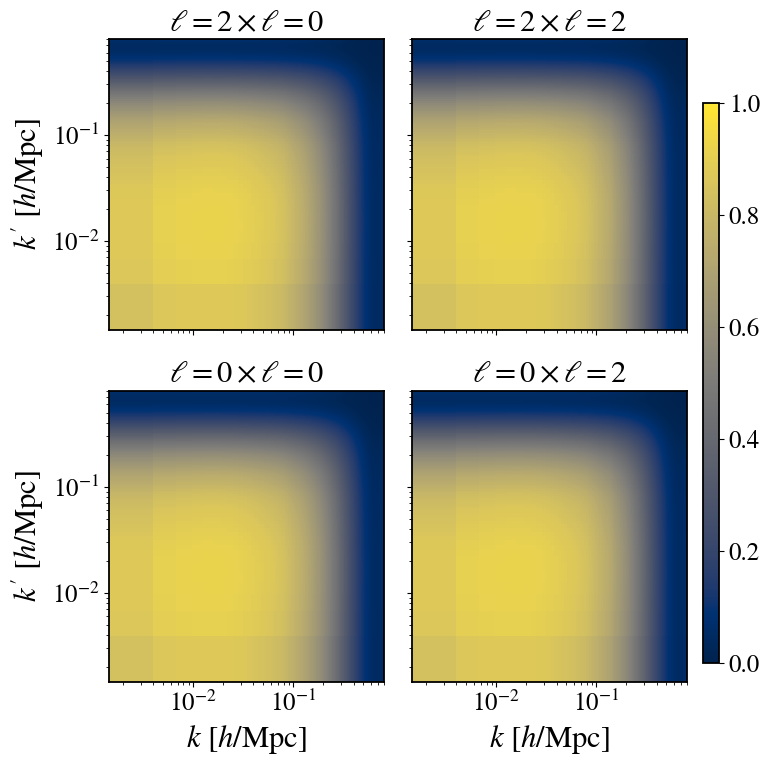}
\hfill
\includegraphics[width=0.48\textwidth]{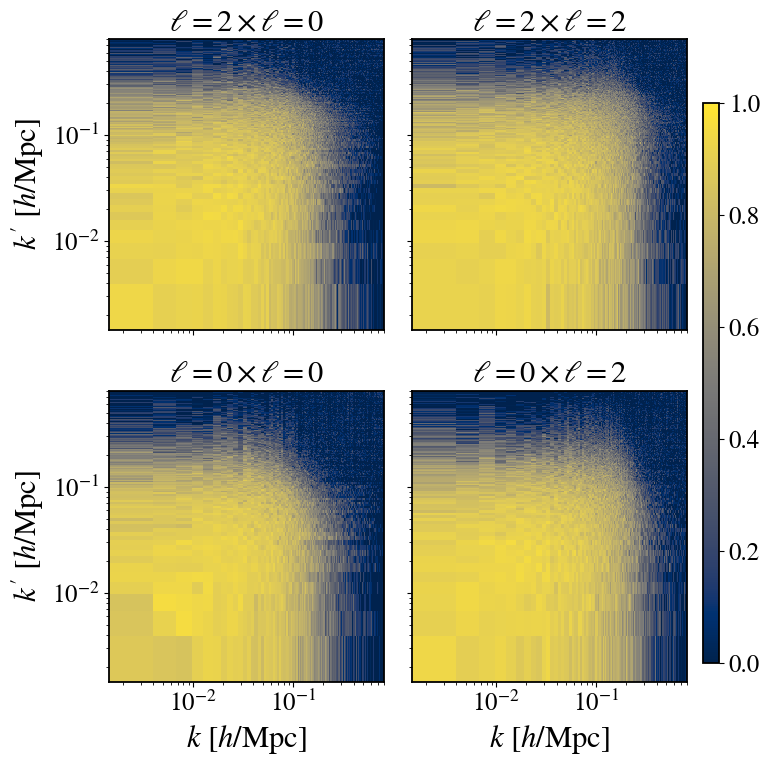}
\caption{%
Comparison of the CV correlation coefficient $\rho(k,\ell;k',\ell')$ in redshift space with survey mask.
{Left:} analytic $\rho$ (computed from the boxed analytic expression; $(0,2)$ and $(2,2)$ populated by copying the $(0,0)$ block).
{Right:} brute-force $\rho$ measured using $10{,}000$ realizations split into $100$ batches of $100$.
Each panel shows the same $(2\times 2)$ block ordering $(\ell,\ell')=(0,0),(0,2),(2,0),(2,2)$.
The large-scale agreement is nearly exact; the correlation falls rapidly around $k\simeq 0.2\,h\,{\rm Mpc}^{-1}$, which directly governs the transition from strong variance reduction to negligible improvement.  }
\label{fig:rho_analytic_vs_brute}
\end{figure*}

\subsection{Analytic versus brute-force $\beta$}
\label{sec:beta_rsd_mask}

Fig.~\ref{fig:beta_analytic_vs_brute} compares the analytic $\beta$ prediction with a brute-force estimate.

The analytic $\beta$ is obtained from the boxed expression for $\beta$ derived earlier in the paper.
However, due to the difficulty of computing the corresponding moments for $\ell\neq 0$ in the redshift-space setting (see Section~\ref{sec:multipoles}), we adopt the approximation
\begin{equation}
\beta_{0,2} \approx \beta_{0,0}\,,\qquad
\beta_{2,2} \approx \beta_{0,0}\,,
\end{equation}
(i.e.\ we ``copy'' the $(0,0)$ analytic block to populate the $(0,2)$ and $(2,2)$ blocks).
This is not expected to be perfectly optimal, but should yield the correct order of magnitude and, crucially, should not bias the CV result.

The brute-force $\beta$ is computed from the estimates of $\mathrm{Cov}(X,C)$ and $\mathrm{Var}(C)$ where we split the $10{,}000$ simulation realizations into $100$ batches of $100$.  We compute the batch-wise estimates of $\mathrm{Cov}(X,C)/\mathrm{Var}(C)$, and assemble the full $\beta$ matrix from these batch statistics.

Overall agreement between analytic and brute-force $\beta$ is very good.
While the brute-force map exhibits visible noise on small scales (notably in the upper-right corner of the brute-force panels), the large-scale structure matches closely and is sufficiently accurate for our purposes: even if the weight is not perfectly optimal, the CV procedure does not bias $Y$ as long as $\mu_C$ is correct.

To further validate the analytic approximation (including the reuse of the $(0,0)$ block), Fig.~\ref{fig:beta_diagonals} compares analytic and brute-force $\beta$ on the diagonals.

Each figure shows three diagonal bands: the main diagonal, the first off-diagonal, and the second off-diagonal in $(k_i,k_j)$.
Within each band, we plot analytic and brute-force $\beta$ curves.
The three groups of panels correspond to $(\ell,\ell')=(0,0)$, $(0,2)$, and $(2,2)$, displayed side by side.

On large scales the analytic and brute-force estimates match well.
On smaller scales the brute-force curves become noisier, while the analytic curves remain smooth.
Notably, the $(0,2)$ and $(2,2)$ diagonal blocks appear slightly less noisy than $(0,0)$ in the brute-force comparison, even though we do not compute them independently analytically; we do not claim a definitive physical reason, but the empirical behavior supports the approximation.

Even if $\beta$ were somewhat imperfect, the CV-corrected covariance $Y$ would remain unbiased provided $\mu_C$ is correct.
However, in practice, using a noisy brute-force $\beta$ can degrade the estimator variance more than using a smooth approximate $\beta$.
Since the analytic $\beta$ is visually and quantitatively very similar to the brute-force result on the relevant (large-scale) modes, we use the analytic $\beta$ as the CV weight in the remainder of the analysis.

It is instructive to quantify how sensitive the variance reduction is to the choice of $\beta$. Writing $\tilde\beta_{ij} \equiv \beta_{ij}/\beta^{\rm opt}_{ij}$, Eq.~\eqref{eq:varY} gives
\beq\label{eq:suboptimal_beta}
\frac{\Var[Y_{ij}]}{\Var[X_{ij}]} = 1 - \rho_{ij}^2\left(2\tilde\beta_{ij} - \tilde\beta_{ij}^2\right)\,,
\eeq
which is quadratic around the optimum: a mildly suboptimal $\beta$ (say, within tens of percent of $\beta^{\rm opt}$) sacrifices little of the variance reduction. On large scales, where $r(k)\to1$ and $\Pab \to \Pbb$, the optimal weight $\beta^{\rm opt}_{ij} = \alpha(k_i)\alpha(k_j) \to 1$, so the simple choice $\beta=1$ is indeed close to optimal there. On small scales, however, $\beta^{\rm opt}\to0$ as the fields decorrelate, and keeping $\beta=1$ adds the (uncorrelated) noise of the control estimator, $\Var[Y]\to\Var[X]+\Var[C] > \Var[X]$. The essential requirement on any practical $\beta$ is therefore that it approach unity on large scales and vanish smoothly where $\rho\to0$; the analytic expression satisfies both automatically, whereas $\beta=1$ would need to be supplemented with an explicit high-$k$ cutoff or taper. We verify this behavior numerically in the batch framework of this section.

Over the range $0.01<k<0.2\,h\,{\rm Mpc}^{-1}$ analyzed here, the three choices are in fact very close: the variance ratio $\Var[X]/\Var[Y]$ obtained with $\beta=1$ agrees with the analytic-$\beta$ result to within $1.8\%$ at all $k$, and the largest penalty incurred by $\beta=1$ is a $3.5\%$ reduction of the gain near $k\simeq0.18\,h\,{\rm Mpc}^{-1}$. We nonetheless retain the analytic $\beta$, since it vanishes automatically where $\rho\to0$ and so remains safe if the analysis range is extended.

\subsection{Analytic versus brute-force $\rho$}
\label{sec:rho_rsd_mask}

Since the effective variance reduction is controlled by the CV correlation coefficient $\rho$, we next compare $\rho$ between analytic and brute-force constructions, shown in Fig.~\ref{fig:rho_analytic_vs_brute}.

As in the $\beta$ case, the analytic $\rho$ is computed from the boxed expression and the $(0,2)$ and $(2,2)$ blocks are populated by reusing the $(0,0)$ analytic result.
The brute-force $\rho$ is obtained from the same batch construction used for $\beta$.

The agreement between analytic and brute-force $\rho$ is very good across all multipole combinations.
Most importantly, $\rho$ is extremely close to $1$ on large scales and then drops sharply around
$k \simeq 0.2\,h\,{\rm Mpc}^{-1}$, becoming close to $0$ beyond this scale.
This is the key reason why the CV improvement saturates at higher $k$: the predicted variance reduction factor,
$\bigl(1-\rho^2\bigr)^{-1}$, becomes large only while $\rho^2$ remains close to $1$.

We note that our analytical $\rho$ depends only on the cross-correlation coefficient, $r(k)$, between our control variate and sample of interest, the covariance matrix of which we are aiming to determine. Thus, as long as $r(k)$ is close to one on large scales, which is guaranteed for cosmological tracers, we should find very good performance. The theoretical variance reduction we can achieve (provided our approximation for $\rho$ is sufficiently accurate) is given by 
\begin{equation}
\frac{\mathrm{Var}[X]}{\mathrm{Var}[Y]} = \frac{1}{1-\rho^2}\,,
\end{equation}
which we study in the next Section.


\subsection{Robustness to target--control velocity mismatch}
\label{sec:2lpt_validation}

As anticipated above, in the fiducial mocks the coherent large-scale velocity of the target is identical to that of the Zeldovich control field, and the decorrelation between the two is driven by the stochastic velocity dispersion, the lognormal transformation, and shot noise. To test the method under a controlled mismatch of the coherent velocity field as well, we repeat the analysis with the 2LPT target mocks of Sec.~\ref{sec:2lpt_mocks}, keeping the control field pure Zeldovich.

We generate $1000$ paired 2LPT realizations. Because the initial conditions are shared realization by realization with the fiducial run, the Zeldovich control spectra are reproduced to better than $10^{-5}$, so the fiducial and 2LPT configurations differ {only} in the target and the comparison below is strictly paired: the same realizations, the same batch draws, and the same $\mu_C$ are used in both cases. The 2LPT displacements do modify the target, shifting the monopole by $+0.8\%$ near $k\simeq0.05$--$0.1\,h\,{\rm Mpc}^{-1}$ and by $-6.2\%$ at $k=0.19\,h\,{\rm Mpc}^{-1}$.

\begin{figure*}
\centering
\includegraphics[width=0.48\textwidth]{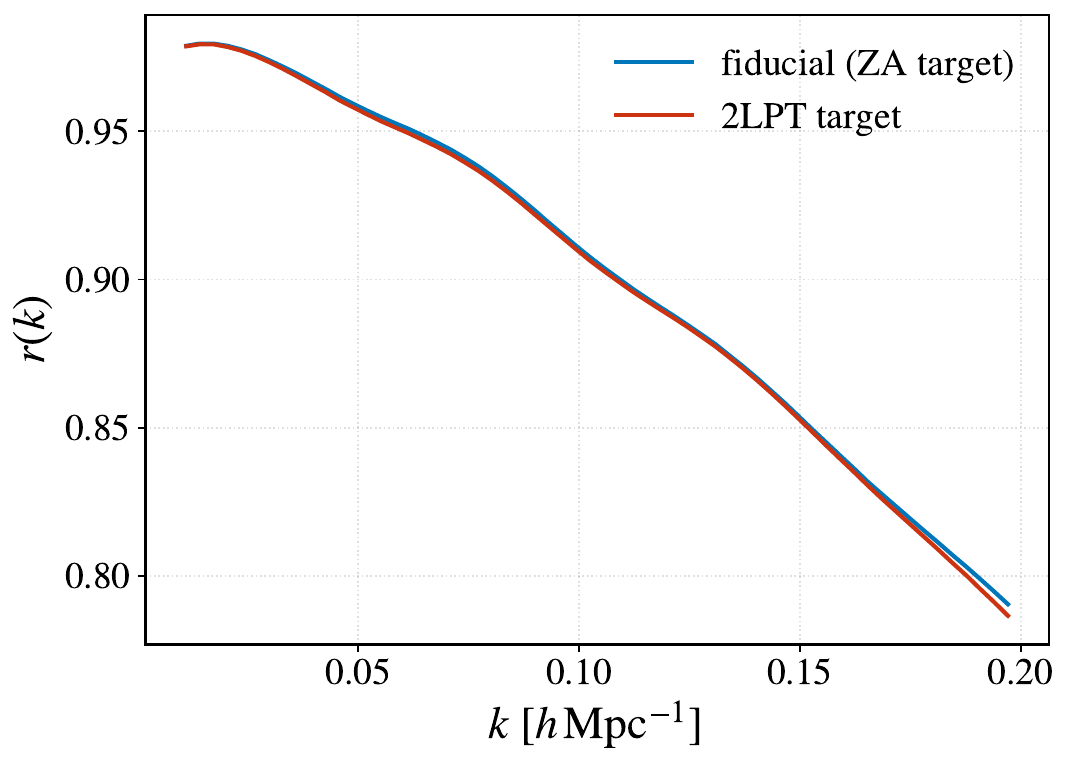}
\hfill
\includegraphics[width=0.48\textwidth]{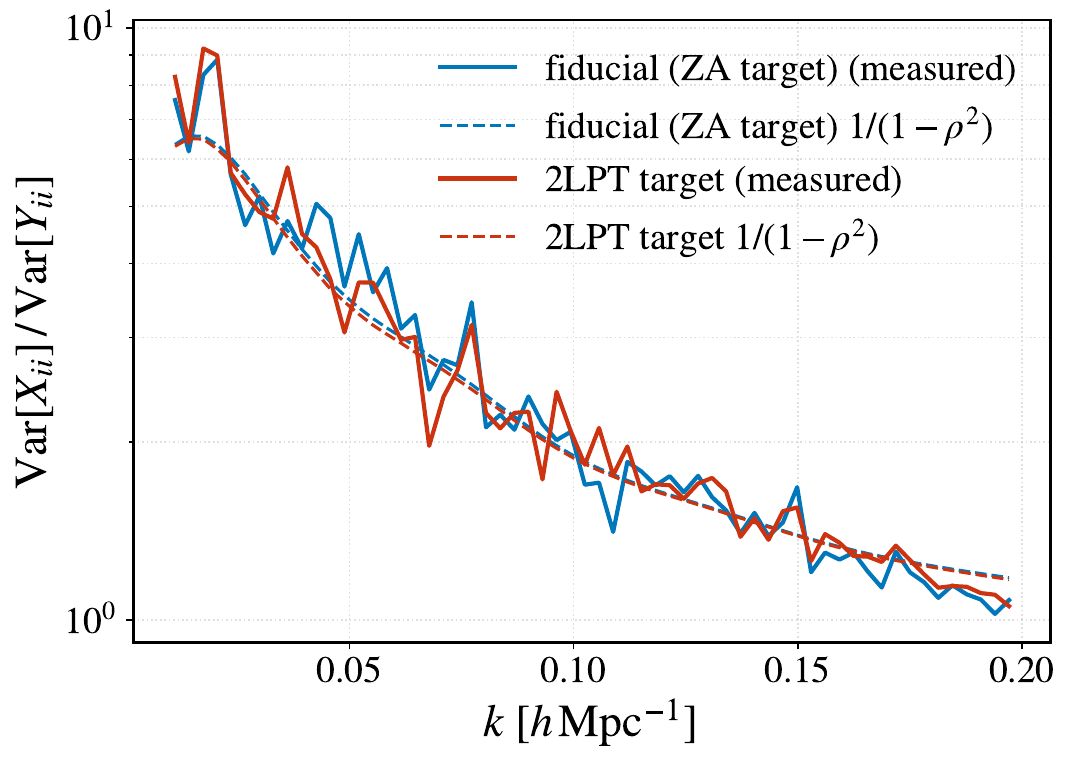}
\caption{Robustness of the control-variate gain to a mismatch between the coherent velocity fields of the target and the control. In both panels the fiducial configuration (Zeldovich targets) is compared with targets advanced to second order in Lagrangian perturbation theory, with the control held fixed at pure Zeldovich; the two runs share initial conditions realization by realization.
Left: the monopole cross-correlation coefficient $r(k)$.
Right: the measured variance reduction $\Var(X)/\Var(Y)$ (solid) compared with the optimal prediction $(1-\rho^2)^{-1}$ (dashed), evaluated with the analytic $\rho_{ii}=r^4(k_i)$, for each configuration.
The 2LPT targets leave both $r(k)$ and the achieved gain essentially unchanged: a paired bootstrap gives a gain ratio of $1.02\pm0.07$ on large scales.}
\label{fig:2lpt_robustness}
\end{figure*}

The effect on the target--control cross-correlation is nevertheless very small. At $k=0.011\,h\,{\rm Mpc}^{-1}$ the monopole correlation coefficient decreases from $r=0.9788$ to $r_{\rm 2LPT}=0.9787$, and its median over $0.01<k<0.2\,h\,{\rm Mpc}^{-1}$ from $0.9057$ to $0.9046$; equivalently, $1-r$ grows by only $\simeq0.5\%$ of itself. The variance reduction is correspondingly unchanged: on the largest scales ($k<0.05\,h\,{\rm Mpc}^{-1}$) the measured gain is $5.05$ in the fiducial case and $5.24$ with 2LPT targets, while for $k>0.10\,h\,{\rm Mpc}^{-1}$ it is $1.4$ in both. A bootstrap over the batch draws, paired between the two configurations, gives a 2LPT-to-fiducial gain ratio of $1.02\pm0.07$ for $k<0.05\,h\,{\rm Mpc}^{-1}$ and $1.03\pm0.03$ for $k>0.10\,h\,{\rm Mpc}^{-1}$, both consistent with unity. This is shown in Fig.~\ref{fig:2lpt_robustness}.

We speculate that this insensitivity has a simple origin: $\vec\Psi^{(2)}$ is a deterministic functional of the {same} linear density field that generates the Zeldovich control, so promoting the target to second order shifts its mean clustering without injecting stochasticity that the control cannot track. Breaking the exact coincidence of the coherent velocity fields is therefore not, by itself, sufficient to degrade the CV; what degrades it is genuine decorrelation, of the kind already supplied in our fiducial setup by the velocity dispersion, the lognormal transformation, and shot noise.

Crucially, the measured gain continues to track the optimal relation $\left(1-\rho^2\right)^{-1}$ to the same accuracy as in the fiducial case. Following the discussion of Sec.~\ref{sec:simulation_gain}, we evaluate the prediction using $\rho$ measured directly from the batches as well as from the analytical formula. The two determinations of $\rho$ agree closely with each other, so neither the near-optimality of the estimator nor the accuracy of the analytic prediction is degraded by the target--control mismatch.

This test illustrates the central practical point: the performance of the estimator degrades gracefully and {predictably} as the target and control fields decorrelate. Since $r(k)$ is directly measurable from a handful of paired realizations at negligible cost, the expected gain for any target mock suite (including full $N$-body mocks, where the ZA velocity field is likewise not exact) can be forecast before committing to the full covariance campaign.

\begin{figure*}
\centering
\includegraphics[width=0.48\textwidth]{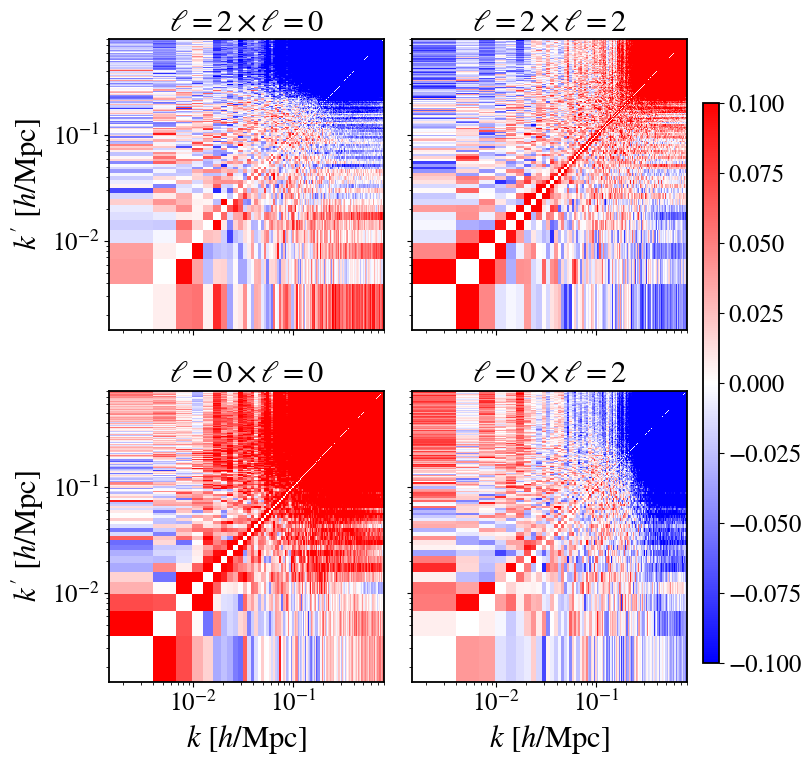}
\hfill
\includegraphics[width=0.48\textwidth]{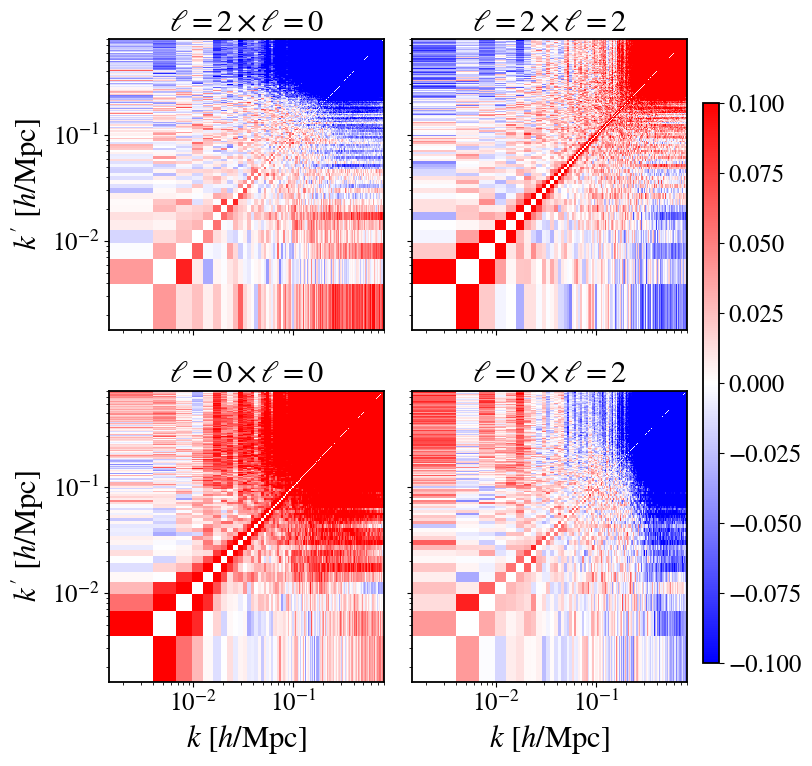}
\hfill
\caption{%
Effect of control variates on redshift-space covariance estimates for the realistic (masked) survey.
Left: correlation matrix (diagonal subtracted) of the original covariance $X$ estimated from $1000$ simulations.
Right: correlation matrix (diagonal subtracted) of the CV-corrected covariance $Y$ (same number of simulations).
Each big panel contains the $2\times 2$ block structure corresponding to $(\ell,\ell')=(0,0),(0,2),(2,0),(2,2)$.
The CV preserves the qualitative correlation structure while visibly reducing the noise (i.e., the map on the right appears smoother and with fewer sharp features).
}
\label{fig:corr_X_Y_reduction}
\end{figure*}

\begin{figure*}
\centering
\includegraphics[width=\textwidth]{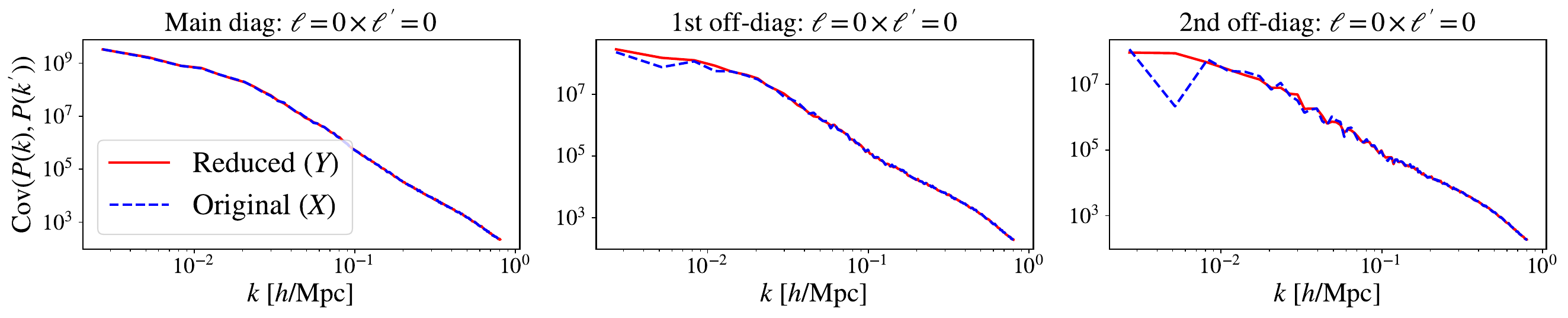}
\\[0.6em]
\includegraphics[width=\textwidth]{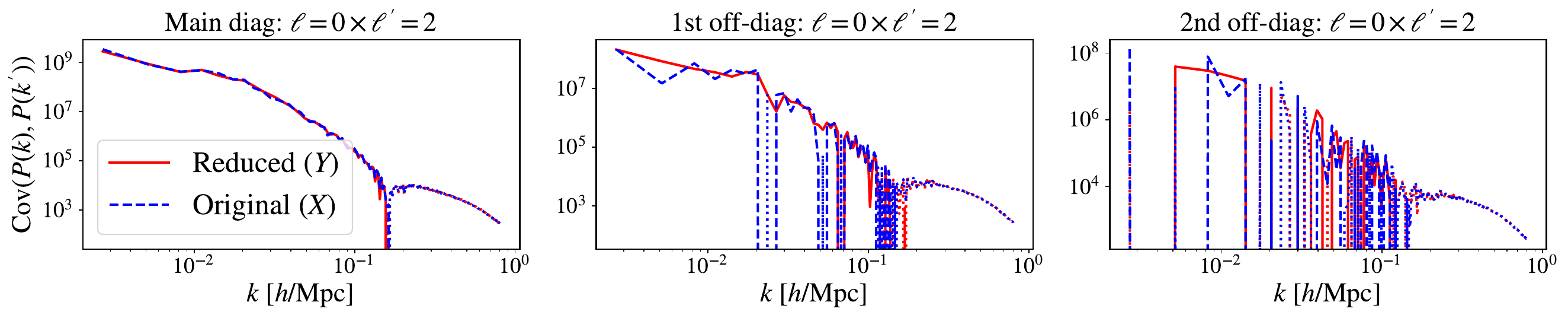}
\\[0.6em]
\includegraphics[width=\textwidth]{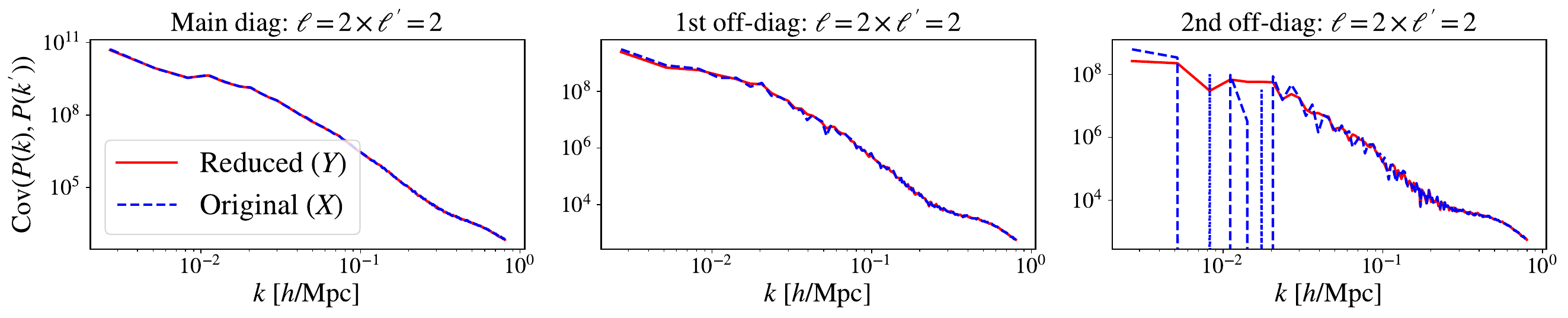}
\caption{%
Diagonal comparison between the original covariance estimate $X$ and the CV-corrected covariance estimate $Y$ in redshift space with mask.
The three rows correspond to $(\ell,\ell')=(0,0)$ (top), $(0,2)$ (middle), and $(2,2)$ (bottom).
Within each row, the left/center/right subpanels correspond to the main diagonal, first off-diagonal, and second off-diagonal in $(k_i,k_j)$.
Each subpanel shows $X$ (original) and $Y$ (CV-reduced).
The CV reduces noise and oscillations while preserving the diagonal values (no evident bias), demonstrating that the control variates indeed act as intended.  }
\label{fig:XY_diagonals}
\end{figure*}

\section{Results}
\label{sec:results_rsd_mask}

In this section, we study the key performance of our CV method. We first visualize the resulting CV-reduced covariance matrix (to highlight noise and off-diagonal structure) and make comparisons with its non-reduced version via diagonal slices (to directly demonstrate variance suppression and lack of bias).
We then show how the variance reduction translates into an effective gain in simulation number.

We largely confine our attention to $k\ge 0.01\,h\,\mathrm{Mpc}^{-1}$.  At $k<0.01\,h\,\mathrm{Mpc}^{-1}$ analytic approximations to the covariance matrix become very accurate, obviating the need for a numerical approach. At the same time, the assumption of a Gaussian likelihood for $P_\ell(k)$ becomes more questionable.

\subsection{Covariance matrices: $X$ versus $Y$}
\label{sec:corr_X_Y}

We assess the CV impact on the covariance estimates by comparing correlation matrices. This comparison is closely related to our broader discussion of testing the accuracy and convergence of our CV-reduced covariance matrices in Section~\ref{sec:other_metrics}.

Fig.~\ref{fig:corr_X_Y_reduction} presents:
(1) the correlation matrix of the original covariance $X$ estimated from $1000$ simulations;
(2) the correlation matrix of the CV-corrected covariance $Y$ estimated from $1000$ simulations. It aims to visually show the improvement of the CV-reduced covariance.

On large scales, correlations are comparatively weak: the correlation matrices are close to diagonal, with only modest mode coupling visible.
At smaller scales, the covariance becomes strongly correlated in $k$-bins, with correlations showing non-trivial signs across different multipole blocks
(e.g.\ anti-correlations for $(0,2)$-type blocks, and positive correlations for $(0,0)$ and $(2,2)$).
Most importantly, $X$ and $Y$ appear qualitatively similar (consistent with the unbiasedness of the method), but $Y$ is visibly cleaner: the off-diagonal bands are smoother and exhibit reduced noise.
The improvement is most visible where the correlation/noise is largest, i.e.\ on large scales where $\rho$ is close to $1$.


Correlation matrices can be visually compelling but they normalize by the diagonal elements, making it difficult to interpret variance reduction directly.
Fig.~\ref{fig:XY_diagonals} therefore shows diagonal slices of $X$ and $Y$ in a more direct way.

For each multipole block pair $(\ell,\ell')=(0,0)$, $(0,2)$, and $(2,2)$, we plot:
the main diagonal, the first off-diagonal, and the second off-diagonal.
In each subpanel, we show two curves: the original covariance $X$ and the CV-corrected covariance $Y$.

The reduced covariance $Y$ is noticeably more stable and less noisy than $X$, with reduced oscillatory behavior at high $k$.
Crucially, $Y$ remains consistent with $X$ in the relevant large-scale regime, supporting the conclusion that the method is not introducing a bias but is primarily suppressing noise.

\subsection{Effective simulation gain}
\label{sec:simulation_gain}

The correlation coefficient $\rho$ can be converted into an effective simulation gain.
In the CV framework, the variance reduction factor is approximately $\mathrm{Var}[Y]/\mathrm{Var}[X]\simeq (1-\rho^2)$, whose inverse we can interpret as an effective gain in the number of paired simulations.

\begin{figure*}
\centering
\includegraphics[width=0.48\textwidth]{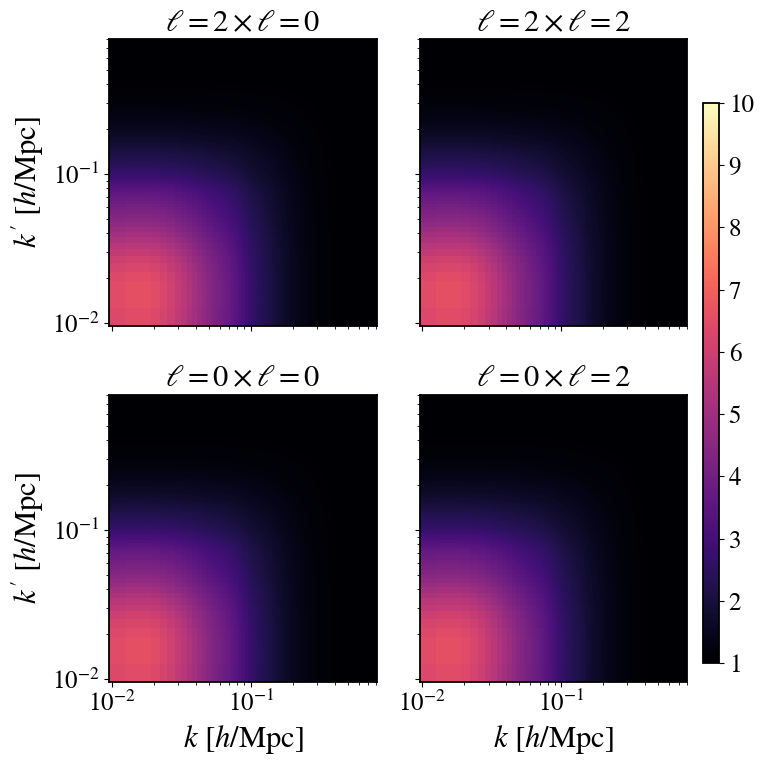}
\hfill
\includegraphics[width=0.48\textwidth]{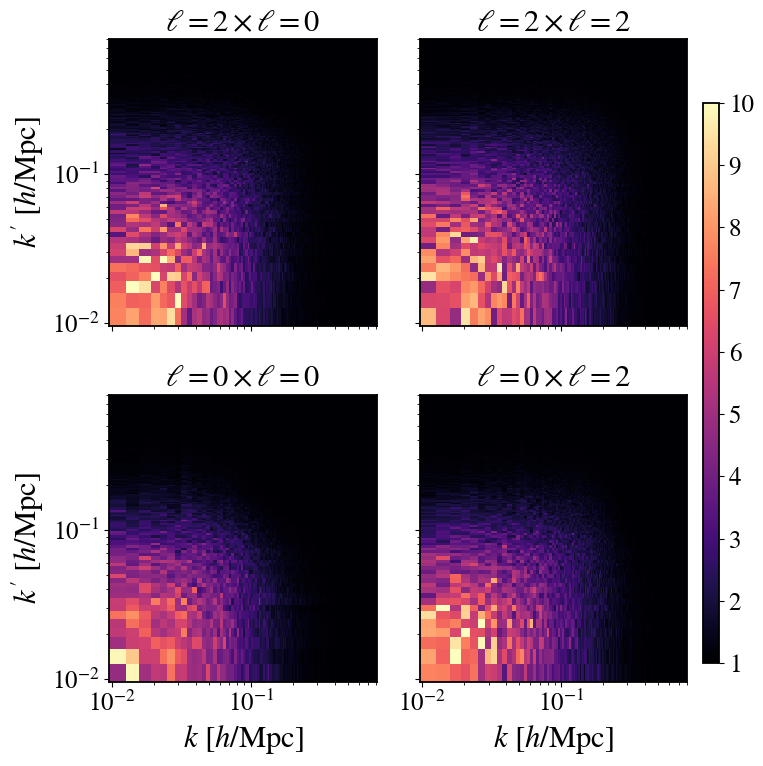}
\caption{%
Effective simulation gain (variance reduction) in redshift space with survey mask.
Left: analytic prediction using the CV relation $\mathrm{Var}[X]/\mathrm{Var}[Y]\simeq (1-\rho^2)^{-1}$ (with multipole blocks populated by reusing the $(0,0)$ block).
Right: brute-force measurement of the corresponding variance reduction factor from $100\times 100$ batch statistics.
Each panel shows the $2\times 2$ block structure $(\ell,\ell')=(0,0),(0,2),(2,0),(2,2)$.
The agreement between analytic and brute-force is very good; gains are strongest on large scales (typically $\sim 8$), and decline toward unity at higher $k$ where $\rho$ becomes small.
}
\label{fig:Nsim_gain_2d}
\end{figure*}

Fig.~\ref{fig:Nsim_gain_2d} shows two maps side by side:
(i) the analytic prediction for the effective simulation gain using $\left(1-\rho^2\right)^{-1}$;
(ii) the brute-force gain measured directly from the batch estimates of $\mathrm{Var}[X]/\mathrm{Var}[Y]$.

Both maps are displayed with the same $2\times 2$ block ordering $(\ell,\ell')=(0,0),(0,2),(2,0),(2,2)$.

On large scales, the gain is close to maximal and is of order $\sim 8$ in this realistic masked redshift-space setup.
The gain decreases toward smaller scales, consistent with the sharp drop of $\rho$ observed around $k\simeq 0.2\,h\,{\rm Mpc}^{-1}$.
This behavior is expected: once $\rho$ approaches $0$, the factor $\left(1-\rho^2\right)^{-1}$ saturates to unity, and no further variance reduction should be expected.

A gain factor of $\sim8$ means that, at fixed estimator variance, the CV-corrected covariance from $\sim 100$ paired simulations achieves the statistical quality that would otherwise require $\sim 800$ unpaired simulations.

Because $\rho$ entering the covariance CV is effectively linked to the (roughly squared) correlation structure of the underlying power-spectrum control field, the covariance-level CV performance tracks the power-spectrum-level performance but with somewhat smaller gains.
Empirically, this means the covariance CV is typically only about a factor of $\sim 2$ worse than what is achievable for the power spectrum itself, which is consistent with large improvements in the covariance even in this highly challenging, realistic masked/redshift-space configuration.

\begin{figure*}
\centering
\includegraphics[width=\textwidth]{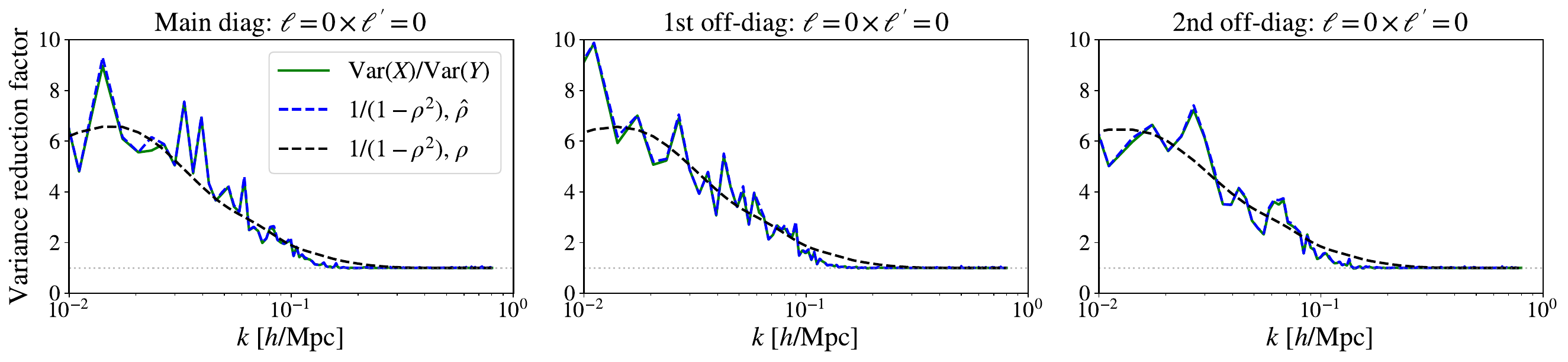}
\\[0.6em]
\includegraphics[width=\textwidth]{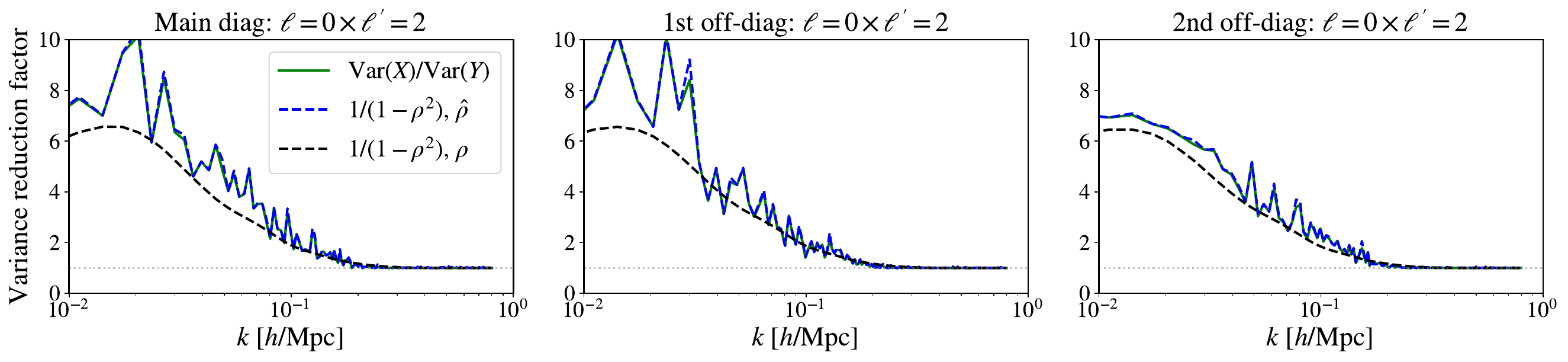}
\\[0.6em]
\includegraphics[width=\textwidth]{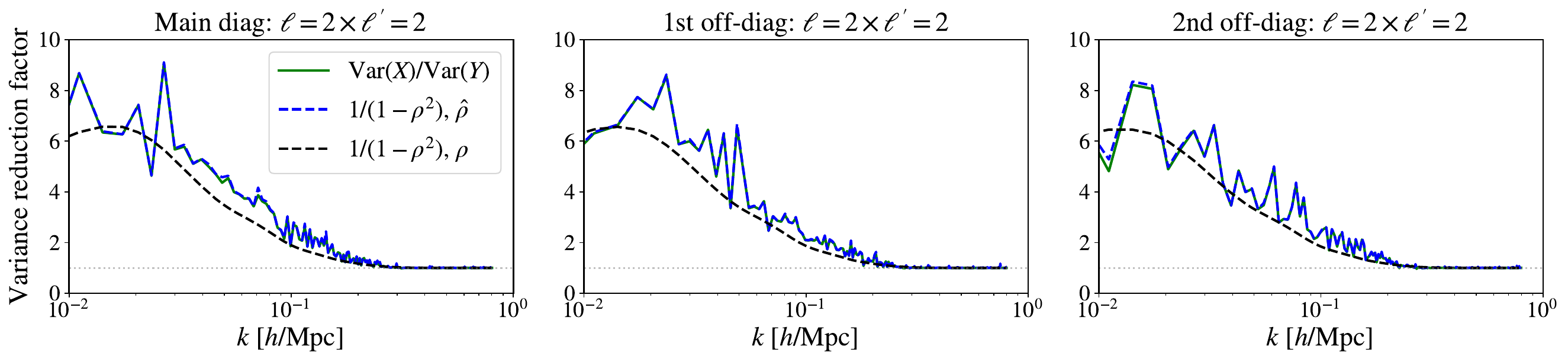}
\caption{%
Variance ratio $\mathrm{Var}[X]/\mathrm{Var}[Y]$ on diagonal slices for the realistic redshift-space, masked survey.
The three rows correspond to $(\ell,\ell')=(0,0)$ (top), $(0,2)$ (middle), and $(2,2)$ (bottom).
Within each row, the main diagonal, first off-diagonal, and second off-diagonal are shown.
Three curves are overplotted: the measured variance reduction from brute-force batching (green curve), the analytic prediction $(1-\rho^2)^{-1}$ using the estimated $\hat \rho$ from the batches (blue), and the analytic prediction $(1-\rho^2)^{-1}$ using the analytic $\rho$ (black).
The analytic prediction from the batches matches the measured variance reduction very well, whereas the analytic $\rho$ slightly underestimates the improvement.
The variance ratio $\sim 7$ up to $k\simeq 0.03\,h\,{\rm Mpc}^{-1}$, decreases to $\sim 5$ until $k\simeq 0.1\,h\,{\rm Mpc}^{-1}$, and approaches unity for $k\gtrsim 0.2\,h\,{\rm Mpc}^{-1}$.
On smaller scales, brute-force indicates a modestly stronger improvement for $(2,0)$ and $(2,2)$ than for $(0,0)$ (by about $\sim 10\%$).
}
\label{fig:variance_reduction_diagonals}
\end{figure*}

To quantify where the variance reduction is concentrated, Fig.~\ref{fig:variance_reduction_diagonals} shows the diagonal variance ratio $\mathrm{Var}[X]/\mathrm{Var}[Y]$ as a function of $k$ for the three block pairs.
Each block includes the main diagonal, first off-diagonal, and second off-diagonal.
We compare:
(i) the brute-force measured variance ratio;
(ii) the analytic prediction $(1-\rho^2)^{-1}$ using the batches
(iii) the analytic prediction $(1-\rho^2)^{-1}$ using our analytical approximation.

Across diagonal bands, the analytic prediction from the batches matches the brute-force measurement very well.
The variance ratio is largest on large scales:
it is about a factor of $\sim 8$ up to $k\simeq 0.03\,h\,{\rm Mpc}^{-1}$, then decreases to $\sim 5$ until $k\simeq 0.1\,h\,{\rm Mpc}^{-1}$.
After $k\simeq 0.2\,h\,{\rm Mpc}^{-1}$, the ratio becomes consistent with unity, reflecting the loss of correlation ($\rho\rightarrow 0$) at those scales.

The discrepancy between the measured gain and the analytic prediction using our approximation for $\rho$ can be attributed to the approximations we make in estimating $\rho$. 
Indeed, in our diagonal blocks we find that the measured gain agrees with the optimal relation $1/(1-\rho^2)$ when $\rho$ is computed directly from the $100\times 100$ batches, for both analytic and brute-force estimates of $\beta$. 
This indicates that the underlying optimal $\beta$--$\rho$ connection remains valid. 
We expect that on large scales the variance reduction should be even larger as the Zeldovich approximation approaches the truth. We attribute the slight dip seen as $k\to0$ to small numerical artifact due to mask edge effects.
The sensitivity is amplified when $\rho\simeq 1$, where the gain scales as
\[
\frac{1}{1-\rho^2}=\frac{1}{(1-\rho)(1+\rho)} \approx \frac{1}{2(1-\rho)} ,
\]
so small differences in $\rho$ translate into relatively large changes in the inferred gain.

An additional feature is that the brute-force comparison suggests slightly better improvement for the $(2,0)$ and $(2,2)$ blocks than for $(0,0)$ on smaller scales (at the level of $\sim 10\%$), even though the analytic construction reuses the $(0,0)$ block.

\subsection*{Summary of findings}
Across the suite of diagnostics (analytic vs brute-force $\beta$ and $\rho$, correlation-matrix comparisons, diagonal stability checks, and effective simulation gain), the CV method behaves consistently with its theoretical design:
(i) $\beta$ and $\rho$ agree well with brute-force measurements on the scales that matter most;
(ii) $Y$ closely matches the structure of $X$ while reducing estimator noise;
(iii) the predicted and measured reduction in variance agree, with large gains on large scales and diminishing returns once correlations vanish around $k\simeq 0.2\,h\,{\rm Mpc}^{-1}$.

\section{Other metrics}
\label{sec:other_metrics}

In this section, we supplement the primary results with accuracy and consistency checks designed to validate that the control variate (CV) does not introduce biases in the recovered covariance. In particular, we study the behavior of the CV-corrected observables in several complementary comparisons to confirm agreement with the reference predictions within the expected finite-realization scatter. We then assess convergence by varying the number of mock realizations used to estimate the covariance (and related quantities). These tests address how quickly the covariance estimate approaches a reference value and whether the CV method improves the required number of mocks, especially in the precision-matrix–sensitive regime.

We confine the range of analyzed modes to
$0.01\,h\,\mathrm{Mpc}^{-1}<k<0.2\,h\,\mathrm{Mpc}^{-1}$. 
This choice is motivated by two considerations. First, it targets precisely the scales most relevant for large-scale structure inference, where perturbative and analytically controlled approaches, such as EFT-based methods, remain well supported, and where upcoming surveys including DESI (and likely Euclid, PFS, \textit{SPHEREx}, and LSST) are expected to carry substantial constraining power. 
Second, extending the analysis to smaller scales introduces significant finite-sample noise that manifests as numerical instabilities in covariance-matrix operations, including poor conditioning for matrix inversions and the appearance of nonphysical behavior (e.g.\ negative eigenvalues). 
With the above cut, the covariance matrix is correspondingly more stable and better aligned with the modes actually used for cosmological inference.
Applying this restriction leaves a total of $N_{\rm data}=120$ data points, corresponding to $60$ $k$-bins and two multipoles, $\ell=0,2$.

\subsection{Accuracy tests}
\label{sec:accuracy_tests}

Here, we probe whether the control-variate covariance matrix exhibits only per-element noise reduction, or whether it also features an improvement in its matrix properties. Overall we show that the CV-reduced estimate converges faster and reproduces the covariance geometry (and the corresponding precision-matrix behavior entering likelihoods) more accurately than the original covariance estimate.

\begin{figure}[t]
  \centering
  \includegraphics[width=\linewidth]{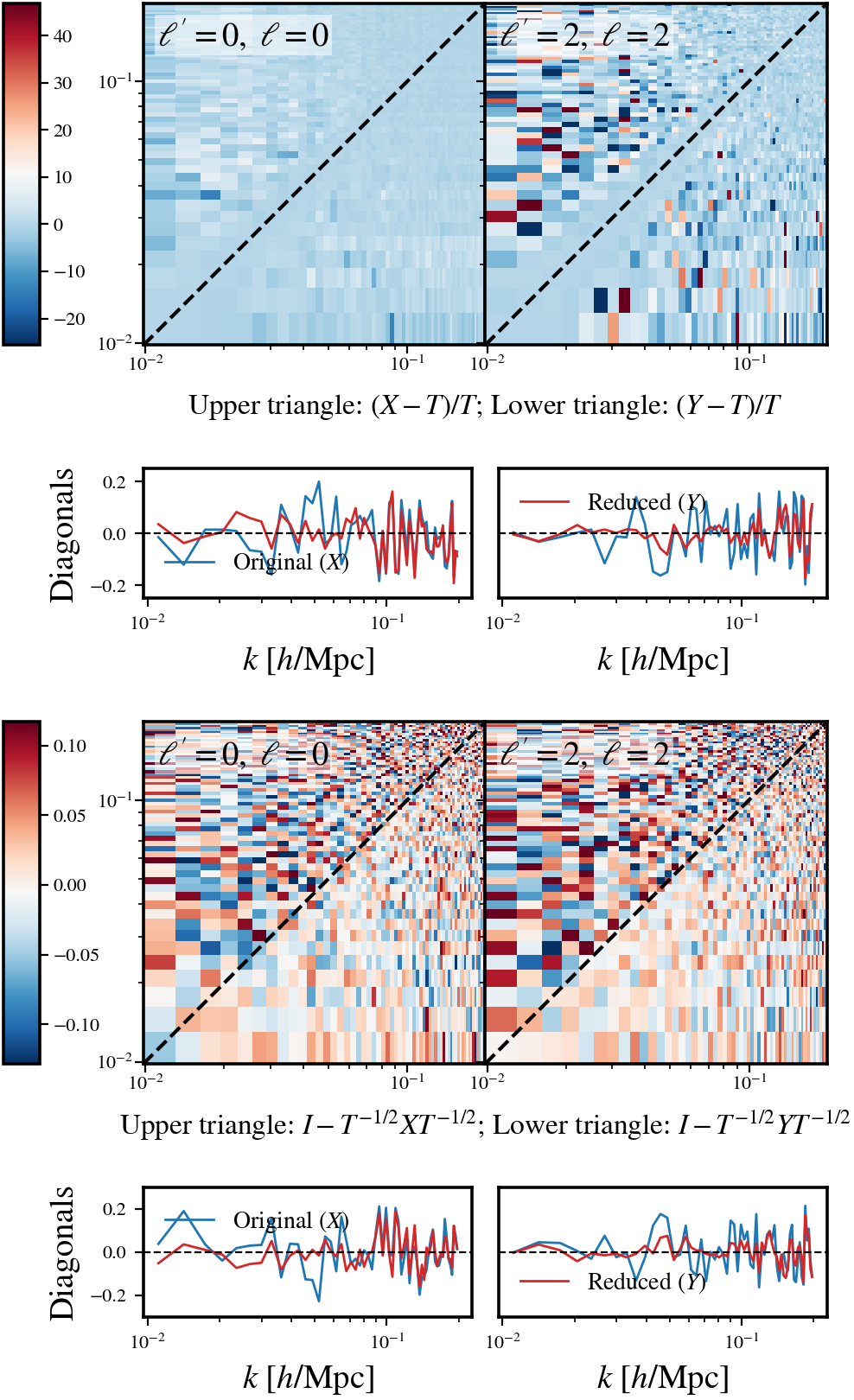}
  \caption{
  Accuracy comparison between the original covariance estimate $X$ and the control–variate (CV) covariance estimate $Y$, both for $N_{\rm sim}=200$, relative to the reference ``true'' covariance $T$ constructed from $10{,}000$ realizations.
  Top panel: relative deviations $(X_{ij}-T_{ij})/T_{ij}$ (upper triangle) and $(Y_{ij}-T_{ij})/T_{ij}$ (lower triangle).
  Bottom panel: inverse-covariance geometry test
  $O \equiv I - T^{-1/2}\,X\,T^{-1/2}$ (upper triangle) and
  $O \equiv I - T^{-1/2}\,Y\,T^{-1/2}$ (lower triangle).
  The remaining 1D panels show the diagonal elements: the blue curve corresponds to $X$ and the red curve corresponds to $Y$.
  We see that the CV-reduced covariance is less noisy and closer to the truth on large scales, and recovers the same behavior as the original on small scales. 
  }
  \label{fig:cv_accuracy_XYT_200}
\end{figure}

Throughout this subsection we define the ``true'' covariance as
\begin{equation}
T \equiv \hat{C}_{\rm true},
\end{equation}
obtained from $10{,}000$ independent realizations. The covariance estimated from $N_{\rm sim}$ realizations is denoted by
\begin{equation}
X \equiv \hat{C}_{\rm original}(N_{\rm sim}), 
\qquad
Y \equiv \hat{C}_{\rm CV}(N_{\rm sim}),
\end{equation}
and we keep $N_{\rm sim}=200$ for the main visual comparison in Fig.~\ref{fig:cv_accuracy_XYT_200}. Qualitatively, we have checked that the behavior is unchanged for a larger number of simulations.

Fig.~\ref{fig:cv_accuracy_XYT_200} shows a comparison between $Y$ and $X$ relative to $T$ in terms of two metrics: relative deviations $(X_{ij}-T_{ij})/T_{ij}$ (upper triangle) and $(Y_{ij}-T_{ij})/T_{ij}$ (lower triangle) and an inverse-covariance geometry test
$O \equiv I - T^{-1/2}\,X\,T^{-1/2}$ (upper triangle) and
$O \equiv I - T^{-1/2}\,Y\,T^{-1/2}$. We find that the CV-reduced covariance $Y$ is systematically closer to $T$ than the original covariance $X$, with the difference concentrated on large scales, $k \lesssim 0.1 h/{\rm Mpc}$, where the covariance is most expensive to estimate numerically.

On large scales (corresponding to $k \lesssim 0.1\,h/{\rm Mpc}$ in the plotted diagonal index), the diagonal deviations of $X$ from $T$ are substantially noisier than those of $Y$.
On small scales the two estimates are indistinguishable at the level of their mutual agreement.

Although not shown in the figure we note that while $X$ and $Y$ agree closely with each other on small scales, both can still disagree with $T$ at the $\sim 20\%$ level for the top-panel diagnostic. This is not a CV-induced bias as it features in the comparison of both $X$ and $Y$ with T. It is a convergence `artifact' of the reference covariance used for $T$ in that particular diagnostic setting. Concretely, when we increase the number of mocks used in the estimated covariance from $200$ to $5{,}000$, the apparent small-scale bias is reduced to the $\sim 1\%$ level, while leaving the large-scale (noise-driven) behavior unchanged. The large-scale deviations are not biased in this sense; rather, they are dominated by finite-sample noise, which CV substantially reduces.

The bottom-panel diagnostics are particularly reassuring for likelihood applications.
In the $O$-matrix test we do not see a corresponding bias in the inverse-covariance geometry on either large or small scales.
Instead, what stands out is that the magnitude of any large-scale deviation is larger for $X$ than for $Y$: the upper triangle (original $X$) exhibits stronger departure from the reference geometry, while the lower triangle (CV-reduced $Y$) stays closer to zero deviation.
The 1D panel for the diagonal of $O$ shows the same qualitative story as the top panel: $Y$ matches the reference more closely on large scales, while both are similar on small scales.

We also performed a closure test of the finite-sample effect:
starting from a high-accuracy reference covariance $T$ from a large set of mocks ($10{,}000$ realizations),
we repeated the covariance estimation with more mocks ($1{,}000$ realizations) for both the original estimator $X$ and the CV estimator $Y$,
and compared the resulting deviations relative to the same $T$ reference.
As expected, using $1{,}000$ mocks decreases realization-to-realization bias in general.
Deviations obtained with $Y$ are systematically smaller than those obtained with $X$, indicating that CV reduces finite-realization fluctuations without producing a detectable systematic bias in the recovered covariance.

Finally, we tested an eigenvalue-ordering diagnostic (which we ultimately did not include as a figure).
By ordering the eigenvalues of both $X$ and $Y$ and comparing them to the corresponding reference eigenvalues from $T$, we found that the CV estimator does not bias the recovered eigenvalue spectrum and is {less noisy} than the original $X$ estimator.
This provides further support that $Y$ preserves the matrix behavior relevant for inference while improving stability.

\subsection{Convergence tests}
\label{sec:convergence_tests}

We now quantify how covariance- and precision-sensitive metrics improve as the number of mocks used to estimate the covariance increases. We show that the CV-reduced covariance $Y$ converges both faster and more accurately toward the correct matrix behavior than the original covariance $X$.

\begin{figure}[t]
  \centering
  \includegraphics[width=\linewidth]{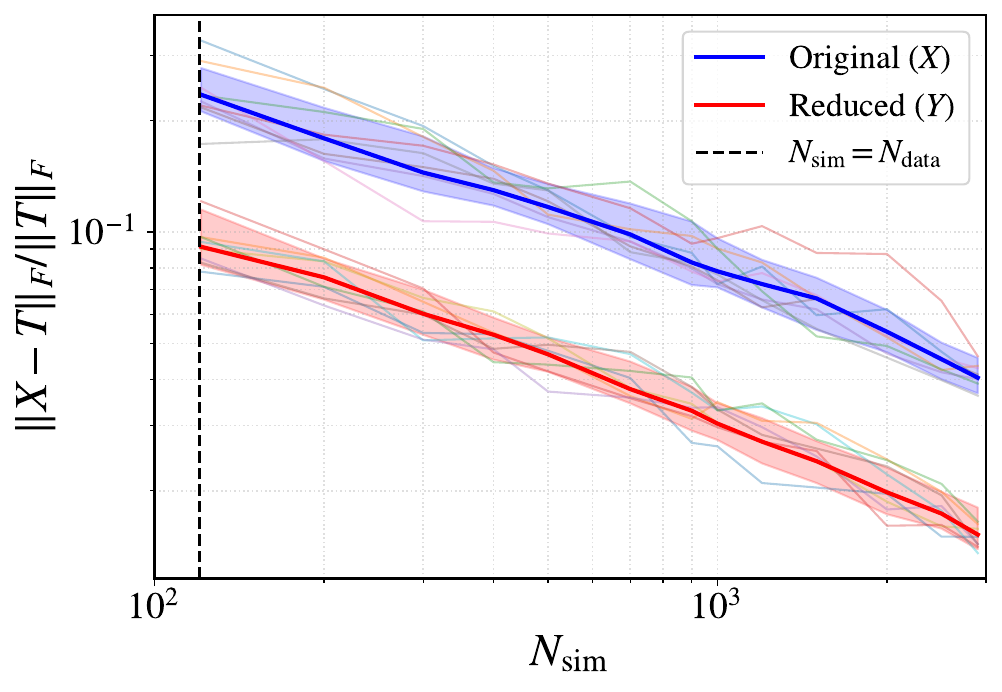}
  \vspace{0.25em}
  \includegraphics[width=\linewidth]{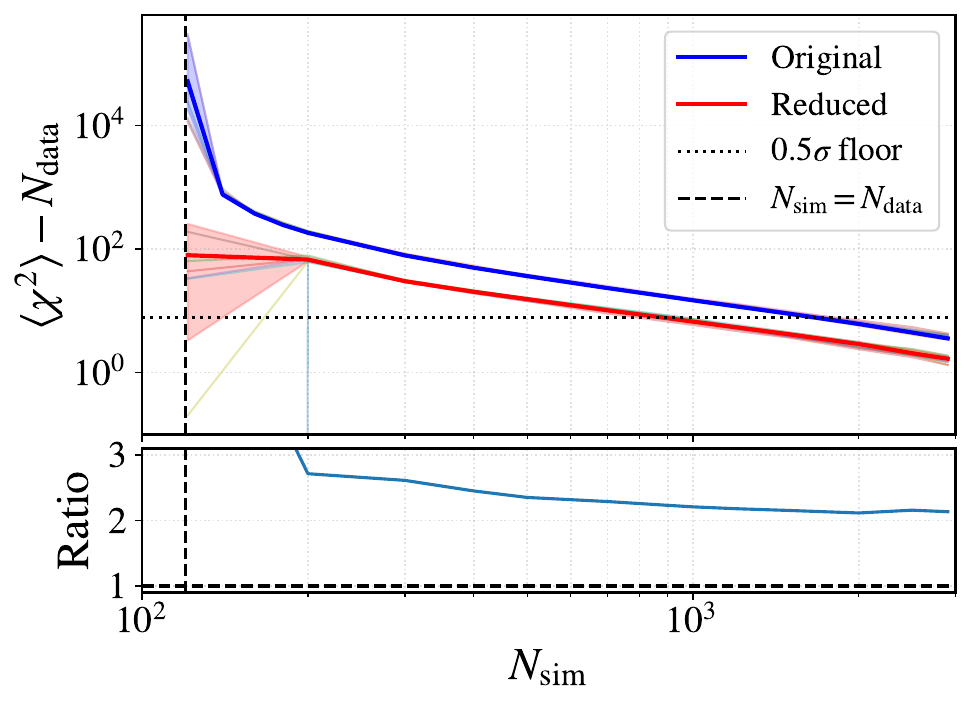}
  \caption{
  Convergence of covariance and precision-matrix–sensitive diagnostics as a function of the number of mock realizations $N_{\rm sim}$ used to construct the covariance estimate.
  Top panel: the Frobenius distance between the estimated covariance (original $X$ or CV-reduced $Y$) and the reference covariance $T$ (constructed from $10{,}000$ realizations).
  Bottom panel: the average $\chi^2$ computed from the estimated precision matrices, showing $\langle \chi^2 \rangle = {\rm Tr}(T\,X^{-1})$ for the original estimator and ${\rm Tr}(T\,Y^{-1})$ for the CV estimator,
  alongside a fractional-difference subpanel comparing the two.  The dotted horizontal line shows $\langle \chi^2\rangle-N_{\rm data}<0.5\,\sigma(\chi^2)$, which we find is achieved at $N_{\rm sim}\!\sim\!900$ with the CV-reduced matrix and at $N_{\rm sim}\!\sim\!1800$ with the non-CV matrix.
  The original estimator $X$ cannot be reliably inverted for small $N_{\rm sim}$ (specifically when $N_{\rm sim}\sim N_{\rm data}$),
  so it appears noisy in that regime.
  By contrast, $Y$ remains invertible and produces a stable, sensible estimate already at $N_{\rm sim} = N_{\rm data}$, indicating that CV effectively behaves like a shrinkage estimator.
  }
  \label{fig:cv_convergence_frob_chi2}
\end{figure}

The top panel of Fig.~\ref{fig:cv_convergence_frob_chi2} demonstrates a substantial improvement in the covariance estimate itself.
Across the range of $N_{\rm sim}$ explored, the CV-reduced estimate $Y$ achieves a Frobenius distance to $T$ smaller than that of $X$ by roughly a factor of $2$--$3$ (approximately constant with $N_{\rm sim}$). This is also shown in Table~\ref{tab:XY_byN} alongside other matrix comparison metrics.

\begin{table*}[htbp]
\centering
\resizebox{\textwidth}{!}{%
\begin{tabular}{l|cc|cc|cc|cc|cc}
\hline
Metric & \multicolumn{2}{c}{$125$} & \multicolumn{2}{c}{$250$} & \multicolumn{2}{c}{$500$} & \multicolumn{2}{c}{$1000$} & \multicolumn{2}{c}{$2000$} \\
 & X & Y & X & Y & X & Y & X & Y & X & Y \\
\cline{2-3} \cline{4-5} \cline{6-7} \cline{8-9} \cline{10-11}
\hline
$\log_{10}||C - T||_F$ & $9.12\pm 0.056$ & $8.71\pm 0.055$ & $8.97\pm 0.064$ & $8.54\pm 0.082$ & $8.81\pm 0.054$ & $8.39\pm 0.073$ & $8.64\pm 0.066$ & $8.24\pm 0.057$ & $8.48\pm 0.057$ & $8.06\pm 0.059$ \\
$\log_{10}||C - T||_2$ & $8.93\pm 0.11$ & $8.51\pm 0.11$ & $8.8\pm 0.11$ & $8.36\pm 0.12$ & $8.63\pm 0.1$ & $8.23\pm 0.15$ & $8.49\pm 0.14$ & $8.07\pm 0.1$ & $8.31\pm 0.097$ & $7.87\pm 0.11$ \\
$\log_{10}||C^{-1} - T^{-1}||_F$ & $-1.91\pm 0.23$ & $-2.99\pm 0.39$ & $-3.77\pm 0.034$ & $-4.03\pm 0.041$ & $-4.22\pm 0.021$ & $-4.42\pm 0.024$ & $-4.52\pm 0.019$ & $-4.67\pm 0.026$ & $-4.77\pm 0.02$ & $-4.9\pm 0.021$ \\
$\log_{10}||C^{-1} - T^{-1}||_2$ & $-1.99\pm 0.25$ & $-3.09\pm 0.46$ & $-4.04\pm 0.048$ & $-4.28\pm 0.07$ & $-4.52\pm 0.043$ & $-4.7\pm 0.047$ & $-4.82\pm 0.045$ & $-4.97\pm 0.071$ & $-5.08\pm 0.048$ & $-5.22\pm 0.057$ \\
KL divergence $\times 10^{1}$ & $5.36\pm 0.13$ & $1.92\pm 0.11$ & $1.74\pm 0.038$ & $0.771\pm 0.019$ & $0.763\pm 0.012$ & $0.342\pm 0.0066$ & $0.344\pm 0.0049$ & $0.157\pm 0.0025$ & $0.151\pm 0.0021$ & $0.0696\pm 0.0013$ \\
log-Euclidean $\times 10^{1}$ & $1.89\pm 0.058$ & $1.16\pm 0.069$ & $0.747\pm 0.013$ & $0.468\pm 0.0096$ & $0.443\pm 0.006$ & $0.289\pm 0.0037$ & $0.282\pm 0.0033$ & $0.191\pm 0.0021$ & $0.182\pm 0.0022$ & $0.125\pm 0.0017$ \\
Bures--Wasserstein $\times 10^{4}$ & $1.99\pm 0.12$ & $0.844\pm 0.055$ & $1.38\pm 0.086$ & $0.581\pm 0.041$ & $0.971\pm 0.047$ & $0.405\pm 0.025$ & $0.666\pm 0.041$ & $0.281\pm 0.015$ & $0.447\pm 0.024$ & $0.186\pm 0.0098$ \\
$\langle \chi^2\rangle$ scatter (post-Hartlap) & --- & --- & --- & --- & 0.872 & 0.456 & 0.455 & 0.321 & 0.308 & 0.272 \\
\hline
\end{tabular}
}
\caption{
Comparison of the original covariance estimate $X$ and the control–variate covariance estimate $Y$ across $N_{\rm sim}$.
The reference ``true'' covariance is $T$, constructed from $10{,}000$ independent realizations.
In the rows labeled with $||C-T||$, $C$ denotes the estimated covariance (either $X$ or $Y$).
All quantities are computed by comparing each estimator to $T$, using metrics sensitive to both covariance and matrix-geometry structure. In all cases, we find that the CV-reduced covariance matrix is substantially closer to the `truth' compared with its non-CV-reduced counterpart.
}
\label{tab:XY_byN}
\end{table*}

The bottom panel focuses on a likelihood-relevant diagnostic based on inverse covariance,
\begin{equation}
\langle \chi^2 \rangle \propto {\rm Tr}(T\,C^{-1}),
\end{equation}
with $C=X$ or $C=Y$.
We note that $X^{-1}$ is not well-defined for $N_{\rm sim}<N_{\rm data}$ (the estimated covariance is singular or ill-conditioned),
so the $\langle\chi^2\rangle$ curve for $X$ takes large values at $N_{\rm sim} \sim N_{\rm data}$.
By contrast, the curve for $Y$ yields stable values even when $N_{\rm sim} = N_{\rm data}$, though those values tend to be noisy.
This behavior is the central evidence behind our shrinkage interpretation: CV effectively regularizes the covariance estimate so that its inverse has meaningful, convergent behavior earlier in $N_{\rm sim}$.
We should be careful not to overstate this: ``earlier'' is meant relative to $X$, whose inverse diverges catastrophically as $N_{\rm sim}\to N_{\rm data}$. It does not mean that $Y$ is usable at $N_{\rm sim}\simeq N_{\rm data}$. Being a difference of noisy matrices, $Y$ is in fact frequently not positive definite in that regime, and although ${\rm Tr}(T\,Y^{-1})$ is far less extreme than its $X$ counterpart there, it remains biased (see the caveat in Sec.~\ref{sec:chi2_distribution}).

We note that throughout this work we evaluate the expected chi-square statistic directly using the Monte-Carlo ensemble and the corresponding precision matrices, without applying a theoretical Wishart--Hartlap correction. The Hartlap factor is derived under assumptions that the estimated covariance is Wishart-distributed (and hence that the corresponding finite-sample inverse bias has a simple multiplicative form); these assumptions are not strictly satisfied in our masked, control-variate--modified covariance estimators. We therefore treat $\langle \chi^2\rangle$ as an empirical diagnostic of the finite-realization ``projection effects'' relevant for likelihood inference.

In cosmological inference, it is standard to demand that parameter-projection effects or biases remain below $0.5\,\sigma$ (e.g.\ as adopted in recent DESI inference analyses).  For illustration we adopt an analogous criterion for the expected $\chi^2$ excess: $\langle \chi^2\rangle-N_{\rm data}<\sqrt{N_{\rm data}/2}$. 
For $N_{\rm data}=120$, we obtain this target at $N_{\rm sim}\sim 900$ with the CV-reduced precision, compared to $N_{\rm sim}\sim 1800$ for the non-CV estimator, i.e.\ a factor of $\sim 2$ fewer.

For $N_{\rm sim}\gtrsim 1000$, the CV improvement in $\langle \chi^2 \rangle$ is on the order of a factor of $2$--$3$, and the CV-reduced curve tracks the more correct behavior more closely at small $N_{\rm sim}\sim N_{\rm data}$.
Once $N_{\rm sim}\gg N_{\rm data}$, the $Y$ estimator outdoes $X$ by a factor of 2. This is a substantial improvement that leads to more accurate likelihood evaluations when the CV method is applied in cosmological inference.

\subsection{Fixing the inverse-covariance bias}
\label{sec:chi2_distribution}

We now address directly how CV-corrected covariance matrices estimated from a finite number of mocks should be used in parameter inference. Two distinct finite-sample effects must be distinguished: (i) the bias of the precision matrix, $\mathbb{E}[\hat C^{-1}]\neq T^{-1}$, which for a Wishart-distributed sample covariance is exactly multiplicative and is removed by the Hartlap factor $\alpha_{\rm H}=(N_{\rm sim}-N_{\rm data}-2)/(N_{\rm sim}-1)$ \citep{2007A&A...464..399H}; and (ii) the noise of the precision matrix, which propagates into the widths and scatter of parameter posteriors \citep{2013PhRvD..88f3537D,2014MNRAS.439.2531P,2016MNRAS.456L.132S,2022MNRAS.510.3207P}. Given an accurate model, a multiplicative correction can remove (i) but does nothing for (ii); the value of the CV construction lies precisely in reducing (ii), while (i) must still be corrected for.

Because the CV estimator Y is not Wishart-distributed, the standard correction does not apply. We instead compute a correction using a bootstrap procedure
\beq\label{eq:alpha_boot}
\alpha_{\rm boot}(N_{\rm sim}) \equiv \frac{N_{\rm data}}{\big\langle {\rm Tr}\big(\Sigma^{dd}, \hat Y_\ast^{-1}\big)\big\rangle_{\rm boot}} ,
\eeq
at negligible computational cost (see App.~\ref{sec:boot} for more details). This bootstrap calibration automatically accounts for the scale-dependent variance reduction produced by the element-wise correction $\beta_{ij}(C_{ij}-\mu_{ij})$, without requiring expensive additional mocks.

We find $\alpha_{\rm boot}$ to lie between the two Hartlap bounds above, approaching $\alpha_{\rm H}(N_{\rm sim}^{\rm eff})$ on covariances dominated by well-correlated scales; concretely, for $N_{\rm data}=120$ and a scalar effective gain of $1.94$, at $N_{\rm sim}=1000$ we obtain $\alpha_{\rm boot}=0.943$, compared with $\alpha_{\rm H}(N_{\rm sim})=0.879$ and $\alpha_{\rm H}(N_{\rm sim}^{\rm eff})=0.938$. The bootstrap calibration and the effective-$N_{\rm sim}$ Hartlap factor agree to better than $1\%$ for $N_{\rm sim}\ge500$, as summarized in Table~\ref{tab:alpha_boot}.
One practical caveat is worth recording. The sample covariance $X$ is positive definite as soon as $N_{\rm sim}$ exceeds $N_{\rm data}$, whereas $Y$ is a difference of two noisy matrices and carries no such guarantee. In our configuration $Y$ becomes reliably invertible, and the bootstrap calibration correspondingly stable, once $N_{\rm sim}$ is a modest factor ($\sim2$) times $N_{\rm data}$; we therefore quote $\alpha_{\rm boot}$ only in that regime. This is of little practical consequence, since useful covariance estimation requires $N_{\rm sim}\gg N_{\rm data}$ in any case.

\begin{figure*}
\centering
\includegraphics[width=0.48\textwidth]{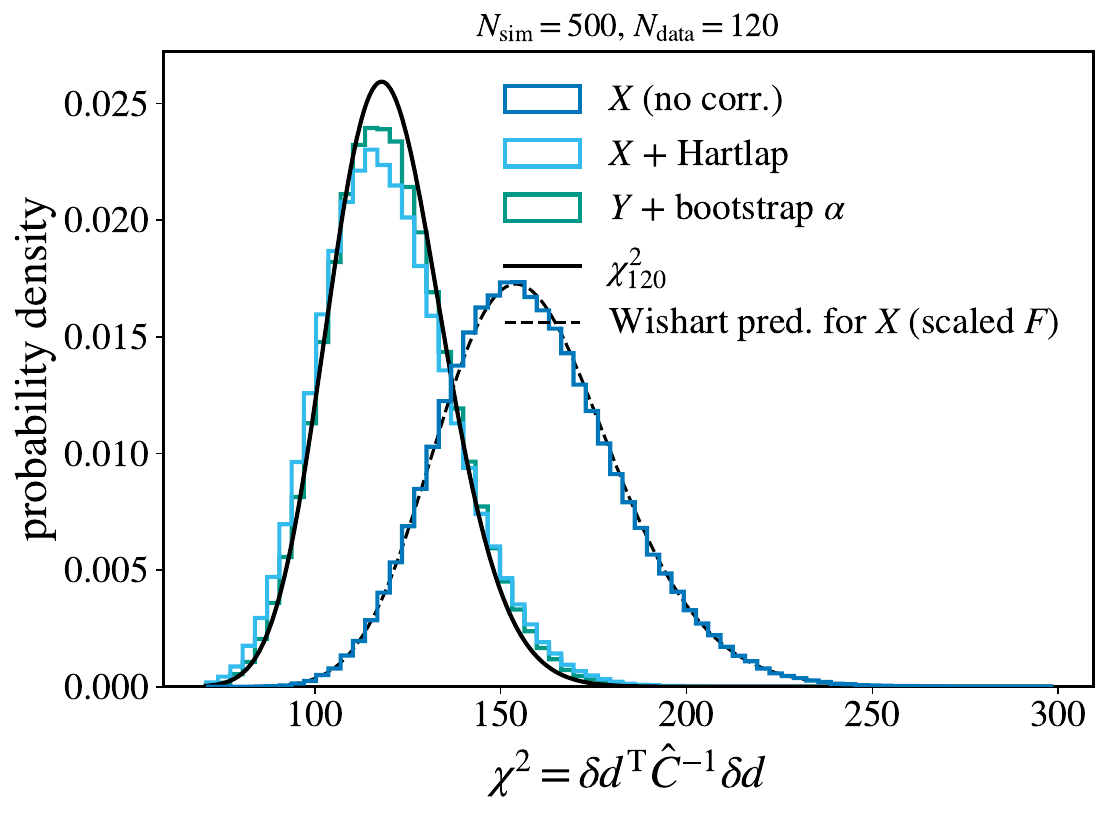}
\hfill
\includegraphics[width=0.48\textwidth]{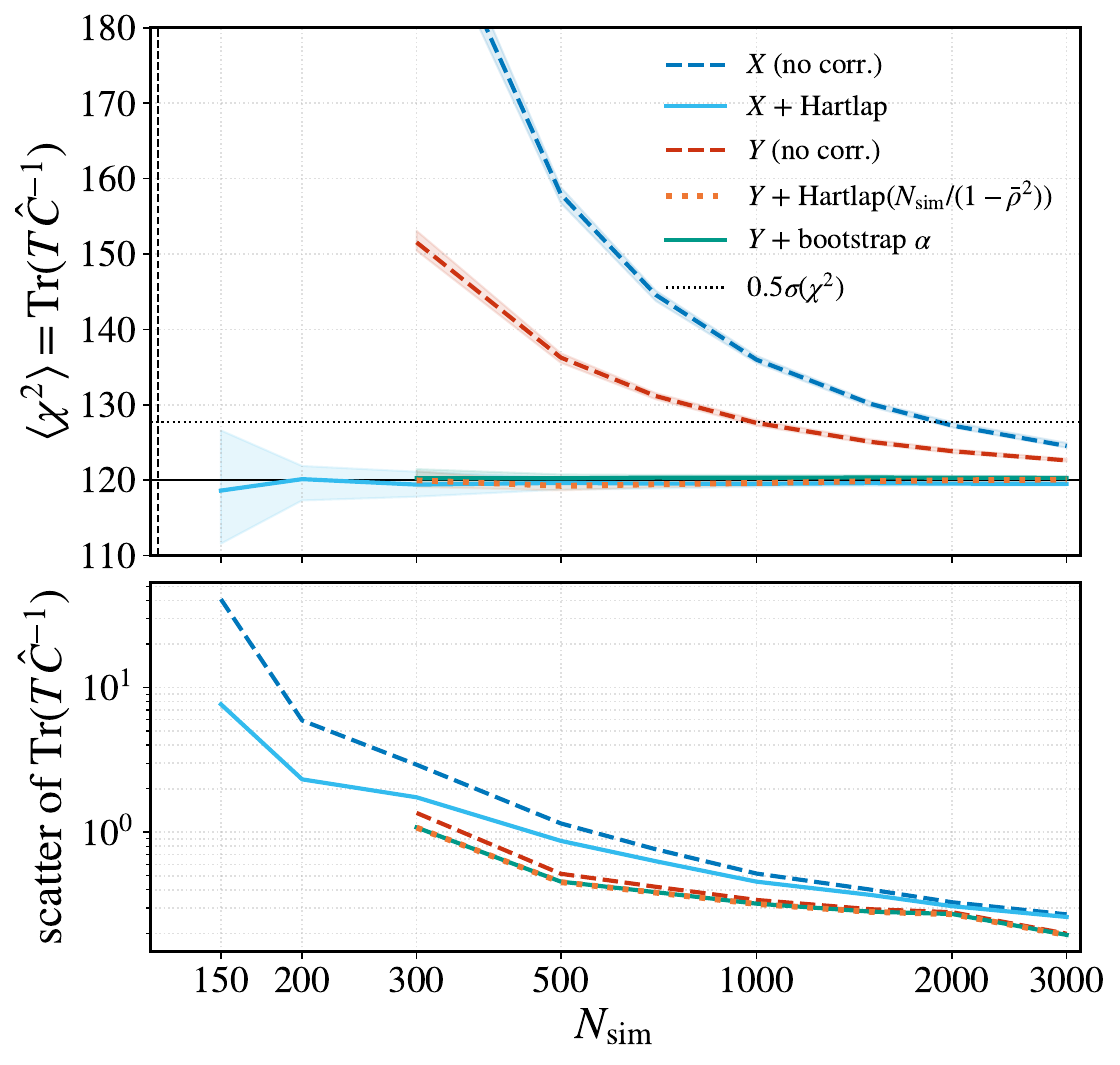}
\caption{ 
Left: distribution of $\chi^2=\delta d^{\rm T}\,\hat C^{-1}\delta d$ over held-out data vectors, for the raw sample covariance $X$, the Hartlap-corrected $X$, and the bootstrap-corrected CV estimator $Y$, at $N_{\rm sim}=500$, compared with the $\chi^2_{N_{\rm data}}$ expectation (solid black) and the exact Wishart prediction for $X$ (dashed black). The improved performance of CV is not legible as the $\delta_d$ draws dominate the width: the intrinsic width of $\chi^2$ (with $N_{\rm data} = 120$) is 15.5 and the covariance noise contributes 0.3\% of the total variance.
Right: mean (top) and realization-to-realization scatter (bottom) of ${\rm Tr}(T\hat C^{-1})$ versus $N_{\rm sim}$ for corrected and uncorrected. The corrections remove the mean bias; CV reduces the scatter. At $N_{\rm sim}=500$, the CV covariance shows 1.9 times less scatter. The `analytic' $N_{\rm sim}/(1-\bar \rho^2)$ lies on top of the bootstrap $\alpha$ curve.
}
\label{fig:chi2_distribution}
\end{figure*}

Fig.~\ref{fig:chi2_distribution} presents the resulting test of the statistic of direct relevance to inference, $\chi^2 = \delta d^{\rm T}\,\hat C^{-1}\delta d$, for data vectors $\delta d$ drawn from held-out realizations, with $\hat C$ estimated from $N_{\rm sim}$ mocks. The left panel shows the full distribution of $\chi^2$ at $N_{\rm sim}=500$ for the raw sample covariance $X$, the Hartlap-corrected $X$, and the bootstrap-corrected CV estimator $Y$, compared with the $\chi^2_{N_{\rm data}}$ expectation and, for $X$, with the exact Wishart prediction (a scaled $F$ distribution \citep{2016MNRAS.456L.132S}). The bootstrap-corrected $Y$ recovers a mean consistent with $N_{\rm data}$, $\langle\chi^2\rangle = 120.5 \pm 17.9$ for $N_{\rm data}=120$ (against a $\chi^2_{120}$ target of $120\pm15.5$), compared with $119.8\pm18.7$ for the Hartlap-corrected sample covariance and $158.2\pm24.7$ with no correction at all --- the last in excellent agreement with the analytic Wishart expectation $N_{\rm data}(N_{\rm sim}-1)/(N_{\rm sim}-N_{\rm data}-2)=158.8$, and a distribution closer to the $\chi^2_{N_{\rm data}}$ target than the Hartlap-corrected sample covariance, whose variance remains inflated by precision-matrix noise. The right panel shows the mean and, importantly, the realization-to-realization scatter of ${\rm Tr}(T\,\hat C^{-1})$ as a function of $N_{\rm sim}$: while the Hartlap-type corrections successfully remove the mean bias for both estimators (by construction), the scatter of the CV-based precision matrix is smaller by a factor of $1.3$--$1.9$ at fixed $N_{\rm sim}$, the improvement being largest ($1.9$, close to the scalar effective gain of $1.94$) in the regime $N_{\rm sim}\sim500$ of practical interest.
 Since it is this scatter that propagates into parameter uncertainties and their errors \citep{2013PhRvD..88f3537D,2022MNRAS.510.3207P}, the $\langle\chi^2\rangle$ comparison alone understates the benefit of the CV construction.

Finally, we address the unbiasedness of Fisher-matrix estimates. We perform the following test. The templates $t_a$ are the derivatives of the model data vector with respect to the parameters, evaluated at the fiducial point. We take the two parameters to be the amplitudes $A_0$ and $A_2$ of the monopole and quadrupole, i.e.\ the trivial model $\vec d(A_0,A_2) = (A_0\bar P_0,\,A_2\bar P_2)$ with $\bar P_\ell$ the ensemble-mean multipole of held-out realizations, so that
\beq\label{eq:fisher_templates}
t_0 = \big(\bar P_0,\,\vec 0\big)\,,\qquad t_2 = \big(\vec 0,\,\bar P_2\big)\,.
\eeq
This choice requires no derivatives of a cosmological model, introduces no additional scale dependence, and probes the broadband directions of $\hat C^{-1}$ that dominate realistic parameter constraints. The single object varied in the test is $\hat C^{-1}$, which we take in turn to be $X^{-1}$, $\alpha_{\rm H}X^{-1}$, $\hat Y^{-1}$ and $\alpha_{\rm boot}\hat Y^{-1}$, and we have a reference $F_{ab}$, which uses the covariance of the $10^4$ held-out realizations.

Using these templates, we form $F_{ab}=t_a^{\rm T}\hat C^{-1} t_b$ and the corresponding parameter variances $(F^{-1})_{aa}$, and compare their expectation values and scatter against the reference computed from $T$. Without any correction, both $X^{-1}$ and $Y^{-1}$ overestimate $F$ (by $\sim(1-\alpha_{\rm H})^{-1}$ for $X$, and by a factor $\sim1.4$ smaller for $Y$: at $N_{\rm sim}=500$ the bias of $F_{00}$ is $+27.0\%$ for $X$ and $+18.8\%$ for $Y$). With the Hartlap factor applied to $X$ and $\alpha_{\rm boot}$ applied to $Y$, both Fisher estimates are unbiased at the few-percent level for $N_{\rm sim}\gtrsim700$ ($-3.9\%$ and $+1.8\%$ respectively at $N_{\rm sim}=700$; $-4.5\%$ and $-0.6\%$ at $N_{\rm sim}=1000$).

We stress that these residuals should be read against the precision with which the reference itself is determined: $F_{00}$ built from a single $10^4$-realization reference ensemble carries a bootstrap uncertainty of $\pm3.4\%$, so biases at this level are not resolvable and only differences between estimators at fixed $N_{\rm sim}$ are meaningful.
The reduction in scatter is parameter dependent. For the quadrupole amplitude the CV-based Fisher estimate is $\sim1.5$ times less scattered than the Hartlap-corrected sample-covariance estimate at fixed $N_{\rm sim}$; for the monopole amplitude the two are comparable ($\sim1.0$). We make the conjecture that this is because the Fisher matrix contracts $\hat C^{-1}$ with a fixed broadband template, so the gain realized on a given parameter reflects the variance reduction weighted along that template's direction rather than the diagonal gain of Fig.~\ref{fig:variance_reduction_diagonals}, and it is the quadrupole that benefits most here.
 In practice, therefore, our recommendation for real analyses is: use the CV-reduced covariance, apply the bootstrap-calibrated $\alpha_{\rm boot}$ (or, more conservatively, propagate the covariance uncertainty following \citealt{2022MNRAS.510.3207P} with the effective $N_{\rm sim}$ implied by the bootstrap), and validate with the $\chi^2$-distribution test described here, which requires no additional expensive simulations.

\begin{table}
\centering
\begin{tabular}{rccc}
\hline\hline
$N_{\rm sim}$ & $\alpha_{\rm boot}$ & $\alpha_{\rm H}(N_{\rm sim})$ & $\alpha_{\rm H}(N_{\rm sim}^{\rm eff})$ \\
\hline
300  & 0.794 & 0.595 & 0.792 \\
500  & 0.883 & 0.758 & 0.875 \\
700  & 0.918 & 0.827 & 0.911 \\
1000 & 0.943 & 0.879 & 0.938 \\
2000 & 0.972 & 0.940 & 0.969 \\
3000 & 0.981 & 0.960 & 0.979 \\
\hline\hline
\end{tabular}
\caption{Bootstrap calibration of the residual inverse bias of the CV covariance, $\alpha_{\rm boot}$ of Eq.~\eqref{eq:alpha_boot}, compared with the two Hartlap bounds it is expected to lie between, for $N_{\rm data}=120$ and a scalar effective gain $N_{\rm sim}^{\rm eff}/N_{\rm sim}=1.94$. The bootstrap calibration and the effective-$N_{\rm sim}$ Hartlap factor agree to better than $1\%$ for $N_{\rm sim}\ge500$.}
\label{tab:alpha_boot}
\end{table}

Table~\ref{tab:XY_byN} provides a compact summary of multiple distance/divergence metrics \cite{bhatia2009,horn2012} comparing both covariance estimators to the reference $T$ at different $N_{\rm sim}$ values.
Here $C$ in the norms denotes the estimated covariance (either $X$ or $Y$), and all comparisons are made against $T$ computed from $10{,}000$ realizations.
Across the listed metrics: 
\begin{itemize}
  \item $\|C-T\|_F$: Frobenius distance; overall covariance error magnitude.
  \item $\|C-T\|_2$: Operator norm or `spectral error'; worst-direction covariance mismatch.
  \item $\|C^{-1}-T^{-1}\|_F$: Frobenius error of precision matrices.
  \item $\|C^{-1}-T^{-1}\|_2$: Spectral error of precision matrices.
  \item KL divergence: Information loss; covariance/precision discrepancy in KL.
  \item log-Euclidean distance; mismatch on the manifold of Symmetric Positive Definite (SPD) matrices.
  \item Bures--Wasserstein: Optimal transport distance between Gaussians.
\end{itemize}
$Y$ is noticeably better than $X$ in all cases. In many cases we observe an improvement by about a factor of 3 or so and particularly better performance when $N_{\rm sim} \sim N_{\rm data}$.
This overall pattern confirms that CV improves covariance-matrix accuracy in a way that is stable under a range of matrix distances and information-geometric divergences, i.e.\ it targets exactly the aspects of the covariance that influence inference.

\section{Applying the method in practice}
\label{sec:practice}

\referee{Our validation used lognormal targets so that a $10^4$-realization reference ensemble could be afforded; the intended targets are the $N$-body-based mock suites of a survey. We summarize here what such an application involves.}

\referee{The control variate neither replaces the $N$-body mocks nor splits the covariance into a large-scale part and a small-scale part that are estimated separately. One runs the $\Nsim$ target mocks one would run anyway, and the full covariance, on all scales and including every cross-scale element, is one single matrix, $Y=X-\beta(C-\mu_C)$ of Eq.~\eqref{eq:Y}. 
}

\referee{
Since $\beta_{ij} \to 0$ where the fields decorrelate, $Y_{ij}\to X_{ij}$ on small scales, and likewise for elements coupling a small scale to a large one: there the estimator is simply the sample covariance of the $N$-body mocks, nonlinear mode coupling included, at no extra noise [Eq.~\eqref{eq:suboptimal_beta}]. On the large, window-dominated scales, $Y_{ij}$ instead carries the noise of $\Nsim/(1-\rho_{ij}^2)$ mocks, which has reduced noise over the raw covariance matrix.}

\referee{The number of mocks is therefore chosen as in any sample-covariance analysis. The fractional noise on each element is $\simeq\sqrt{2/\Nsim}$ [Eq.~\eqref{eq:wishart}], independently of scale and of the number of modes per bin, so the small-scale requirement is set by $N_{\rm data}$ through the usual Hartlap- and Percival-type factors \citep{2007A&A...464..399H,2014MNRAS.439.2531P,2022MNRAS.510.3207P}; what the control variate buys is a reduction of the precision-matrix noise once the large scales are included (for our $N_{\rm data}=120$ data vector, the criterion of Sec.~\ref{sec:convergence_tests} is met with $\Nsim\simeq900$ paired mocks instead of $\simeq1800$). The same logic applies to super-sample covariance \citep{2013PhRvD..87l3504T}: since $\langle Y\rangle=\langle X\rangle$ for any $\beta$, it is contained in $Y$ to exactly the extent that it is contained in the target mocks, and because the control shares the super-survey modes realization by realization we expect it to share in the variance reduction.
}

\referee{The procedure is then as follows.}
\begin{enumerate}
\item \referee{For each target mock, generate a Zeldovich realization from the same initial phases [Eqs.~\eqref{eq:zel}--\eqref{eq:psi} and~\eqref{eq:za_bias}] and pass it through the same geometry, selection, weights, estimator, and binning. The mesh need only resolve the modes used in the analysis, so the cost is a small fraction of one $N$-body mock. Being a field-level construction, the control inherits the window and, for light-cone mocks, the redshift evolution of growth, bias, and velocities as well as wide-angle effects, all of which analytic covariances struggle to include; catalog-level effects such as fiber assignment enter only the target and affect the method only through $r(k)$.}
\item \referee{From the mean spectra of a handful of pairs, compute $r(k)$ and $\alpha(k)$ [Eqs.~\eqref{eq:r_def} and~\eqref{eq:beta1_def}], hence the gain map $(1-\rho_{ij}^2)^{-1}$ with $\rho_{ij}=r^2(k_i)\,r^2(k_j)$ (Fig.~\ref{fig:Nsim_gain_2d}). This forecasts the benefit before the full suite is committed.}
\item \referee{Estimate $\mu_C$ from $N_{\rm th}\gg\Nsim$ independent Zeldovich realizations through the same pipeline (we use $10^4$) and form $Y$ with $\beta_{ij}=\alpha(k_i)\,\alpha(k_j)$, copying the monopole block to the other multipoles (Sec.~\ref{sec:beta_rsd_mask}). Check that $Y$ is positive definite, which holds for $\Nsim\gtrsim2N_{\rm data}$ in our configuration.}
\item \referee{For inference, correct the precision matrix with $\alpha_{\rm boot}(\Nsim)$ of Eq.~\eqref{eq:alpha_boot} and validate with the $\chi^2$ test of Sec.~\ref{sec:chi2_distribution}; neither requires further expensive simulations.}
\end{enumerate}

\section{Conclusions}
\label{sec:conclusions}

We have presented a control-variate method for accelerating the estimation of power spectrum covariance matrices from simulations.
By pairing each target mock (lognormal in this work) with a cheap Zeldovich-approximation realization sharing the same initial conditions, we exploit the strong correlation between the two fields to subtract correlated sample variance from the covariance estimator.
A key practical advantage is that the Zeldovich field is fast to generate while capturing precisely the correlations that dominate the covariance uncertainty on the most challenging, large scales.

The central technical contribution of this work is a fully analytic expression for the optimal control-variate coefficient $\beta(k,\ell; k',\ell')$, derived under the disconnected Gaussian approximation [Eq.~\eqref{eq:beta_analytic}].
The resulting $\beta$ depends only on the mean auto- and cross-power spectra of the target and control fields and the number of modes per $k$-bin, and is therefore inexpensive and numerically stable to evaluate.
We validate the analytic $\beta$ against brute-force estimates from simulation ensembles and find excellent agreement, confirming that the expression captures the correct dependence on the underlying two-point statistics in a noise-free way.

In addition, we derive an analytic expression for the corresponding covariance-variance reduction correlation $\rho(k,\ell; k',\ell')$.
For the monopole case, $\rho$ takes the particularly simple form $\rho(k,k')=r^2(k)\,r^2(k')$, where $r(k)$ is the standard cross-correlation coefficient between the target field and the Zeldovich control field.
This result is important because it implies that the expected performance of the control-variate covariance estimator can be anticipated directly from $r(k)$ measured from the cross correlation, without requiring additional covariance-level sampling.
We validate this prediction against brute-force computation and find that the analytic expression reproduces the trend well (with a small systematic underprediction at the level shown in Fig.~\ref{fig:variance_reduction_diagonals}).

Because $\rho_{ij}=r^2(k_i)\,r^2(k_j)$ factorizes in this way, we can compare directly to the analogous control-variate performance in the power spectrum itself, where the relevant factor is $r^2$.
Consequently, the covariance performance is only about a factor of two ``worse'' than the na\"ive power-spectrum expectation, rather than deteriorating dramatically at the covariance level, explaining why the covariance inherits the strong large-scale correlation efficiently.

Using lognormal mocks with $r(k)$ closely matching that of realistic DESI LRG-like samples, we find that the control-variate estimator reduces the variance of the covariance matrix by approximately an order of magnitude on the most challenging large scales ($k \lesssim 0.05\,h\,{\rm Mpc}^{-1}$), precisely where accurate covariance estimation is most difficult.
In practical terms, this corresponds to an effective simulation-number gain of $\sim 10$ for the same covariance quality, and can be confirmed visually by the substantially reduced covariance noise in the CV-reduced estimate.

Finally, we compared a comprehensive set of additional matrix-distance and likelihood-sensitivity metrics between the CV-reduced covariance and the high-accuracy ``truth'' covariance, as well as between the CV-reduced and non-reduced estimators.
Across these metrics we consistently find that the CV-reduced covariance is closer to $T$ than the original estimate by about a factor of $\sim 3$ in accuracy, while also converging faster as $N_{\rm sim}$ increases.
Notably, in the regime where $N_{\rm sim}$ approaches the number of data points, the CV-reduced matrices remain well-conditioned and invertible, i.e.\ they effectively behave like a shrinkage estimator, yielding stable precision-matrix behavior where the original estimator becomes ill-conditioned.

Our main results can be summarized as follows:

\begin{enumerate}
\item The Zeldovich approximation provides an extremely effective control variate: it is cheap to generate and highly accurate on the same large scales where covariance estimation is most numerically challenging.
\item We derive an analytic expression for the optimal control-variate coefficient $\beta(k,\ell; k',\ell')$, which is smooth and inexpensive; brute-force validation shows excellent agreement with the analytic prediction.
\item We derive an analytic expression for $\rho(k,\ell; k',\ell')$, which depends only on the cross-correlation between the target field and the Zeldovich control field; in the monopole case $\rho(k,k')=r^2(k)r^2(k')$.
We validate this dependence against brute-force computations (with a small systematic underprediction visible in one figure).
\item Despite the covariance involving products of fields, the performance penalty relative to power-spectrum control variates is modest: because $\rho$ factorizes as $r^2(k_i)r^2(k_j)$, the resulting covariance performance is only about a factor of two worse than the naive $r^2$ expectation.
\item For DESI-like masked redshift-space lognormal mocks, the CV-reduced covariance achieves an $\mathcal{O}(10)$ variance-reduction gain on the largest scales, corresponding to an effective $\sim 10$ increase in the number of simulations needed for comparable covariance accuracy.
\item Across a suite of covariance and precision-matrix-sensitive metrics, the CV-reduced estimate is consistently closer to the reference ``truth'' (by about a factor of $\sim 3$) and converges faster; when $N_{\rm sim}$ is close to $N_{\rm data}$, CV behaves like a shrinkage estimator and prevents the onset of ill-conditioning.
\item When the target mocks are displaced with 2LPT positions and velocities while the control remains Zeldovich (thus breaking the exact correspondence of the coherent large-scale velocity fields) the CV gains degrade gracefully and remain accurately predicted by the measured $r(k)$.  This demonstrates robustness to controlled modeling mismatch \referee{(construction in Sec.~\ref{sec:2lpt_mocks}, results in Sec.~\ref{sec:2lpt_validation})}.
\item The distribution of $\chi^2=\delta d^{\rm T}\, \hat C^{-1}\delta d$ obtained with the CV covariance, after applying a cheap parametric-bootstrap calibration of the inverse bias, is consistent with the expected number of degrees of freedom, and the corresponding Fisher-matrix estimates are unbiased within our measurement precision while exhibiting substantially reduced realization-to-realization scatter (Sec.~\ref{sec:chi2_distribution}).
\end{enumerate}

Looking ahead, several extensions are natural.
First, the method should be applied to maximally realistic mocks of the large-scale structure (e.g., DESI, \textit{Euclid}, PFS, LSST, \textit{SPHEREx}) paired with appropriate masking to construct a variance-reduced covariance matrix for a given galaxy sample.
\referee{We regard validation on state-of-the-art $N$-body-based mocks as the essential next step before application to observational data; the procedure, including the role of the small scales and of super-sample covariance, is laid out in Sec.~\ref{sec:practice}. We are currently pursuing this program on the existing DESI mock suites in the context of the DR3 analysis, and will report the results in future work.}

Second, one could imagine finding a more strongly correlated control variate (e.g.\ using 2LPT instead of the Zeldovich approximation), which could yield an even larger numerical gain.
Third, extending the framework to the covariance matrix of marked power spectra, the galaxy bispectrum, or other summary statistics could be especially valuable, though deriving an analytic $\beta$ would require higher-point functions.
While our focus has been on 3D galaxy surveys, there is no reason why it cannot be applied to line intensity mapping surveys, imaging surveys or cosmic shear surveys with minimal changes.
Finally, since the control variate could in principle be applied at the field level rather than only at the summary-statistic level, combining CV with field-level estimators may offer further gains.
In addition, combining the control-variate approach with shrinkage estimators or analytic templates \cite{2008MNRAS.389..766P,2015MNRAS.454.4326P,2016MNRAS.460.1567P,2017MNRAS.466L..83J,2018MNRAS.473.4150F,2020PhRvD.102l3521W} could yield improvements in regimes where none of these methods alone is sufficient.

\begin{acknowledgments}
We thank Stephen Chen, Joe DeRose, Jiamin Hou, and Nikhil Padmanabhan for useful discussions.
MW was supported by the DOE and NASA.
This research was supported in part by grant NSF PHY-2309135 to the Kavli Institute for Theoretical Physics (KITP).
The ideas leading to this work were initiated at the Aspen Center for Physics, which is supported by National Science Foundation grant PHY-2210452.
This research used resources of the National Energy Research Scientific Computing Center (NERSC), a Department of Energy User Facility.
\end{acknowledgments}

\appendix

\section{Bootstrap procedure for Hartlap-like correction}
\label{sec:boot}

We spell out the calibration explicitly, since every ingredient is already available in the CV pipeline. Let $\vec d$ and $\vec c$ denote the target and control data vectors, each of length $N_{\rm data}$, and let
\beq\label{eq:joint_cov}
\Sigma \;=\;
\begin{pmatrix} \Sigma^{dd} & \Sigma^{dc} \\ \Sigma^{cd} & \Sigma^{cc}\end{pmatrix}
\eeq
be their joint $2N_{\rm data}\times2N_{\rm data}$ covariance, estimated once from the full paired ensemble (in a real analysis, from the best available smooth or CV-reduced estimates). Here $\Sigma^{dd}$ is the covariance of the target data vector, i.e.\ the very quantity the estimator is trying to recover, and $\Sigma^{cc}$ is that of the control, which plays the role of $\mu_C$. The calibration then proceeds as follows. (i) Draw $N_{\rm sim}$ independent Gaussian pairs $(\vec d_\ast,\vec c_\ast)\sim\mathcal{N}(0,\Sigma)$, so that the surrogate target and control are correlated exactly as the real ones are. (ii) Form their sample covariances $X_\ast$ and $C_\ast$ and apply the {same} CV estimator with the same element-wise weights, $\hat Y_\ast = X_\ast - \beta(C_\ast-\mu_C)$. (iii) Evaluate ${\rm Tr}(\Sigma^{dd}\hat Y_\ast^{-1})$; were $\hat Y_\ast^{-1}$ an unbiased estimate of the true precision matrix, this trace would equal $N_{\rm data}$ by construction, so its excess measures precisely the inverse bias we wish to remove. (iv) Repeat for $\mathcal{O}(10^2)$ surrogate ensembles and take the ratio in Eq.~\eqref{eq:alpha_boot}; the debiased precision matrix is then $\alpha_{\rm boot}\hat Y^{-1}$. The logic is identical to that underlying the usual factor, with the closed-form Wishart result replaced by a Monte Carlo measurement: applied to $X$ rather than to $\hat Y$, the identical procedure returns $\alpha_{\rm H}(N_{\rm sim})$ to within $0.1\%$ over the whole range of $N_{\rm sim}$ considered here, which is a useful validation of the machinery. Its cost is that of $\mathcal{O}(10^2)$ matrix inversions rather than of any additional mock, and we stress that the Gaussian surrogates are used only to characterize the {noise} of the estimator, not to model the covariance itself, which enters through the measured $\Sigma$.

\bibliography{main}

@ARTICLE{2016arXiv161100036D,
       author = {{DESI Collaboration} and {Aghamousa}, Amir and {Aguilar}, Jessica and {Ahlen}, Steve and {Alam}, Shadab and {Allen}, Lori E. and {Allende Prieto}, Carlos and {Annis}, James and {Bailey}, Stephen and {Balland}, Christophe and {Ballester}, Otger and {Baltay}, Charles and {Beaufore}, Lucas and {Bebek}, Chris and {Beers}, Timothy C. and {Bell}, Eric F. and {Bernal}, Jos{\'e} Luis and {Besuner}, Robert and {Beutler}, Florian and {Blake}, Chris and {Bleuler}, Hannes and {Blomqvist}, Michael and {Blum}, Robert and {Bolton}, Adam S. and {Briceno}, Cesar and {Brooks}, David and {Brownstein}, Joel R. and {Buckley-Geer}, Elizabeth and {Burden}, Angela and {Burtin}, Etienne and {Busca}, Nicolas G. and {Cahn}, Robert N. and {Cai}, Yan-Chuan and {Cardiel-Sas}, Laia and {Carlberg}, Raymond G. and {Carton}, Pierre-Henri and {Casas}, Ricard and {Castander}, Francisco J. and {Cervantes-Cota}, Jorge L. and {Claybaugh}, Todd M. and {Close}, Madeline and {Coker}, Carl T. and {Cole}, Shaun and {Comparat}, Johan and {Cooper}, Andrew P. and {Cousinou}, M.-C. and {Crocce}, Martin and {Cuby}, Jean-Gabriel and {Cunningham}, Daniel P. and {Davis}, Tamara M. and {Dawson}, Kyle S. and {de la Macorra}, Axel and {De Vicente}, Juan and {Delubac}, Timoth{\'e}e and {Derwent}, Mark and {Dey}, Arjun and {Dhungana}, Govinda and {Ding}, Zhejie and {Doel}, Peter and {Duan}, Yutong T. and {Ealet}, Anne and {Edelstein}, Jerry and {Eftekharzadeh}, Sarah and {Eisenstein}, Daniel J. and {Elliott}, Ann and {Escoffier}, St{\'e}phanie and {Evatt}, Matthew and {Fagrelius}, Parker and {Fan}, Xiaohui and {Fanning}, Kevin and {Farahi}, Arya and {Farihi}, Jay and {Favole}, Ginevra and {Feng}, Yu and {Fernandez}, Enrique and {Findlay}, Joseph R. and {Finkbeiner}, Douglas P. and {Fitzpatrick}, Michael J. and {Flaugher}, Brenna and {Flender}, Samuel and {Font-Ribera}, Andreu and {Forero-Romero}, Jaime E. and {Fosalba}, Pablo and {Frenk}, Carlos S. and {Fumagalli}, Michele and {Gaensicke}, Boris T. and {Gallo}, Giuseppe and {Garcia-Bellido}, Juan and {Gaztanaga}, Enrique and {Pietro Gentile Fusillo}, Nicola and {Gerard}, Terry and {Gershkovich}, Irena and {Giannantonio}, Tommaso and {Gillet}, Denis and {Gonzalez-de-Rivera}, Guillermo and {Gonzalez-Perez}, Violeta and {Gott}, Shelby and {Graur}, Or and {Gutierrez}, Gaston and {Guy}, Julien and {Habib}, Salman and {Heetderks}, Henry and {Heetderks}, Ian and {Heitmann}, Katrin and {Hellwing}, Wojciech A. and {Herrera}, David A. and {Ho}, Shirley and {Holland}, Stephen and {Honscheid}, Klaus and {Huff}, Eric and {Hutchinson}, Timothy A. and {Huterer}, Dragan and {Hwang}, Ho Seong and {Illa Laguna}, Joseph Maria and {Ishikawa}, Yuzo and {Jacobs}, Dianna and {Jeffrey}, Niall and {Jelinsky}, Patrick and {Jennings}, Elise and {Jiang}, Linhua and {Jimenez}, Jorge and {Johnson}, Jennifer and {Joyce}, Richard and {Jullo}, Eric and {Juneau}, St{\'e}phanie and {Kama}, Sami and {Karcher}, Armin and {Karkar}, Sonia and {Kehoe}, Robert and {Kennamer}, Noble and {Kent}, Stephen and {Kilbinger}, Martin and {Kim}, Alex G. and {Kirkby}, David and {Kisner}, Theodore and {Kitanidis}, Ellie and {Kneib}, Jean-Paul and {Koposov}, Sergey and {Kovacs}, Eve and {Koyama}, Kazuya and {Kremin}, Anthony and {Kron}, Richard and {Kronig}, Luzius and {Kueter-Young}, Andrea and {Lacey}, Cedric G. and {Lafever}, Robin and {Lahav}, Ofer and {Lambert}, Andrew and {Lampton}, Michael and {Landriau}, Martin and {Lang}, Dustin and {Lauer}, Tod R. and {Le Goff}, Jean-Marc and {Le Guillou}, Laurent and {Le Van Suu}, Auguste and {Lee}, Jae Hyeon and {Lee}, Su-Jeong and {Leitner}, Daniela and {Lesser}, Michael and {Levi}, Michael E. and {L'Huillier}, Benjamin and {Li}, Baojiu and {Liang}, Ming and {Lin}, Huan and {Linder}, Eric and {Loebman}, Sarah R. and {Luki{\'c}}, Zarija and {Ma}, Jun and {MacCrann}, Niall and {Magneville}, Christophe and {Makarem}, Laleh and {Manera}, Marc and {Manser}, Christopher J. and {Marshall}, Robert and {Martini}, Paul and {Massey}, Richard and {Matheson}, Thomas and {McCauley}, Jeremy and {McDonald}, Patrick and {McGreer}, Ian D. and {Meisner}, Aaron and {Metcalfe}, Nigel and {Miller}, Timothy N. and {Miquel}, Ramon and {Moustakas}, John and {Myers}, Adam and {Naik}, Milind and {Newman}, Jeffrey A. and {Nichol}, Robert C. and {Nicola}, Andrina and {Nicolati da Costa}, Luiz and {Nie}, Jundan and {Niz}, Gustavo and {Norberg}, Peder and {Nord}, Brian and {Norman}, Dara and {Nugent}, Peter and {O'Brien}, Thomas and {Oh}, Minji and {Olsen}, Knut A.~G.},
        title = "{The DESI Experiment Part I: Science,Targeting, and Survey Design}",
      journal = {arXiv e-prints},
         year = 2016,
        month = oct,
          eid = {arXiv:1611.00036},
        pages = {arXiv:1611.00036},
          doi = {10.48550/arXiv.1611.00036},
archivePrefix = {arXiv},
       eprint = {1611.00036},
 primaryClass = {astro-ph.IM},
       adsurl = {https://ui.adsabs.harvard.edu/abs/2016arXiv161100036D}
}

@ARTICLE{2009arXiv0912.0201L,
       author = {{LSST Science Collaboration} and {Abell}, Paul A. and {Allison}, Julius and {Anderson}, Scott F. and {Andrew}, John R. and {Angel}, J. Roger P. and {Armus}, Lee and {Arnett}, David and {Asztalos}, S.~J. and {Axelrod}, Tim S. and {Bailey}, Stephen and {Ballantyne}, D.~R. and {Bankert}, Justin R. and {Barkhouse}, Wayne A. and {Barr}, Jeffrey D. and {Barrientos}, L. Felipe and {Barth}, Aaron J. and {Bartlett}, James G. and {Becker}, Andrew C. and {Becla}, Jacek and {Beers}, Timothy C. and {Bernstein}, Joseph P. and {Biswas}, Rahul and {Blanton}, Michael R. and {Bloom}, Joshua S. and {Bochanski}, John J. and {Boeshaar}, Pat and {Borne}, Kirk D. and {Bradac}, Marusa and {Brandt}, W.~N. and {Bridge}, Carrie R. and {Brown}, Michael E. and {Brunner}, Robert J. and {Bullock}, James S. and {Burgasser}, Adam J. and {Burge}, James H. and {Burke}, David L. and {Cargile}, Phillip A. and {Chandrasekharan}, Srinivasan and {Chartas}, George and {Chesley}, Steven R. and {Chu}, You-Hua and {Cinabro}, David and {Claire}, Mark W. and {Claver}, Charles F. and {Clowe}, Douglas and {Connolly}, A.~J. and {Cook}, Kem H. and {Cooke}, Jeff and {Cooray}, Asantha and {Covey}, Kevin R. and {Culliton}, Christopher S. and {de Jong}, Roelof and {de Vries}, Willem H. and {Debattista}, Victor P. and {Delgado}, Francisco and {Dell'Antonio}, Ian P. and {Dhital}, Saurav and {Di Stefano}, Rosanne and {Dickinson}, Mark and {Dilday}, Benjamin and {Djorgovski}, S.~G. and {Dobler}, Gregory and {Donalek}, Ciro and {Dubois-Felsmann}, Gregory and {Durech}, Josef and {Eliasdottir}, Ardis and {Eracleous}, Michael and {Eyer}, Laurent and {Falco}, Emilio E. and {Fan}, Xiaohui and {Fassnacht}, Christopher D. and {Ferguson}, Harry C. and {Fernandez}, Yanga R. and {Fields}, Brian D. and {Finkbeiner}, Douglas and {Figueroa}, Eduardo E. and {Fox}, Derek B. and {Francke}, Harold and {Frank}, James S. and {Frieman}, Josh and {Fromenteau}, Sebastien and {Furqan}, Muhammad and {Galaz}, Gaspar and {Gal-Yam}, A. and {Garnavich}, Peter and {Gawiser}, Eric and {Geary}, John and {Gee}, Perry and {Gibson}, Robert R. and {Gilmore}, Kirk and {Grace}, Emily A. and {Green}, Richard F. and {Gressler}, William J. and {Grillmair}, Carl J. and {Habib}, Salman and {Haggerty}, J.~S. and {Hamuy}, Mario and {Harris}, Alan W. and {Hawley}, Suzanne L. and {Heavens}, Alan F. and {Hebb}, Leslie and {Henry}, Todd J. and {Hileman}, Edward and {Hilton}, Eric J. and {Hoadley}, Keri and {Holberg}, J.~B. and {Holman}, Matt J. and {Howell}, Steve B. and {Infante}, Leopoldo and {Ivezic}, Zeljko and {Jacoby}, Suzanne H. and {Jain}, Bhuvnesh and {R} and {Jedicke} and {Jee}, M. James and {Garrett Jernigan}, J. and {Jha}, Saurabh W. and {Johnston}, Kathryn V. and {Jones}, R. Lynne and {Juric}, Mario and {Kaasalainen}, Mikko and {Styliani} and {Kafka} and {Kahn}, Steven M. and {Kaib}, Nathan A. and {Kalirai}, Jason and {Kantor}, Jeff and {Kasliwal}, Mansi M. and {Keeton}, Charles R. and {Kessler}, Richard and {Knezevic}, Zoran and {Kowalski}, Adam and {Krabbendam}, Victor L. and {Krughoff}, K. Simon and {Kulkarni}, Shrinivas and {Kuhlman}, Stephen and {Lacy}, Mark and {Lepine}, Sebastien and {Liang}, Ming and {Lien}, Amy and {Lira}, Paulina and {Long}, Knox S. and {Lorenz}, Suzanne and {Lotz}, Jennifer M. and {Lupton}, R.~H. and {Lutz}, Julie and {Macri}, Lucas M. and {Mahabal}, Ashish A. and {Mandelbaum}, Rachel and {Marshall}, Phil and {May}, Morgan and {McGehee}, Peregrine M. and {Meadows}, Brian T. and {Meert}, Alan and {Milani}, Andrea and {Miller}, Christopher J. and {Miller}, Michelle and {Mills}, David and {Minniti}, Dante and {Monet}, David and {Mukadam}, Anjum S. and {Nakar}, Ehud and {Neill}, Douglas R. and {Newman}, Jeffrey A. and {Nikolaev}, Sergei and {Nordby}, Martin and {O'Connor}, Paul and {Oguri}, Masamune and {Oliver}, John and {Olivier}, Scot S. and {Olsen}, Julia K. and {Olsen}, Knut and {Olszewski}, Edward W. and {Oluseyi}, Hakeem and {Padilla}, Nelson D. and {Parker}, Alex and {Pepper}, Joshua and {Peterson}, John R. and {Petry}, Catherine and {Pinto}, Philip A. and {Pizagno}, James L. and {Popescu}, Bogdan and {Prsa}, Andrej and {Radcka}, Veljko and {Raddick}, M. Jordan and {Rasmussen}, Andrew and {Rau}, Arne and {Rho}, Jeonghee and {Rhoads}, James E. and {Richards}, Gordon T. and {Ridgway}, Stephen T. and {Robertson}, Brant E. and {Roskar}, Rok and {Saha}, Abhijit and {Sarajedini}, Ata and {Scannapieco}, Evan and {Schalk}, Terry and {Schindler}, Rafe and {Schmidt}, Samuel},
        title = "{LSST Science Book, Version 2.0}",
      journal = {arXiv e-prints},
         year = 2009,
        month = dec,
          eid = {arXiv:0912.0201},
        pages = {arXiv:0912.0201},
          doi = {10.48550/arXiv.0912.0201},
archivePrefix = {arXiv},
       eprint = {0912.0201},
 primaryClass = {astro-ph.IM},
       adsurl = {https://ui.adsabs.harvard.edu/abs/2009arXiv0912.0201L}
}

@ARTICLE{2025A&A...697A...1E,
       author = {{Euclid Collaboration} and {Mellier}, Y. and {Abdurro'uf} and {Acevedo Barroso}, J.~A. and {Ach{\'u}carro}, A. and {Adamek}, J. and {Adam}, R. and {Addison}, G.~E. and {Aghanim}, N. and {Aguena}, M. and {Ajani}, V. and {Akrami}, Y. and {Al-Bahlawan}, A. and {Alavi}, A. and {Albuquerque}, I.~S. and {Alestas}, G. and {Alguero}, G. and {Allaoui}, A. and {Allen}, S.~W. and {Allevato}, V. and {Alonso-Tetilla}, A.~V. and {Altieri}, B. and {Alvarez-Candal}, A. and {Alvi}, S. and {Amara}, A. and {Amendola}, L. and {Amiaux}, J. and {Andika}, I.~T. and {Andreon}, S. and {Andrews}, A. and {Angora}, G. and {Angulo}, R.~E. and {Annibali}, F. and {Anselmi}, A. and {Anselmi}, S. and {Arcari}, S. and {Archidiacono}, M. and {Aric{\`o}}, G. and {Arnaud}, M. and {Arnouts}, S. and {Asgari}, M. and {Asorey}, J. and {Atayde}, L. and {Atek}, H. and {Atrio-Barandela}, F. and {Aubert}, M. and {Aubourg}, E. and {Auphan}, T. and {Auricchio}, N. and {Aussel}, B. and {Aussel}, H. and {Avelino}, P.~P. and {Avgoustidis}, A. and {Avila}, S. and {Awan}, S. and {Azzollini}, R. and {Baccigalupi}, C. and {Bachelet}, E. and {Bacon}, D. and {Baes}, M. and {Bagley}, M.~B. and {Bahr-Kalus}, B. and {Balaguera-Antolinez}, A. and {Balbinot}, E. and {Balcells}, M. and {Baldi}, M. and {Baldry}, I. and {Balestra}, A. and {Ballardini}, M. and {Ballester}, O. and {Balogh}, M. and {Ba{\~n}ados}, E. and {Barbier}, R. and {Bardelli}, S. and {Baron}, M. and {Barreiro}, T. and {Barrena}, R. and {Barriere}, J.-C. and {Barros}, B.~J. and {Barthelemy}, A. and {Bartolo}, N. and {Basset}, A. and {Battaglia}, P. and {Battisti}, A.~J. and {Baugh}, C.~M. and {Baumont}, L. and {Bazzanini}, L. and {Beaulieu}, J.-P. and {Beckmann}, V. and {Belikov}, A.~N. and {Bel}, J. and {Bellagamba}, F. and {Bella}, M. and {Bellini}, E. and {Benabed}, K. and {Bender}, R. and {Benevento}, G. and {Bennett}, C.~L. and {Benson}, K. and {Bergamini}, P. and {Bermejo-Climent}, J.~R. and {Bernardeau}, F. and {Bertacca}, D. and {Berthe}, M. and {Berthier}, J. and {Bethermin}, M. and {Beutler}, F. and {Bevillon}, C. and {Bhargava}, S. and {Bhatawdekar}, R. and {Bianchi}, D. and {Bisigello}, L. and {Biviano}, A. and {Blake}, R.~P. and {Blanchard}, A. and {Blazek}, J. and {Blot}, L. and {Bosco}, A. and {Bodendorf}, C. and {Boenke}, T. and {B{\"o}hringer}, H. and {Boldrini}, P. and {Bolzonella}, M. and {Bonchi}, A. and {Bonici}, M. and {Bonino}, D. and {Bonino}, L. and {Bonvin}, C. and {Bon}, W. and {Booth}, J.~T. and {Borgani}, S. and {Borlaff}, A.~S. and {Borsato}, E. and {Bose}, B. and {Botticella}, M.~T. and {Boucaud}, A. and {Bouche}, F. and {Boucher}, J.~S. and {Boutigny}, D. and {Bouvard}, T. and {Bouwens}, R. and {Bouy}, H. and {Bowler}, R.~A.~A. and {Bozza}, V. and {Bozzo}, E. and {Branchini}, E. and {Brando}, G. and {Brau-Nogue}, S. and {Brekke}, P. and {Bremer}, M.~N. and {Brescia}, M. and {Breton}, M.-A. and {Brinchmann}, J. and {Brinckmann}, T. and {Brockley-Blatt}, C. and {Brodwin}, M. and {Brouard}, L. and {Brown}, M.~L. and {Bruton}, S. and {Bucko}, J. and {Buddelmeijer}, H. and {Buenadicha}, G. and {Buitrago}, F. and {Burger}, P. and {Burigana}, C. and {Busillo}, V. and {Busonero}, D. and {Cabanac}, R. and {Cabayol-Garcia}, L. and {Cagliari}, M.~S. and {Caillat}, A. and {Caillat}, L. and {Calabrese}, M. and {Calabro}, A. and {Calderone}, G. and {Calura}, F. and {Camacho Quevedo}, B. and {Camera}, S. and {Campos}, L. and {Ca{\~n}as-Herrera}, G. and {Candini}, G.~P. and {Cantiello}, M. and {Capobianco}, V. and {Cappellaro}, E. and {Cappelluti}, N. and {Cappi}, A. and {Caputi}, K.~I. and {Cara}, C. and {Carbone}, C. and {Cardone}, V.~F. and {Carella}, E. and {Carlberg}, R.~G. and {Carle}, M. and {Carminati}, L. and {Caro}, F. and {Carrasco}, J.~M. and {Carretero}, J. and {Carrilho}, P. and {Carron Duque}, J. and {Carry}, B.},
        title = "{Euclid: I. Overview of the Euclid mission}",
      journal = {\aap},
         year = 2025,
        month = may,
       volume = {697},
          eid = {A1},
        pages = {A1},
          doi = {10.1051/0004-6361/202450810},
archivePrefix = {arXiv},
       eprint = {2405.13491},
 primaryClass = {astro-ph.CO},
       adsurl = {https://ui.adsabs.harvard.edu/abs/2025A&A...697A...1E}
}

@ARTICLE{1999ApJ...527....1S,
       author = {{Scoccimarro}, Rom{\'a}n and {Zaldarriaga}, Matias and {Hui}, Lam},
        title = "{Power Spectrum Correlations Induced by Nonlinear Clustering}",
      journal = {\apj},
         year = 1999,
        month = dec,
       volume = {527},
       number = {1},
        pages = {1-15},
          doi = {10.1086/308059},
archivePrefix = {arXiv},
       eprint = {astro-ph/9901099},
 primaryClass = {astro-ph},
       adsurl = {https://ui.adsabs.harvard.edu/abs/1999ApJ...527....1S}
}

@ARTICLE{1999MNRAS.308.1179M,
       author = {{Meiksin}, A. and {White}, Martin},
        title = "{The growth of correlations in the matter power spectrum}",
      journal = {\mnras},
         year = 1999,
        month = oct,
       volume = {308},
       number = {4},
        pages = {1179-1184},
          doi = {10.1046/j.1365-8711.1999.02825.x},
archivePrefix = {arXiv},
       eprint = {astro-ph/9812129},
 primaryClass = {astro-ph},
       adsurl = {https://ui.adsabs.harvard.edu/abs/1999MNRAS.308.1179M}
}

@ARTICLE{2018A&A...615A...1L,
       author = {{Lacasa}, Fabien},
        title = "{Covariance of the galaxy angular power spectrum with the halo model}",
      journal = {\aap},
         year = 2018,
        month = jul,
       volume = {615},
          eid = {A1},
        pages = {A1},
          doi = {10.1051/0004-6361/201732343},
archivePrefix = {arXiv},
       eprint = {1711.07372},
 primaryClass = {astro-ph.CO},
       adsurl = {https://ui.adsabs.harvard.edu/abs/2018A&A...615A...1L}
}

@ARTICLE{2002PhR...372....1C,
       author = {{Cooray}, Asantha and {Sheth}, Ravi},
        title = "{Halo models of large scale structure}",
      journal = {\physrep},
         year = 2002,
        month = dec,
       volume = {372},
       number = {1},
        pages = {1-129},
          doi = {10.1016/S0370-1573(02)00276-4},
archivePrefix = {arXiv},
       eprint = {astro-ph/0206508},
 primaryClass = {astro-ph},
       adsurl = {https://ui.adsabs.harvard.edu/abs/2002PhR...372....1C}
}

@ARTICLE{2017JCAP...11..051B,
       author = {{Barreira}, Alexandre and {Schmidt}, Fabian},
        title = "{Response approach to the matter power spectrum covariance}",
      journal = {\jcap},
         year = 2017,
        month = nov,
       volume = {2017},
       number = {11},
          eid = {051},
        pages = {051},
          doi = {10.1088/1475-7516/2017/11/051},
archivePrefix = {arXiv},
       eprint = {1705.01092},
 primaryClass = {astro-ph.CO},
       adsurl = {https://ui.adsabs.harvard.edu/abs/2017JCAP...11..051B}
}

@ARTICLE{2009MNRAS.396...19N,
       author = {{Norberg}, P. and {Baugh}, C.~M. and {Gazta{\~n}aga}, E. and {Croton}, D.~J.},
        title = "{Statistical analysis of galaxy surveys - I. Robust error estimation for two-point clustering statistics}",
      journal = {\mnras},
         year = 2009,
        month = jun,
       volume = {396},
       number = {1},
        pages = {19-38},
          doi = {10.1111/j.1365-2966.2009.14389.x},
archivePrefix = {arXiv},
       eprint = {0810.1885},
 primaryClass = {astro-ph},
       adsurl = {https://ui.adsabs.harvard.edu/abs/2009MNRAS.396...19N}
}

@book{10.5555/515699,
author = {Ross, Sheldon M.},
title = {Simulation},
year = {2002},
isbn = {0125980531},
publisher = {Academic Press, Inc.},
address = {USA},
edition = {3rd}
}

@BOOK{1981csup.book.....H,
       author = {{Hockney}, R.~W. and {Eastwood}, J.~W.},
        title = "{Computer Simulation Using Particles}",
         year = 1981,
       adsurl = {https://ui.adsabs.harvard.edu/abs/1981csup.book.....H}
}

@ARTICLE{2021MNRAS.503.1897C,
       author = {{Chartier}, Nicolas and {Wandelt}, Benjamin and {Akrami}, Yashar and {Villaescusa-Navarro}, Francisco},
        title = "{CARPool: fast, accurate computation of large-scale structure statistics by pairing costly and cheap cosmological simulations}",
      journal = {\mnras},
         year = 2021,
        month = may,
       volume = {503},
       number = {2},
        pages = {1897-1914},
          doi = {10.1093/mnras/stab430},
archivePrefix = {arXiv},
       eprint = {2009.08970},
 primaryClass = {astro-ph.CO},
       adsurl = {https://ui.adsabs.harvard.edu/abs/2021MNRAS.503.1897C}
}

@ARTICLE{2022MNRAS.509.2220C,
       author = {{Chartier}, Nicolas and {Wandelt}, Benjamin D.},
        title = "{CARPool covariance: fast, unbiased covariance estimation for large-scale structure observables}",
      journal = {\mnras},
         year = 2022,
        month = jan,
       volume = {509},
       number = {2},
        pages = {2220-2233},
          doi = {10.1093/mnras/stab3097},
archivePrefix = {arXiv},
       eprint = {2106.11718},
 primaryClass = {astro-ph.CO},
       adsurl = {https://ui.adsabs.harvard.edu/abs/2022MNRAS.509.2220C}
}

@ARTICLE{2022MNRAS.515.1296C,
       author = {{Chartier}, Nicolas and {Wandelt}, Benjamin D.},
        title = "{Bayesian control variates for optimal covariance estimation with pairs of simulations and surrogates}",
      journal = {\mnras},
         year = 2022,
        month = sep,
       volume = {515},
       number = {1},
        pages = {1296-1315},
          doi = {10.1093/mnras/stac1837},
archivePrefix = {arXiv},
       eprint = {2204.03070},
 primaryClass = {astro-ph.CO},
       adsurl = {https://ui.adsabs.harvard.edu/abs/2022MNRAS.515.1296C}
}

@ARTICLE{2022MNRAS.514.1289M,
       author = {{Mohammad}, Faizan G. and {Percival}, Will J.},
        title = "{Creating jackknife and bootstrap estimates of the covariance matrix for the two-point correlation function}",
      journal = {\mnras},
         year = 2022,
        month = jul,
       volume = {514},
       number = {1},
        pages = {1289-1301},
          doi = {10.1093/mnras/stac1458},
archivePrefix = {arXiv},
       eprint = {2109.07071},
 primaryClass = {astro-ph.CO},
       adsurl = {https://ui.adsabs.harvard.edu/abs/2022MNRAS.514.1289M}
}

@ARTICLE{2019MNRAS.482.1786L,
       author = {{Lippich}, Martha and {S{\'a}nchez}, Ariel G. and {Colavincenzo}, Manuel and {Sefusatti}, Emiliano and {Monaco}, Pierluigi and {Blot}, Linda and {Crocce}, Martin and {Alvarez}, Marcelo A. and {Agrawal}, Aniket and {Avila}, Santiago and {Balaguera-Antol{\'\i}nez}, Andr{\'e}s and {Bond}, Richard and {Codis}, Sandrine and {Dalla Vecchia}, Claudio and {Dorta}, Antonio and {Fosalba}, Pablo and {Izard}, Albert and {Kitaura}, Francisco-Shu and {Pellejero-Ibanez}, Marcos and {Stein}, George and {Vakili}, Mohammadjavad and {Yepes}, Gustavo},
        title = "{Comparing approximate methods for mock catalogues and covariance matrices - I. Correlation function}",
      journal = {\mnras},
         year = 2019,
        month = jan,
       volume = {482},
       number = {2},
        pages = {1786-1806},
          doi = {10.1093/mnras/sty2757},
archivePrefix = {arXiv},
       eprint = {1806.09477},
 primaryClass = {astro-ph.CO},
       adsurl = {https://ui.adsabs.harvard.edu/abs/2019MNRAS.482.1786L}
}

@ARTICLE{2019MNRAS.485.2806B,
       author = {{Blot}, Linda and {Crocce}, Martin and {Sefusatti}, Emiliano and {Lippich}, Martha and {S{\'a}nchez}, Ariel G. and {Colavincenzo}, Manuel and {Monaco}, Pierluigi and {Alvarez}, Marcelo A. and {Agrawal}, Aniket and {Avila}, Santiago and {Balaguera-Antol{\'\i}nez}, Andr{\'e}s and {Bond}, Richard and {Codis}, Sandrine and {Dalla Vecchia}, Claudio and {Dorta}, Antonio and {Fosalba}, Pablo and {Izard}, Albert and {Kitaura}, Francisco-Shu and {Pellejero-Ibanez}, Marcos and {Stein}, George and {Vakili}, Mohammadjavad and {Yepes}, Gustavo},
        title = "{Comparing approximate methods for mock catalogues and covariance matrices II: power spectrum multipoles}",
      journal = {\mnras},
         year = 2019,
        month = may,
       volume = {485},
       number = {2},
        pages = {2806-2824},
          doi = {10.1093/mnras/stz507},
archivePrefix = {arXiv},
       eprint = {1806.09497},
 primaryClass = {astro-ph.CO},
       adsurl = {https://ui.adsabs.harvard.edu/abs/2019MNRAS.485.2806B}
}

@ARTICLE{2009ApJ...700..479T,
       author = {{Takahashi}, Ryuichi and {Yoshida}, Naoki and {Takada}, Masahiro and {Matsubara}, Takahiko and {Sugiyama}, Naoshi and {Kayo}, Issha and {Nishizawa}, Atsushi J. and {Nishimichi}, Takahiro and {Saito}, Shun and {Taruya}, Atsushi},
        title = "{Simulations of Baryon Acoustic Oscillations. II. Covariance Matrix of the Matter Power Spectrum}",
      journal = {\apj},
         year = 2009,
        month = jul,
       volume = {700},
       number = {1},
        pages = {479-490},
          doi = {10.1088/0004-637X/700/1/479},
archivePrefix = {arXiv},
       eprint = {0902.0371},
 primaryClass = {astro-ph.CO},
       adsurl = {https://ui.adsabs.harvard.edu/abs/2009ApJ...700..479T}
}

@ARTICLE{2020PhRvD.102l3517W,
       author = {{Wadekar}, Digvijay and {Scoccimarro}, Rom{\'a}n},
        title = "{Galaxy power spectrum multipoles covariance in perturbation theory}",
      journal = {\prd},
         year = 2020,
        month = dec,
       volume = {102},
       number = {12},
          eid = {123517},
        pages = {123517},
          doi = {10.1103/PhysRevD.102.123517},
archivePrefix = {arXiv},
       eprint = {1910.02914},
 primaryClass = {astro-ph.CO},
       adsurl = {https://ui.adsabs.harvard.edu/abs/2020PhRvD.102l3517W}
}

@ARTICLE{2021PhRvD.103b3501T,
       author = {{Taruya}, Atsushi and {Nishimichi}, Takahiro and {Jeong}, Donghui},
        title = "{Covariance of the matter power spectrum including the survey window function effect: N -body simulations versus fifth-order perturbation theory on grids}",
      journal = {\prd},
         year = 2021,
        month = jan,
       volume = {103},
       number = {2},
          eid = {023501},
        pages = {023501},
          doi = {10.1103/PhysRevD.103.023501},
archivePrefix = {arXiv},
       eprint = {2007.05504},
 primaryClass = {astro-ph.CO},
       adsurl = {https://ui.adsabs.harvard.edu/abs/2021PhRvD.103b3501T}
}

@ARTICLE{2013PhRvD..88f3537D,
       author = {{Dodelson}, Scott and {Schneider}, Michael D.},
        title = "{The effect of covariance estimator error on cosmological parameter constraints}",
      journal = {\prd},
         year = 2013,
        month = sep,
       volume = {88},
       number = {6},
          eid = {063537},
        pages = {063537},
          doi = {10.1103/PhysRevD.88.063537},
archivePrefix = {arXiv},
       eprint = {1304.2593},
 primaryClass = {astro-ph.CO},
       adsurl = {https://ui.adsabs.harvard.edu/abs/2013PhRvD..88f3537D}
}

@ARTICLE{2007A&A...464..399H,
       author = {{Hartlap}, J. and {Simon}, P. and {Schneider}, P.},
        title = "{Why your model parameter confidences might be too optimistic. Unbiased estimation of the inverse covariance matrix}",
      journal = {\aap},
         year = 2007,
        month = mar,
       volume = {464},
       number = {1},
        pages = {399-404},
          doi = {10.1051/0004-6361:20066170},
archivePrefix = {arXiv},
       eprint = {astro-ph/0608064},
 primaryClass = {astro-ph},
       adsurl = {https://ui.adsabs.harvard.edu/abs/2007A&A...464..399H}
}

@ARTICLE{2013MNRAS.432.1928T,
       author = {{Taylor}, Andy and {Joachimi}, Benjamin and {Kitching}, Thomas},
        title = "{Putting the precision in precision cosmology: How accurate should your data covariance matrix be?}",
      journal = {\mnras},
         year = 2013,
        month = jul,
       volume = {432},
       number = {3},
        pages = {1928-1946},
          doi = {10.1093/mnras/stt270},
archivePrefix = {arXiv},
       eprint = {1212.4359},
 primaryClass = {astro-ph.CO},
       adsurl = {https://ui.adsabs.harvard.edu/abs/2013MNRAS.432.1928T}
}

@ARTICLE{2014MNRAS.439.2531P,
       author = {{Percival}, Will J. and {Ross}, Ashley J. and {S{\'a}nchez}, Ariel G. and {Samushia}, Lado and {Burden}, Angela and {Crittenden}, Robert and {Cuesta}, Antonio J. and {Magana}, Mariana Vargas and {Manera}, Marc and {Beutler}, Florian and {Chuang}, Chia-Hsun and {Eisenstein}, Daniel J. and {Ho}, Shirley and {McBride}, Cameron K. and {Montesano}, Francesco and {Padmanabhan}, Nikhil and {Reid}, Beth and {Saito}, Shun and {Schneider}, Donald P. and {Seo}, Hee-Jong and {Tojeiro}, Rita and {Weaver}, Benjamin A.},
        title = "{The clustering of Galaxies in the SDSS-III Baryon Oscillation Spectroscopic Survey: including covariance matrix errors}",
      journal = {\mnras},
         year = 2014,
        month = apr,
       volume = {439},
       number = {3},
        pages = {2531-2541},
          doi = {10.1093/mnras/stu112},
archivePrefix = {arXiv},
       eprint = {1312.4841},
 primaryClass = {astro-ph.CO},
       adsurl = {https://ui.adsabs.harvard.edu/abs/2014MNRAS.439.2531P}
}

@ARTICLE{2016MNRAS.456L.132S,
       author = {{Sellentin}, Elena and {Heavens}, Alan F.},
        title = "{Parameter inference with estimated covariance matrices}",
      journal = {\mnras},
         year = 2016,
        month = feb,
       volume = {456},
       number = {1},
        pages = {L132-L136},
          doi = {10.1093/mnrasl/slv190},
archivePrefix = {arXiv},
       eprint = {1511.05969},
 primaryClass = {astro-ph.CO},
       adsurl = {https://ui.adsabs.harvard.edu/abs/2016MNRAS.456L.132S}
}

@ARTICLE{2022JCAP...09..059K,
       author = {{Kokron}, Nickolas and {Chen}, Shi-Fan and {White}, Martin and {DeRose}, Joseph and {Maus}, Mark},
        title = "{Accurate predictions from small boxes: variance suppression via the Zel'dovich approximation}",
      journal = {\jcap},
         year = 2022,
        month = sep,
       volume = {2022},
       number = {9},
          eid = {059},
        pages = {059},
          doi = {10.1088/1475-7516/2022/09/059},
archivePrefix = {arXiv},
       eprint = {2205.15327},
 primaryClass = {astro-ph.CO},
       adsurl = {https://ui.adsabs.harvard.edu/abs/2022JCAP...09..059K}
}

@ARTICLE{2023JCAP...02..008D,
       author = {{DeRose}, Joseph and {Chen}, Shi-Fan and {Kokron}, Nickolas and {White}, Martin},
        title = "{Precision redshift-space galaxy power spectra using Zel'dovich control variates}",
      journal = {\jcap},
         year = 2023,
        month = feb,
       volume = {2023},
       number = {2},
          eid = {008},
        pages = {008},
          doi = {10.1088/1475-7516/2023/02/008},
archivePrefix = {arXiv},
       eprint = {2210.14239},
 primaryClass = {astro-ph.CO},
       adsurl = {https://ui.adsabs.harvard.edu/abs/2023JCAP...02..008D}
}

@ARTICLE{2023OJAp....6E..38H,
       author = {{Hadzhiyska}, Boryana and {White}, Martin J. and {Chen}, Xinyi and {Garrison}, Lehman H. and {DeRose}, Joseph and {Padmanabhan}, Nikhil and {Garcia-Quintero}, Cristhian and {Mena-Fern{\'a}ndez}, Juan and {Chen}, Shi-Fan and {Seo}, Hee-Jong and {McDonald}, Patrick and {Aguilar}, Jessica and {Ahlen}, Steven and {Brooks}, David and {Claybaugh}, Todd and {de la Macorra}, Axel and {Doel}, Peter and {Font-Ribera}, Andreu and {Forero-Romero}, Jaime E. and {Gontcho}, Satya Gontcho A. and {Honscheid}, Klaus and {Kremin}, Anthony and {Landriau}, Martin and {Manera}, Marc and {Miquel}, Ramon and {Nie}, Jundan and {Palanque-Delabrouille}, Nathalie and {Rezaie}, Mehdi and {Rossi}, Graziano and {Sanchez}, Eusebio and {Schubnell}, Michael and {Tarl{\'e}}, Gregoy and {Zhou}, Zhimin},
        title = "{Mitigating the noise of DESI mocks using analytic control variates}",
      journal = {The Open Journal of Astrophysics},
         year = 2023,
        month = oct,
       volume = {6},
          eid = {38},
        pages = {38},
          doi = {10.21105/astro.2308.12343},
archivePrefix = {arXiv},
       eprint = {2308.12343},
 primaryClass = {astro-ph.CO},
       adsurl = {https://ui.adsabs.harvard.edu/abs/2023OJAp....6E..38H}
}

@ARTICLE{2025arXiv251007375K,
       author = {{Kokron}, Nickolas and {Chen}, Shi-Fan},
        title = "{Control variates from Eulerian and Lagrangian perturbation theory: Application to the bispectrum}",
      journal = {arXiv e-prints},
         year = 2025,
        month = oct,
          eid = {arXiv:2510.07375},
        pages = {arXiv:2510.07375},
          doi = {10.48550/arXiv.2510.07375},
archivePrefix = {arXiv},
       eprint = {2510.07375},
 primaryClass = {astro-ph.CO},
       adsurl = {https://ui.adsabs.harvard.edu/abs/2025arXiv251007375K}
}

@ARTICLE{2026JCAP...03..078B,
       author = {{Bartlett}, Alexa and {DeRose}, Joseph and {White}, Martin},
        title = "{Simulation budgeting for hybrid effective field theories}",
      journal = {\jcap},
         year = 2026,
        month = mar,
       volume = {2026},
       number = {3},
          eid = {078},
        pages = {078},
          doi = {10.1088/1475-7516/2026/03/078},
archivePrefix = {arXiv},
       eprint = {2510.13962},
 primaryClass = {astro-ph.CO},
       adsurl = {https://ui.adsabs.harvard.edu/abs/2026JCAP...03..078B}
}

@ARTICLE{2011JCAP...07..034B,
       author = {{Blas}, Diego and {Lesgourgues}, Julien and {Tram}, Thomas},
        title = "{The Cosmic Linear Anisotropy Solving System (CLASS). Part II: Approximation schemes}",
      journal = {\jcap},
         year = 2011,
        month = jul,
       volume = {2011},
       number = {7},
          eid = {034},
        pages = {034},
          doi = {10.1088/1475-7516/2011/07/034},
archivePrefix = {arXiv},
       eprint = {1104.2933},
 primaryClass = {astro-ph.CO},
       adsurl = {https://ui.adsabs.harvard.edu/abs/2011JCAP...07..034B}
}

@ARTICLE{1970A&A.....5...84Z,
       author = {{Zel'dovich}, Ya. B.},
        title = "{Gravitational instability: An approximate theory for large density perturbations.}",
      journal = {\aap},
         year = 1970,
        month = mar,
       volume = {5},
        pages = {84-89},
       adsurl = {https://ui.adsabs.harvard.edu/abs/1970A&A.....5...84Z}
}

@ARTICLE{2003ApJ...595..577Y,
       author = {{Yamamoto}, Kazuhiro},
        title = "{Optimal Weighting Scheme in Redshift-Space Power Spectrum Analysis and a Prospect for Measuring the Cosmic Equation of State}",
      journal = {\apj},
         year = 2003,
        month = oct,
       volume = {595},
       number = {2},
        pages = {577-588},
          doi = {10.1086/377488},
archivePrefix = {arXiv},
       eprint = {astro-ph/0208139},
 primaryClass = {astro-ph},
       adsurl = {https://ui.adsabs.harvard.edu/abs/2003ApJ...595..577Y}
}

@ARTICLE{1991MNRAS.248....1C,
       author = {{Coles}, Peter and {Jones}, Bernard},
        title = "{A lognormal model for the cosmological mass distribution.}",
      journal = {\mnras},
         year = 1991,
        month = jan,
       volume = {248},
        pages = {1-13},
          doi = {10.1093/mnras/248.1.1},
       adsurl = {https://ui.adsabs.harvard.edu/abs/1991MNRAS.248....1C}
}

@ARTICLE{2023AJ....165...58Z,
       author = {{Zhou}, Rongpu and {Dey}, Biprateep and {Newman}, Jeffrey A. and {Eisenstein}, Daniel J. and {Dawson}, K. and {Bailey}, S. and {Berti}, A. and {Guy}, J. and {Lan}, Ting-Wen and {Zou}, H. and {Aguilar}, J. and {Ahlen}, S. and {Alam}, Shadab and {Brooks}, D. and {de la Macorra}, A. and {Dey}, A. and {Dhungana}, G. and {Fanning}, K. and {Font-Ribera}, A. and {Gontcho}, S. Gontcho A. and {Honscheid}, K. and {Ishak}, Mustapha and {Kisner}, T. and {Kov{\'a}cs}, A. and {Kremin}, A. and {Landriau}, M. and {Levi}, Michael E. and {Magneville}, C. and {Manera}, Marc and {Martini}, P. and {Meisner}, Aaron M. and {Miquel}, R. and {Moustakas}, J. and {Myers}, Adam D. and {Nie}, Jundan and {Palanque-Delabrouille}, N. and {Percival}, W.~J. and {Poppett}, C. and {Prada}, F. and {Raichoor}, A. and {Ross}, A.~J. and {Schlafly}, E. and {Schlegel}, D. and {Schubnell}, M. and {Tarl{\'e}}, Gregory and {Weaver}, B.~A. and {Wechsler}, R.~H. and {Y{\'e}che}, Christophe and {Zhou}, Zhimin},
        title = "{Target Selection and Validation of DESI Luminous Red Galaxies}",
      journal = {\aj},
         year = 2023,
        month = feb,
       volume = {165},
       number = {2},
          eid = {58},
        pages = {58},
          doi = {10.3847/1538-3881/aca5fb},
archivePrefix = {arXiv},
       eprint = {2208.08515},
 primaryClass = {astro-ph.CO},
       adsurl = {https://ui.adsabs.harvard.edu/abs/2023AJ....165...58Z}
}

@ARTICLE{2023JCAP...11..097Z,
       author = {{Zhou}, Rongpu and {Ferraro}, Simone and {White}, Martin and {DeRose}, Joseph and {Sailer}, Noah and {Aguilar}, Jessica and {Ahlen}, Steven and {Bailey}, Stephen and {Brooks}, David and {Claybaugh}, Todd and {Dawson}, Kyle and {de la Macorra}, Axel and {Dey}, Biprateep and {Doel}, Peter and {Font-Ribera}, Andreu and {Forero-Romero}, Jaime E. and {Gontcho A Gontcho}, Satya and {Guy}, Julien and {Kremin}, Anthony and {Lambert}, Andrew and {Le Guillou}, Laurent and {Levi}, Michael and {Magneville}, Christophe and {Manera}, Marc and {Meisner}, Aaron and {Miquel}, Ramon and {Moustakas}, John and {Myers}, Adam D. and {Newman}, Jeffrey A. and {Nie}, Jundan and {Percival}, Will and {Rezaie}, Mehdi and {Rossi}, Graziano and {Sanchez}, Eusebio and {Schlegel}, David and {Schubnell}, Michael and {Seo}, Hee-Jong and {Tarl{\'e}}, Gregory and {Zhou}, Zhimin},
        title = "{DESI luminous red galaxy samples for cross-correlations}",
      journal = {\jcap},
         year = 2023,
        month = nov,
       volume = {2023},
       number = {11},
          eid = {097},
        pages = {097},
          doi = {10.1088/1475-7516/2023/11/097},
archivePrefix = {arXiv},
       eprint = {2309.06443},
 primaryClass = {astro-ph.CO},
       adsurl = {https://ui.adsabs.harvard.edu/abs/2023JCAP...11..097Z}
}

@ARTICLE{1994ApJ...426...23F,
       author = {{Feldman}, Hume A. and {Kaiser}, Nick and {Peacock}, John A.},
        title = "{Power-Spectrum Analysis of Three-dimensional Redshift Surveys}",
      journal = {\apj},
         year = 1994,
        month = may,
       volume = {426},
        pages = {23},
          doi = {10.1086/174036},
archivePrefix = {arXiv},
       eprint = {astro-ph/9304022},
 primaryClass = {astro-ph},
       adsurl = {https://ui.adsabs.harvard.edu/abs/1994ApJ...426...23F}
}

@ARTICLE{2008MNRAS.389..766P,
       author = {{Pope}, Adrian C. and {Szapudi}, Istv{\'a}n},
        title = "{Shrinkage estimation of the power spectrum covariance matrix}",
      journal = {\mnras},
         year = 2008,
        month = sep,
       volume = {389},
       number = {2},
        pages = {766-774},
          doi = {10.1111/j.1365-2966.2008.13561.x},
archivePrefix = {arXiv},
       eprint = {0711.2509},
 primaryClass = {astro-ph},
       adsurl = {https://ui.adsabs.harvard.edu/abs/2008MNRAS.389..766P}
}

@ARTICLE{2020PhRvD.102l3521W,
       author = {{Wadekar}, Digvijay and {Ivanov}, Mikhail M. and {Scoccimarro}, Roman},
        title = "{Cosmological constraints from BOSS with analytic covariance matrices}",
      journal = {\prd},
         year = 2020,
        month = dec,
       volume = {102},
       number = {12},
          eid = {123521},
        pages = {123521},
          doi = {10.1103/PhysRevD.102.123521},
archivePrefix = {arXiv},
       eprint = {2009.00622},
 primaryClass = {astro-ph.CO},
       adsurl = {https://ui.adsabs.harvard.edu/abs/2020PhRvD.102l3521W}
}

@ARTICLE{2015MNRAS.454.4326P,
       author = {{Paz}, Dante J. and {S{\'a}nchez}, Ariel G.},
        title = "{Improving the precision matrix for precision cosmology}",
      journal = {\mnras},
         year = 2015,
        month = dec,
       volume = {454},
       number = {4},
        pages = {4326-4334},
          doi = {10.1093/mnras/stv2259},
archivePrefix = {arXiv},
       eprint = {1508.03162},
 primaryClass = {astro-ph.CO},
       adsurl = {https://ui.adsabs.harvard.edu/abs/2015MNRAS.454.4326P}
}

@ARTICLE{2016MNRAS.460.1567P,
       author = {{Padmanabhan}, Nikhil and {White}, Martin and {Zhou}, Harrison H. and {O'Connell}, Ross},
        title = "{Estimating sparse precision matrices}",
      journal = {\mnras},
         year = 2016,
        month = aug,
       volume = {460},
       number = {2},
        pages = {1567-1576},
          doi = {10.1093/mnras/stw1042},
archivePrefix = {arXiv},
       eprint = {1512.01241},
 primaryClass = {astro-ph.IM},
       adsurl = {https://ui.adsabs.harvard.edu/abs/2016MNRAS.460.1567P}
}

@ARTICLE{2017MNRAS.466L..83J,
       author = {{Joachimi}, Benjamin},
        title = "{Non-linear shrinkage estimation of large-scale structure covariance}",
      journal = {\mnras},
         year = 2017,
        month = mar,
       volume = {466},
       number = {1},
        pages = {L83-L87},
          doi = {10.1093/mnrasl/slw240},
archivePrefix = {arXiv},
       eprint = {1612.00752},
 primaryClass = {astro-ph.IM},
       adsurl = {https://ui.adsabs.harvard.edu/abs/2017MNRAS.466L..83J}
}

@ARTICLE{2026arXiv260618446R,
       author = {{Raffaele}, Mattera},
        title = "{Covariance shrinkage for cosmological inference with Sellentin-Heavens-type likelihoods}",
      journal = {arXiv e-prints},
         year = 2026,
        month = jun,
          eid = {arXiv:2606.18446},
        pages = {arXiv:2606.18446},
archivePrefix = {arXiv},
       eprint = {2606.18446},
 primaryClass = {astro-ph.CO},
       adsurl = {https://ui.adsabs.harvard.edu/abs/2026arXiv260618446R}
}

@ARTICLE{2026arXiv260413851F,
       author = {{Farina}, Antonio and {Guidi}, Massimo and {Veropalumbo}, Alfonso and {Guida}, Claudio},
        title = "{Denoising clustering covariance matrices with Rotational Invariant Estimators}",
      journal = {arXiv e-prints},
         year = 2026,
        month = apr,
          eid = {arXiv:2604.13851},
        pages = {arXiv:2604.13851},
          doi = {10.48550/arXiv.2604.13851},
archivePrefix = {arXiv},
       eprint = {2604.13851},
 primaryClass = {astro-ph.CO},
       adsurl = {https://ui.adsabs.harvard.edu/abs/2026arXiv260413851F}
}

@ARTICLE{2019MNRAS.483..189H,
       author = {{Hall}, Alex and {Taylor}, Andy},
        title = "{A Bayesian method for combining theoretical and simulated covariance matrices for large-scale structure surveys}",
      journal = {\mnras},
         year = 2019,
        month = feb,
       volume = {483},
       number = {1},
        pages = {189-207},
          doi = {10.1093/mnras/sty3102},
archivePrefix = {arXiv},
       eprint = {1807.06875},
 primaryClass = {astro-ph.CO},
       adsurl = {https://ui.adsabs.harvard.edu/abs/2019MNRAS.483..189H}
}

@ARTICLE{2014PASJ...66R...1T,
       author = {{Takada}, Masahiro and {Ellis}, Richard S. and {Chiba}, Masashi and {Greene}, Jenny E. and {Aihara}, Hiroaki and {Arimoto}, Nobuo and {Bundy}, Kevin and {Cohen}, Judith and {Dor{\'e}}, Olivier and {Graves}, Genevieve and {Gunn}, James E. and {Heckman}, Timothy and {Hirata}, Christopher M. and {Ho}, Paul and {Kneib}, Jean-Paul and {Le F{\`e}vre}, Olivier and {Lin}, Lihwai and {More}, Surhud and {Murayama}, Hitoshi and {Nagao}, Tohru and {Ouchi}, Masami and {Seiffert}, Michael and {Silverman}, John D. and {Sodr{\'e}}, Laerte and {Spergel}, David N. and {Strauss}, Michael A. and {Sugai}, Hajime and {Suto}, Yasushi and {Takami}, Hideki and {Wyse}, Rosemary},
        title = "{Extragalactic science, cosmology, and Galactic archaeology with the Subaru Prime Focus Spectrograph}",
      journal = {\pasj},
         year = 2014,
        month = feb,
       volume = {66},
       number = {1},
          eid = {R1},
        pages = {R1},
          doi = {10.1093/pasj/pst019},
archivePrefix = {arXiv},
       eprint = {1206.0737},
 primaryClass = {astro-ph.CO},
       adsurl = {https://ui.adsabs.harvard.edu/abs/2014PASJ...66R...1T}
}

@ARTICLE{2014arXiv1412.4872D,
       author = {{Dor{\'e}}, Olivier and {Bock}, Jamie and {Ashby}, Matthew and {Capak}, Peter and {Cooray}, Asantha and {de Putter}, Roland and {Eifler}, Tim and {Flagey}, Nicolas and {Gong}, Yan and {Habib}, Salman and {Heitmann}, Katrin and {Hirata}, Chris and {Jeong}, Woong-Seob and {Katti}, Raj and {Korngut}, Phil and {Krause}, Elisabeth and {Lee}, Dae-Hee and {Masters}, Daniel and {Mauskopf}, Phil and {Melnick}, Gary and {Mennesson}, Bertrand and {Nguyen}, Hien and {{\"O}berg}, Karin and {Pullen}, Anthony and {Raccanelli}, Alvise and {Smith}, Roger and {Song}, Yong-Seon and {Tolls}, Volker and {Unwin}, Steve and {Venumadhav}, Tejaswi and {Viero}, Marco and {Werner}, Mike and {Zemcov}, Mike},
        title = "{Cosmology with the SPHEREX All-Sky Spectral Survey}",
      journal = {arXiv e-prints},
         year = 2014,
        month = dec,
          eid = {arXiv:1412.4872},
        pages = {arXiv:1412.4872},
          doi = {10.48550/arXiv.1412.4872},
archivePrefix = {arXiv},
       eprint = {1412.4872},
 primaryClass = {astro-ph.CO},
       adsurl = {https://ui.adsabs.harvard.edu/abs/2014arXiv1412.4872D}
}

@ARTICLE{2018MNRAS.473.4150F,
       author = {{Friedrich}, Oliver and {Eifler}, Tim},
        title = "{Precision matrix expansion - efficient use of numerical simulations in estimating errors on cosmological parameters}",
      journal = {\mnras},
         year = 2018,
        month = jan,
       volume = {473},
       number = {3},
        pages = {4150-4163},
          doi = {10.1093/mnras/stx2566},
archivePrefix = {arXiv},
       eprint = {1703.07786},
 primaryClass = {astro-ph.IM},
       adsurl = {https://ui.adsabs.harvard.edu/abs/2018MNRAS.473.4150F}
}

@book{bhatia2009,
  title={Positive Definite Matrices},
  author={Bhatia, R.},
  isbn={9781400827787},
  lccn={2006050375},
  series={Princeton Series in Applied Mathematics},
  url={https://books.google.com/books?id=-KIFglY18nYC},
  year={2009},
  publisher={Princeton University Press}
}

@book{horn2012,
  title     = {Matrix Analysis},
  author    = {Horn, Roger A. and Johnson, Charles R.},
  edition   = {2nd},
  year      = {2012},
  publisher = {Cambridge University Press},
  address   = {Cambridge, UK},
  isbn      = {9780521839402}
}

@ARTICLE{1995A&A...296..575B,
       author = {{Bouchet}, F.~R. and {Colombi}, S. and {Hivon}, E. and {Juszkiewicz}, R.},
        title = "{Perturbative Lagrangian approach to gravitational instability.}",
      journal = {\aap},
         year = 1995,
        month = apr,
       volume = {296},
        pages = {575},
          doi = {10.48550/arXiv.astro-ph/9406013},
archivePrefix = {arXiv},
       eprint = {astro-ph/9406013},
 primaryClass = {astro-ph},
       adsurl = {https://ui.adsabs.harvard.edu/abs/1995A&A...296..575B}
}

@ARTICLE{1998MNRAS.299.1097S,
       author = {{Scoccimarro}, Roman},
        title = "{Transients from initial conditions: a perturbative analysis}",
      journal = {\mnras},
         year = 1998,
        month = oct,
       volume = {299},
       number = {4},
        pages = {1097-1118},
          doi = {10.1046/j.1365-8711.1998.01845.x},
archivePrefix = {arXiv},
       eprint = {astro-ph/9711187},
 primaryClass = {astro-ph},
       adsurl = {https://ui.adsabs.harvard.edu/abs/1998MNRAS.299.1097S}
}

@ARTICLE{2006MNRAS.373..369C,
       author = {{Crocce}, Mart{\'\i}n and {Pueblas}, Sebasti{\'a}n and {Scoccimarro}, Rom{\'a}n},
        title = "{Transients from initial conditions in cosmological simulations}",
      journal = {\mnras},
         year = 2006,
        month = nov,
       volume = {373},
       number = {1},
        pages = {369-381},
          doi = {10.1111/j.1365-2966.2006.11040.x},
archivePrefix = {arXiv},
       eprint = {astro-ph/0606505},
 primaryClass = {astro-ph},
       adsurl = {https://ui.adsabs.harvard.edu/abs/2006MNRAS.373..369C}
}

@ARTICLE{2022MNRAS.510.3207P,
       author = {{Percival}, Will J. and {Friedrich}, Oliver and {Sellentin}, Elena and {Heavens}, Alan},
        title = "{Matching Bayesian and frequentist coverage probabilities when using an approximate data covariance matrix}",
      journal = {\mnras},
         year = 2022,
        month = mar,
       volume = {510},
       number = {3},
        pages = {3207-3221},
          doi = {10.1093/mnras/stab3540},
archivePrefix = {arXiv},
       eprint = {2108.10402},
 primaryClass = {astro-ph.IM},
       adsurl = {https://ui.adsabs.harvard.edu/abs/2022MNRAS.510.3207P}
}

@ARTICLE{2013PhRvD..87l3504T,
       author = {{Takada}, Masahiro and {Hu}, Wayne},
        title = "{Power spectrum super-sample covariance}",
      journal = {\prd},
         year = 2013,
        month = jun,
       volume = {87},
       number = {12},
          eid = {123504},
        pages = {123504},
          doi = {10.1103/PhysRevD.87.123504},
archivePrefix = {arXiv},
       eprint = {1302.6994},
 primaryClass = {astro-ph.CO},
       adsurl = {https://ui.adsabs.harvard.edu/abs/2013PhRvD..87l3504T}
}
\end{document}